\documentclass[twocolumn,trackchanges]{aastex701}

\usepackage{comment}

\begin{document}

\title{FEAST: Spatially Resolved Star Formation History of the NGC~4485/NGC~4490 System and its Tidal Bridge from JWST/NIRCam Color–Magnitude Diagrams}

\correspondingauthor{Giacomo Bortolini}
\author[0009-0003-6182-8928]{Giacomo Bortolini}
\affiliation{Department of Astronomy, The Oskar Klein Centre, Stockholm University, AlbaNova, SE-10691 Stockholm, Sweden}
\email[show]{giacomo.bortolini@astro.su.se}  

\author[0000-0001-6464-3257]{Matteo Correnti}
\affiliation{INAF Osservatorio Astronomico di Roma, Via Frascati 33, 00078, Monteporzio Catone, Rome, Italy}
\affiliation{ASI-Space Science Data Center, Via del Politecnico, I-00133, Rome, Italy}
\email[]{matteo.correnti@inaf.it}

\author[0000-0002-8192-8091]{Angela Adamo}
\affiliation{Department of Astronomy, The Oskar Klein Centre, Stockholm University, AlbaNova, SE-10691 Stockholm, Sweden}
\email[]{angela.adamo@astro.su.se}

\author[0000-0001-6291-6813]{Michele Cignoni}
\affiliation{Dipartimento di Fisica, Università di Pisa, Largo Bruno Pontecorvo 3, 56127, Pisa, Italy}
\affiliation{INAF - Osservatorio di Astrofisica e Scienza dello Spazio di Bologna, via Piero Gobetti 93/3, 40129 Bologna, Italy}
\affiliation{INFN, Largo B. Pontecorvo 3, 56127, Pisa, Italy}
\email[]{michele.cignoni@unipi.it}

\author[0000-0002-0986-4759]{Monica Tosi}
\affiliation{INAF - Osservatorio di Astrofisica e Scienza dello Spazio di Bologna, via Piero Gobetti 93/3, 40129 Bologna, Italy}
\email[]{monica.tosi@inaf.it}

\author[0009-0003-6182-8928]{Göran Östlin}
\affiliation{Department of Astronomy, The Oskar Klein Centre, Stockholm University, AlbaNova, SE-10691 Stockholm, Sweden}
\email[]{ostlin@astro.su.se}

\author[0000-0001-8068-0891]{Arjan Bik}
\affiliation{Department of Astronomy, The Oskar Klein Centre, Stockholm University, AlbaNova, SE-10691 Stockholm, Sweden}
\email[]{arjan.bik@astro.su.se}

\author[0000-0002-5189-8004]{Daniela Calzetti}
\affiliation{Department of Astronomy, University of Massachusetts, 710 North Pleasant Street, Amherst, MA 01003, USA}
\email[]{calzetti@astro.umass.edu}

\author[0000-0002-1821-7019]{John M. Cannon}
\affiliation{Macalester College, 1600 Grand Avenue, Saint Paul, MN 55105, USA}
\email[]{jcannon@macalester.edu}

\author[0000-0003-2442-6981]{Flavia Dell'Agli}
\affiliation{INAF Osservatorio Astronomico di Roma, Via Frascati 33, 00078, Monteporzio Catone, Rome, Italy}
\email[]{flavia.dellagli@inaf.it}

\author[0000-0002-1723-6330]{Bruce G. Elmegreen}
\affiliation{Katonah, NY 10536}
\email[]{belmegreen@gmail.com}

\author[0000-0002-2199-0977]{Helena Faustino Vieira}
\affiliation{Department of Astronomy, The Oskar Klein Centre, Stockholm University, AlbaNova, SE-10691 Stockholm, Sweden}
\email[]{helena.faustinovieira@astro.su.se}

\author[0000-0002-3247-5321]{Kathryn~Grasha}
\altaffiliation{ARC DECRA Fellow}
\affiliation{Research School of Astronomy and Astrophysics, Australian National University, Canberra, ACT 2611, Australia}
\affiliation{ARC Centre of Excellence for All Sky Astrophysics in 3 Dimensions (ASTRO 3D), Australia}
\email[]{Kathryn.Grasha@anu.edu.au}

\author[0000-0001-8348-2671]{Kelsey E. Johnson}
\affiliation{Department of Astronomy, University of Virginia, Charlottesville, VA, USA}
\email[]{kej7a@virginia.edu}

\author[0000-0001-8490-6632]{Thomas S.-Y. Lai}
\affil{IPAC, California Institute of Technology, 1200 E. California Blvd., Pasadena, CA 91125, USA}
\email[]{ThomasLai.astro@gmail.com}

\author[0009-0009-5509-4706]{Drew Lapeer}
\affiliation{Department of Astronomy, University of Massachusetts, 710 North Pleasant Street, Amherst, MA 01003, USA}
\email{dlapeer@umass.edu}

\author[0000-0002-1000-6081]{Sean T. Linden}
\affiliation{Steward Observatory, University of Arizona, 933 N. Cherry Avenue, Tucson, AZ 85719, USA}
\email{stlsps@rit.edu}

\author[0000-0003-1427-2456]{Matteo Messa}
\affiliation{INAF - Osservatorio di Astrofisica e Scienza dello Spazio di Bologna, via Piero Gobetti 93/3, 40129 Bologna, Italy}
\email[]{matteo.messa@inaf.it}

\author[0000-0002-8222-8986]{Alex Pedrini}
\affiliation{Department of Astronomy, The Oskar Klein Centre, Stockholm University, AlbaNova, SE-10691 Stockholm, Sweden}
\email[]{alex.pedrini@astro.su.se}

\author[0000-0001-6326-7069]{Julia Roman-Duval}
\affiliation{Space Telescope Science Institute, 3700 San Martin Drive, Baltimore, MD 21218, USA}
\email[]{duval@stsci.edu}

\author[0000-0003-2954-7643]{Elena Sabbi}
\affiliation{International Space Science Institute, Hallerstrasse 6, 3012 Bern, Switzerland}
\affiliation{Physical Institute, University of Bern, Sidlerstrasse 5, 3012, Bern, Switzerland}
\email[]{elena.sabbi@issibern.ch}

\author[0000-0002-0806-168X]{Linda J. Smith}
\affiliation{Space Telescope Science Institute, 3700 San Martin Drive, Baltimore, MD 21218, USA}
\email[]{lsmith@stsci.edu}

\begin{abstract}

We present the spatially resolved star formation history (SFH) of the interacting, low-mass, gas-rich galaxy pair NGC~4485/NGC~4490, based on deep JWST/NIRCam imaging obtained as part of the FEAST program. Using resolved stars' near-infrared color-magnitude diagrams (CMDs) and the \texttt{SFERA 2.0} CMD fitting code, we reconstruct the SFH of the system, including the tidal bridge connecting the two galaxies. We find that recent star formation is widespread across the system, extending beyond the main bodies of the galaxies into the tidal bridge and other off-center areas. In particular, a significant fraction of the recent star formation in NGC~4485 occurs outside its central region. All major regions of the system exhibit two prominent episodes of enhanced star formation, occurring around $0.5\mbox{--}1$ Gyr and $10\mbox{--}30$ Myr ago, with the tidal bridge showing the largest increase relative to its average star formation rate. These peaks in star formation activity occur a few hundred million years after the epochs of the first and second pericenter passages predicted by independent dynamical models, supporting a scenario in which repeated tidal interactions triggered the observed bursts through gas redistribution and compression. Our results place direct observational constraints on the formation history of NGC~4485/NGC~4490 and provide evidence for the important role of galaxy interactions in driving spatially extended, bursty star formation in low-mass galaxies.

\end{abstract}



\section{Introduction}
\label{sec:intro}
Understanding how galaxy interactions and mergers regulate the conversion of gas into stars remains a fundamental challenge in the field of galaxy formation and evolution. Within the hierarchical framework of the Lambda-cold dark matter ($\Lambda$CDM) cosmological model, galaxies are predicted to grow through accretion of smaller satellites \citep{White1978,Peebles1982,White1991,Bullock2005,Bullock2017}. Interactions and mergers are therefore expected to play a key role in shaping galaxy evolution across a wide range of masses. They can alter galaxy morphology and kinematics, redistribute angular momentum, drive gas inflows, increase gas densities, and trigger enhanced episodes of star formation \citep[see e.g.,][among others]{Barnes1991,Barnes1996,Ostlin2001,Li2008,Bournaud2011,Ellison2013,Pearson2016,Paudel2018,Karachentsev2018,Thorp2019,Pan2019,Bortolini2024b}. While the impact of interactions and mergers has been extensively studied in the Milky Way and other massive galaxies \citep{Ibata2001,Belokurov2006,McConnachie2009,Martinez-Delgado2010,Crnojevic2016,Merritt2016}, the role of these processes in the low-mass dwarf galaxy regime ($\mathrm{M_{\star} \lesssim 10^{9} \, M_{\odot}}$) has historically been investigated primarily through individual systems \citep[see e.g.,][]{Ostlin2001}, while only in the past decade have dedicated observational surveys enabled a systematic exploration of interacting dwarf galaxy populations \citep[see, e.g.,][]{Stierwalt2015,Carlin2016,Higgs2016,Stierwalt2017,Paudel2018,Annibali2020,Sacchi2024,Correnti2025b,Stierwalt2026,Li2026}.

Observations of isolated low-mass galaxy pairs have revealed systematically elevated star formation rates relative to those of isolated dwarfs, with the enhancement becoming stronger as the separation between the galaxies decreases \citep{Stierwalt2015}. Unlike mergers involving massive galaxies, where star formation is often concentrated in the nuclear regions as tidal torques funnel gas toward the center \citep{Armus1989,Mihos1994,Barnes1996,Ellison2008,Hopkins2008,Hopkins2013,Thorp2019,Hani2020}, interacting dwarf galaxies frequently exhibit spatially extended and clumpy star formation distributed throughout their disks and tidal structures \citep{Stierwalt2015,Stierwalt2017,Privon2017,Paudel2018,Mandal2024,correnti2025,Bortolini2025}. This suggests that the response of the interstellar medium to tidal interactions depends strongly on galaxy mass, with low-mass systems being highly susceptible to efficient gas stripping and structural transformation, while higher-mass galaxies predominantly exhibit gas redistribution and tidally induced morphological features rather than large-scale gas removal \citep{Fattahi2018,Pettitt2018,Henkel2022}.


Recent hydrodynamical simulations further indicate that mergers between gas-rich, low-mass galaxies can produce the high gas densities and pressures required to trigger elevated star formation efficiencies, and the formation of young massive star clusters \citep[see e.g.,][]{Bekki2008,Ejdetjarn2025,Lahen2019,Lahen2020}.  \citet{Lahen2020} found that during the interaction between two gas-rich galaxies with the same stellar masses ($\mathrm{M_{\star} \sim 10^{7} \, M_{\odot}}$), the total star formation rate peaks shortly after each pericenter passage (see Figure 5 in \citealt{Lahen2020}). These bursts of star formation are predicted to be several times stronger than the average SFR, enough to leave strong and observable traces in the stellar populations of such systems.


In this scenario, interacting low-mass galaxies in the nearby Universe (i.e., within $\sim20$~Mpc from us) provide a unique opportunity to investigate these processes at high spatial resolutions. At such distances, individual stars and star clusters can be resolved with modern space-based facilities \citep{Tolstoy2009,McQuinn2010,Weisz2014,Annibali2022,Weisz2024,Smercina2025}, enabling the reconstruction of spatially resolved star formation histories (SFHs) from deep color--magnitude diagrams (CMDs, see \citealt{Cignoni2010} and references therein). Because resolved stellar populations retain a fossil record of the past star formation activity, they provide a direct means of probing how interactions influence star formation across the entire merger history. These nearby systems therefore serve as fundamental laboratories for understanding galaxy assembly in the early Universe, when interactions and mergers were more common and hierarchical growth proceeded more rapidly \citep{Conselice2014}, but where comparable spatial resolution cannot be achieved.

\begin{figure*}
    \centering
    \includegraphics[width=1.0\linewidth]{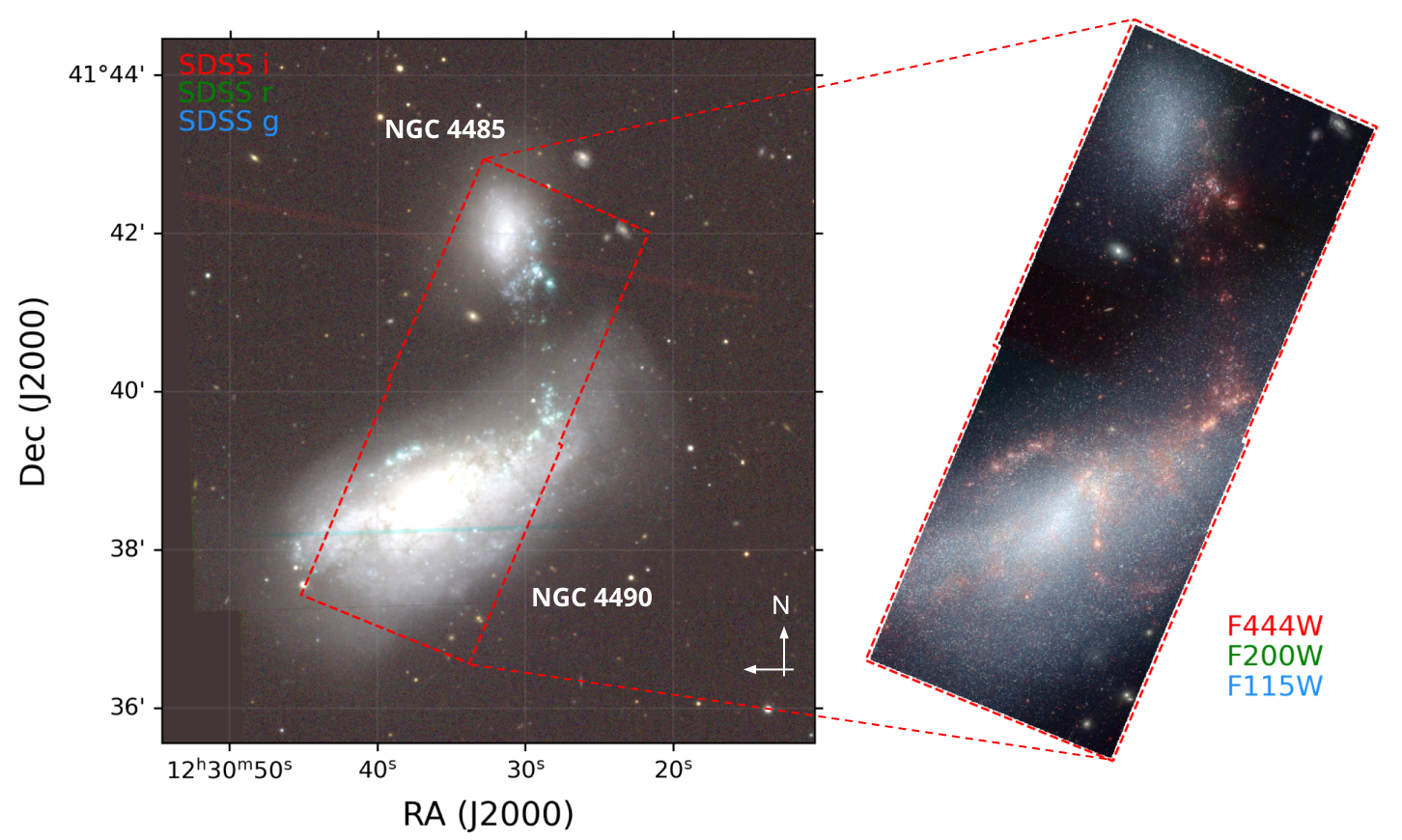}
    \caption{Left panel: SDSS \citep{SDSS2000} optical and infrared RGB image of NGC~4485 and NGC~4490 (red channel = i, green channel = r, blue channel = g). NGC~4485 and NGC~4490 are labeled. The red-dashed box marks the field of view of FEAST JWST/NIRCam observations. North is up, East to the left. Right panel: FEAST JWST/NIRCam color composite image of NGC~4485/NGC~4490 in the F444W (red channel) tracing the hot dust associated with the ISM, F200W (green channel), and F115W (blue channel) filters tracing the stellar continuum. Again North is up, East to the left.}
    \label{fig:SDSS_plus_FEAST_images}
\end{figure*}

One of the best nearby laboratories for studying low-mass, gas-rich galaxy interaction is the NGC~4485/4490 system (Arp 269). NGC~4485 and NGC~4490 form a pair of interacting galaxies with stellar masses of $\sim3.2\times10^8 \,\mathrm{M_\odot}$ and $\sim2.5\times10^9 \,\mathrm{M_\odot}$, respectively. The system is located at a distance of $\sim7.4$~Mpc \citep{Calzetti2015,Sabbi2018}, and the galaxy centers are separated by $\sim7.5$~kpc in projection. They exhibit a prograde motion and a velocity difference of approximately $30 \, \mathrm{km \, s^{-1}}$ \citep{Pearson2018}. NGC~4490 has an oxygen abundance of $\mathrm{12+\log(O/H)=8.37}$ derived from H{\small II} regions \citep{Pilyugin2007}. The pair is also relatively isolated, with the nearest massive galaxy, NGC~4369, located at a projected distance of $\sim300$~kpc and differing in radial velocity by more than $400 \, \mathrm{km \, s^{-1}}$ \citep{Pearson2016,Pearson2018}. These properties make NGC~4485/NGC~4490 particularly interesting, as it is one of the nearby systems in which individual stars can be resolved across a tidally induced, actively star-forming bridge (\citealt{Bortolini2025}, see also Figure~\ref{fig:SDSS_plus_FEAST_images}). Indeed, although a similar bridge can be studied in greater detail between the Large and Small Magellanic Clouds, disentangling the effects of their mutual interaction from those induced by the gravitational influence of the Milky Way remains challenging \citep{Bekki2005,Besla2012,Ficara2026,Burhenne2026}.

\begin{figure*}
    \centering
    \includegraphics[width=0.95\linewidth]{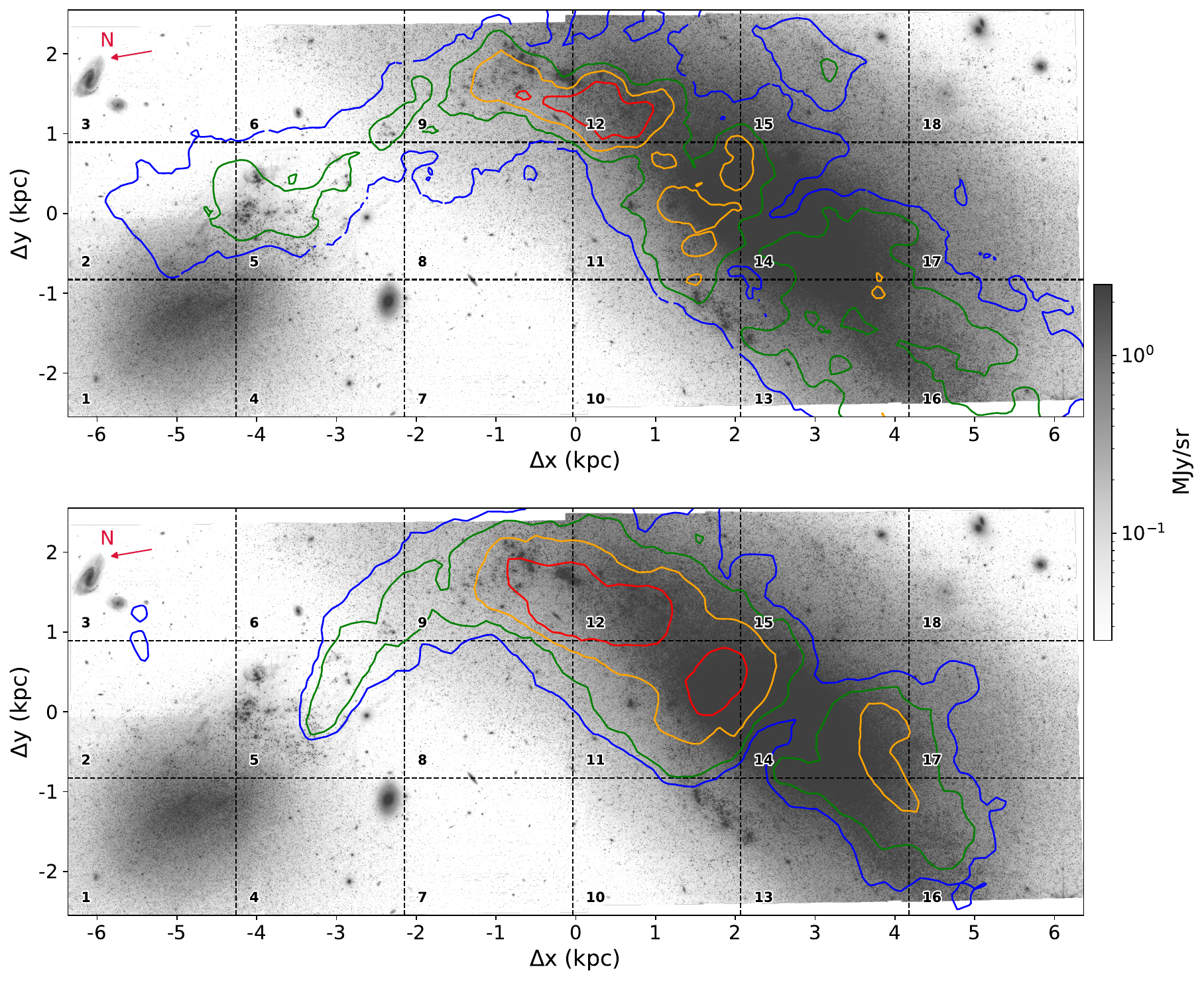}
    \caption{FEAST JWST/NIRCam F200W mosaic of NGC~4485 and NGC~4490 with VLA deep B-configuration H I contours (Cannon et al. in prep.) at column densities of $[2.0, 4.5, 7.0, 9.5] \times 10^{21} \, \mathrm{cm}^{-2}$ (top panel), and IRAM 30 m CO (1–0) contours (S. Linden, private communication) at integrated intensities of $[500, 1000, 2000, 3500] \,\mathrm{K\,km\,s^{-1}}$ (bottom panel). The field of view is divided into 18 equally sized tiles by the black dashed lines, each measuring $\sim60'' \times 48''$ (or $\sim2.1 \times 1.7$ kpc), within which we independently performed the PSF photometry and artificial star tests using DOLPHOT. The reference number of each tile is reported in its bottom-left corner. We adopt the same tile-naming convention as in \citet{Bortolini2025}. The red arrow indicates North.} 
    \label{fig:F200W_image_tiles}
\end{figure*}

The tidal bridge connecting NGC~4485 and NGC~4490 exhibits a non-uniform morphology (see Figures \ref{fig:SDSS_plus_FEAST_images} and \ref{fig:F200W_image_tiles}): star formation is strongly enhanced near the two parent galaxies \citep{Roberts2002,Smith2010,Mandal2024,Bortolini2025}, while the central region contains abundant atomic and molecular gas (J. Cannon, S. Linden, private communication), but comparatively little recent star formation. Some studies have suggested that NGC~4490 is a relatively young galaxy, approximately $2$~Gyr old, that has sustained nearly continuous star formation over at least the past $100$~Myr \citep{Elmegreen1998,Clemens1999,Clemens2000}. In this context, N-body and test-particle simulations by \citet{Pearson2018} provide further insight into the dynamical history of the system, suggesting two pericenter passages occurring approximately $1.3$~Gyr and $200$~Myr ago. If this scenario is correct, such encounters should have left observable signatures in the SFHs of both galaxies and their tidal bridge \citep{Lahen2019,Lahen2020,Ejdetjarn2025}. 

In \citet{Bortolini2025}, we presented new JWST/NIRCam observations of  NGC~4485/NGC~4490, obtained as part of the Cycle~1 program \textit{Feedback in Emerging extrAgalactic Star clusTers} (FEAST; PI: A. Adamo). Leveraging the unprecedented sensitivity and spatial resolution of JWST in the near-infrared, we characterized the resolved stellar populations of NGC~4485, NGC~4490, and their tidal bridge down to several magnitudes below the tip of the red giant branch. Building on that work, in this follow-up study we derive the spatially resolved SFHs of the system using the JWST/NIRCam CMDs and our CMD-fitting code \texttt{SFERA 2.0} \citep{Bortolini2024a}. Our main goal is to determine whether distinct bursts of star formation can be recovered and associated with the interaction history of the system, and to investigate how star formation has proceeded across the galaxies and their tidal bridge over cosmic time.

The paper is organized as follows. In Section~\ref{sec:data}, we describe the observations, data reduction, and point-spread function (PSF) fitting photometry. The observed CMDs and main stellar evolutionary phases are presented in Section~\ref{sec: resolved stellar pops}. In Section~\ref{sec:SFHs} we outline \texttt{SFERA 2.0} CMD-fitting procedure and methods. In Section \ref{sec:Results and Discussion} we present the derived star formation histories, while in Section~\ref{sec:summary and conclusions} we discuss the implications of our results and present a summary of our findings and conclusions.

\section{Data reduction}\label{sec:data}
The observations analyzed in this study have been performed with the NIRCam instrument \citep{Rieke2023} onboard the JWST, as part of the cycle 1 GO 1783 program FEAST (PI: A. Adamo) \footnote{The publicly available FEAST data products can be downloaded from \url{https://feast-survey.github.io/}}. The observational strategy and all data products of the FEAST collaboration are presented in detail in \citet{gregg2024}, \citet{pedrini2024}, \citet{Knutas2025}, \citet{correnti2025}, and \citet{Bortolini2025}. Here we will give a brief recap of the strategy and reduction of the observations used in this article. 

The NIRCam observations (footprint shown in Figure \ref{fig:SDSS_plus_FEAST_images}) were taken with a FULLBOX 4TIGHT primary dither and SHALLOW4 readout pattern, along with 2 small dithers resulting in 8 single exposures covering $6' \times 2.2'$ ($\sim 13 \times 5$ kpc) when combined. This resulted in deeper coverage in some regions of the mosaic with respect to others. Of all the NIRCam filters available for this target in FEAST, the two short wavelength filters F115W and F200W, with a total exposure time of $2920$ s, represent the deepest and best combination to study the resolved stellar populations \citep{Bortolini2025}. In this paper we used both the drizzled, distortion-corrected mosaics (i.e., {\it *i2d.fits}), which result from stage three of the JWST pipeline, and the calibrated science images (i.e., {\it *cal.fits}), which are the output of the stage two JWST pipeline. Figure \ref{fig:SDSS_plus_FEAST_images} shows our NIRCam mosaic in the F115W (blue channel), F200W (green channel) and F444W (red channel) filters and its field of view compared to the SDSS \citep{SDSS2000} optical-infrared image of the same target. The positions of the two dwarf galaxies are labeled. The two shorter wavelength filters probe mainly the stellar continuum, while the F444W is sensitive to the hot dust component of the interstellar medium (ISM) present around young star forming regions.

\subsection{PSF-fitting photometry} \label{sec:PSF-fitting photometry}
A detailed description of the FEAST PSF-fitting photometric reduction pipeline is provided in \citet{correnti2025} and \citet{Bortolini2025}. Here, we briefly summarize the main steps. To speed up the photometry reduction we divided the field of view into 18 equally sized ($\sim 60'' \times 48'' $, or $\sim 2.1 \times 1.7$ kpc) boxes (hereafter `tiles', see Figure \ref{fig:F200W_image_tiles}). In this paper, we adopt the same naming convention for the different tiles introduced in \citet{Bortolini2025}.  

We used the 2.1  version of the photometry package DOLPHOT \citep{Dolphin2000,Dolphin2016}, and its JWST/NIRCam module \citep{Weisz2024} to reduce simultaneously the F115W and F200W filters to reach the maximum depth possible in each tile. The photometry was run with DOLPHOT parameters set to values optimized for JWST/NIRCam crowded images \citep{Weisz2024,correnti2025,Bortolini2025}. The output catalogs contain the measured positions, apparent magnitude in the F115W and F200W filters, signal to noise ratio of each source, plus some diagnostic parameters such as \texttt{sharpness}, \texttt{crowding}, \texttt{roundness}, \texttt{chi2}, \texttt{quality\_flag}, and \texttt{object\_type} (we refer the reader to the DOLPHOT manual for an in-depth description of the meaning of these quantities). 

\begin{figure*}
    \centering
    \includegraphics[width=1.0\linewidth]{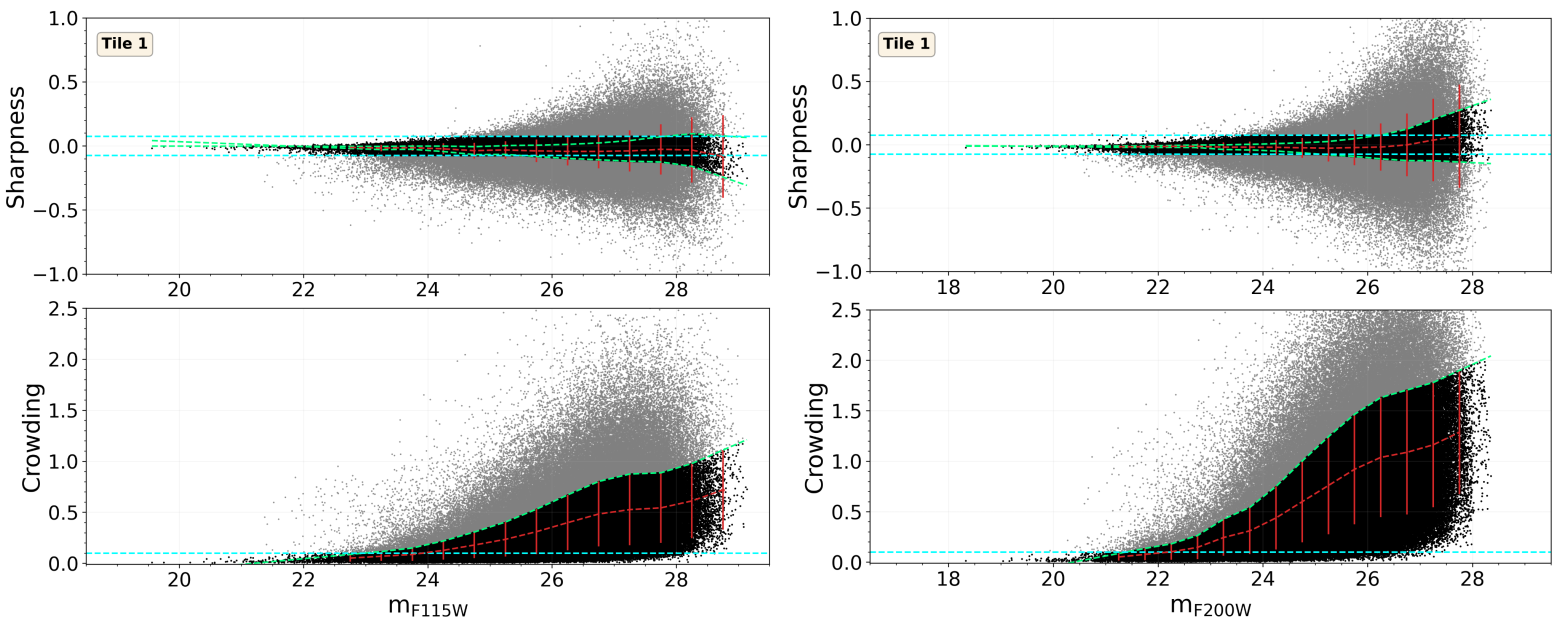}
    \caption{DOLPHOT \texttt{sharpness} (top panels) and \texttt{crowding} (bottom panels) as a function of magnitude in the F115W (left panels) and F200W (right panels) NIRCam filters for the central region of NGC~4485 (i.e., tile 1). Gray points mark the positions of all the sources that passed the preliminary cuts in \texttt{object\_type}, \texttt{quality\_flag}, and SNR. Black dots mark the sources that passed the \texttt{sharpness} and \texttt{crowding} selection. The red dashed line indicates the running \texttt{sharpness} and \texttt{crowding} mean with the estimated $\sigma$. The green and cyan dashed lines represent the selection limit, as described in Section \ref{sec:PSF-fitting photometry}.} 
    \label{fig:Tile1_F115W_F200W_sharpness_crowding_cuts}
\end{figure*}

To maximize the number of good candidate stars, while at the same time reduce the number of spurious detections (i.e. background galaxies, blends) we culled all 18 tile catalogs separately. We required all retained sources in each catalog to have: \texttt{object\_type} = 1, \texttt{quality\_flag} $\leq$ 2, SNR $> 2$ in both the F115W and F200W filters. In each tile, the \texttt{sharpness} cuts have been derived from the \texttt{sharpness} vs. magnitude distribution of the detected sources, keeping only the sources with $|${\tt sharpness}$|$ $<$ 0.075 or within $\pm 0.5 \sigma$ of their running mean sampled on 0.5 mag intervals. For the \texttt{crowding} we selected only sources with {\tt crowding} $<$ 0.1 or within $1\sigma$ from the running mean of their distribution. Figure \ref{fig:Tile1_F115W_F200W_sharpness_crowding_cuts} shows an example of the \texttt{sharpness} and \texttt{crowding} cuts we applied for the sources detected in tile 1. The sources that passed the initial \texttt{object\_type}, \texttt{quality\_flag}, and SNR cuts are shown with gray dots, while the sources that pass the \texttt{sharpness} and \texttt{crowding} selections for each filter are marked in black. The \texttt{sharpness} and \texttt{crowding} limits are marked with green and cyan dashed lines, while the running means of the distributions are shown in red.
\par

\subsection{Artificial star tests}\label{sec:Artificial star tests}
\begin{figure}
    \centering
    \includegraphics[width=1.0\linewidth]{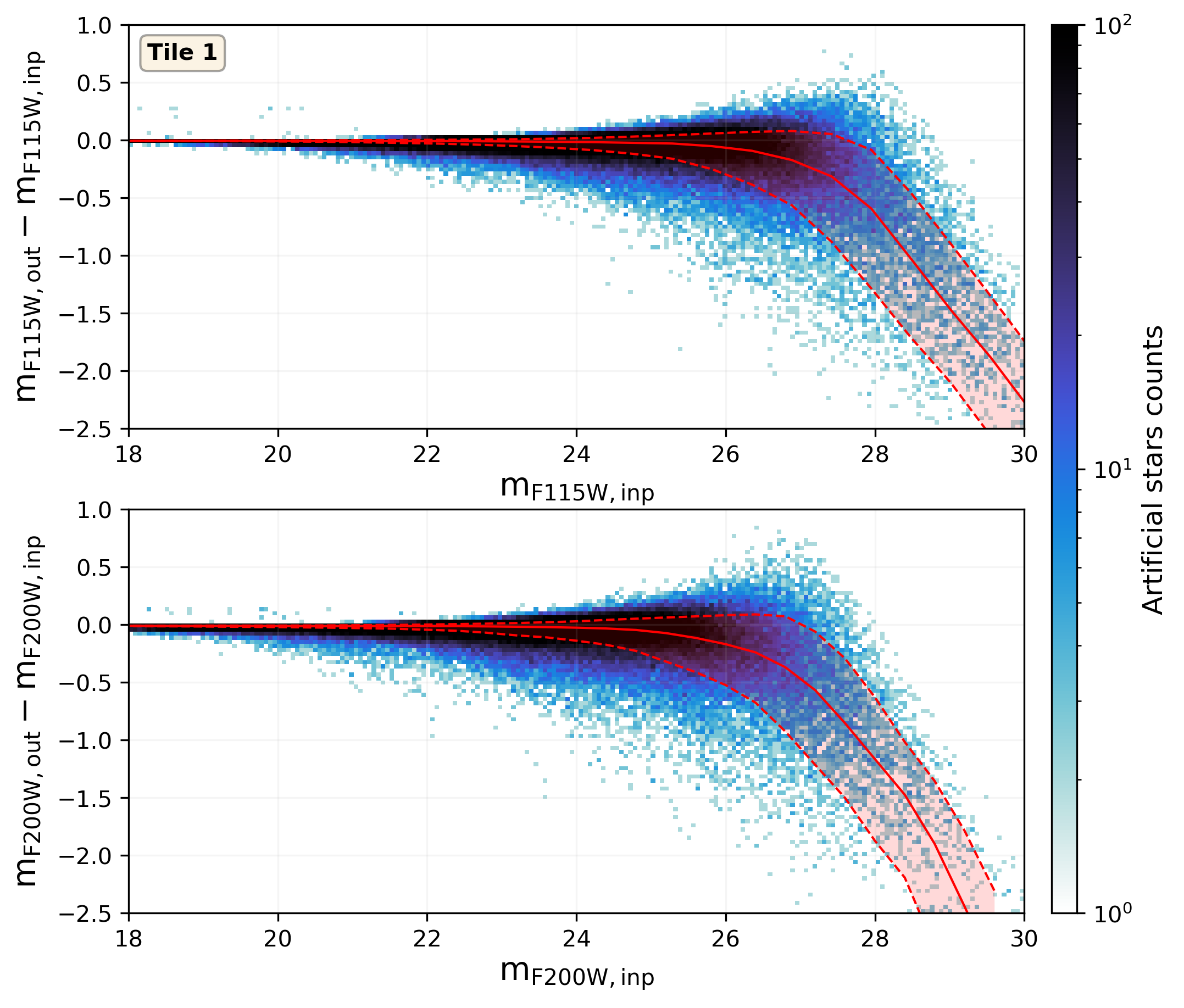}
    \caption{Distribution of the difference between the recovered (i.e., output) and the original (i.e., input) magnitude vs. input magnitude in the F115W (top panel) and F200W (bottom panel) filters of the artificial stars that have passed the quality cuts in tile 1. The distributions are binned and mapped to a logarithmic scale according to the number of artificial stars that fall in each bin.  The red solid line represents the running median of the distributions, while the dashed lines represent the running $16$th and $84$th percentile.}
    \label{fig:tile1_artificial_magout_vs_maginp}
\end{figure}

A careful estimation of the photometric errors and depth of the photometry is an essential step for a correct analysis and application of the synthetic CMD method \citep{Cignoni2010} that we will describe in Section \ref{sec:SFHs}. To estimate the photometric errors and completeness of our catalogs, we simulated 1 million artificial stars for each tile (for a total 18 million artificial stars), following the light profile of the image and carefully sampling all the color and magnitude range covered by our sources. Each artificial star was injected in each tile separately using the DOLPHOT built-in routine for artificial star tests, and the reduction was performed following the same steps and using the same parameters of the original photometry. The final artificial stars catalog for each tile is obtained after culling it with the same cuts that have been applied to the original photometry, and contains the original and recovered positions as well as the magnitudes in each filter. For instance in tile 1, $\sim 550,000$ artificial stars have passed the quality cuts, out of the $\sim 1$ million injected. The difference between the recovered (i.e., output) and the original (i.e. input) magnitude gives us an estimation of the photometric errors as a function of magnitude and color. The number of recovered vs. the number of injected artificial stars gives us an estimation of the completeness of the photometry, that takes into account the natural varying crowding conditions of the galaxy. As an example of the artificial star test results, Figure \ref{fig:tile1_artificial_magout_vs_maginp} shows the output--input vs. input magnitude distribution of the artificial stars in tile 1 that passed the quality cuts for both the F115W and F200W filters. Due to the high concentration of sources the distribution is binned and color-coded according to the number of artificial stars that fall in each bin. The red dashed lines represent the running $16$th, $50$th, and $84$th percentile of the distributions.

\begin{figure*}
    \centering
    \includegraphics[width=1.0\linewidth]{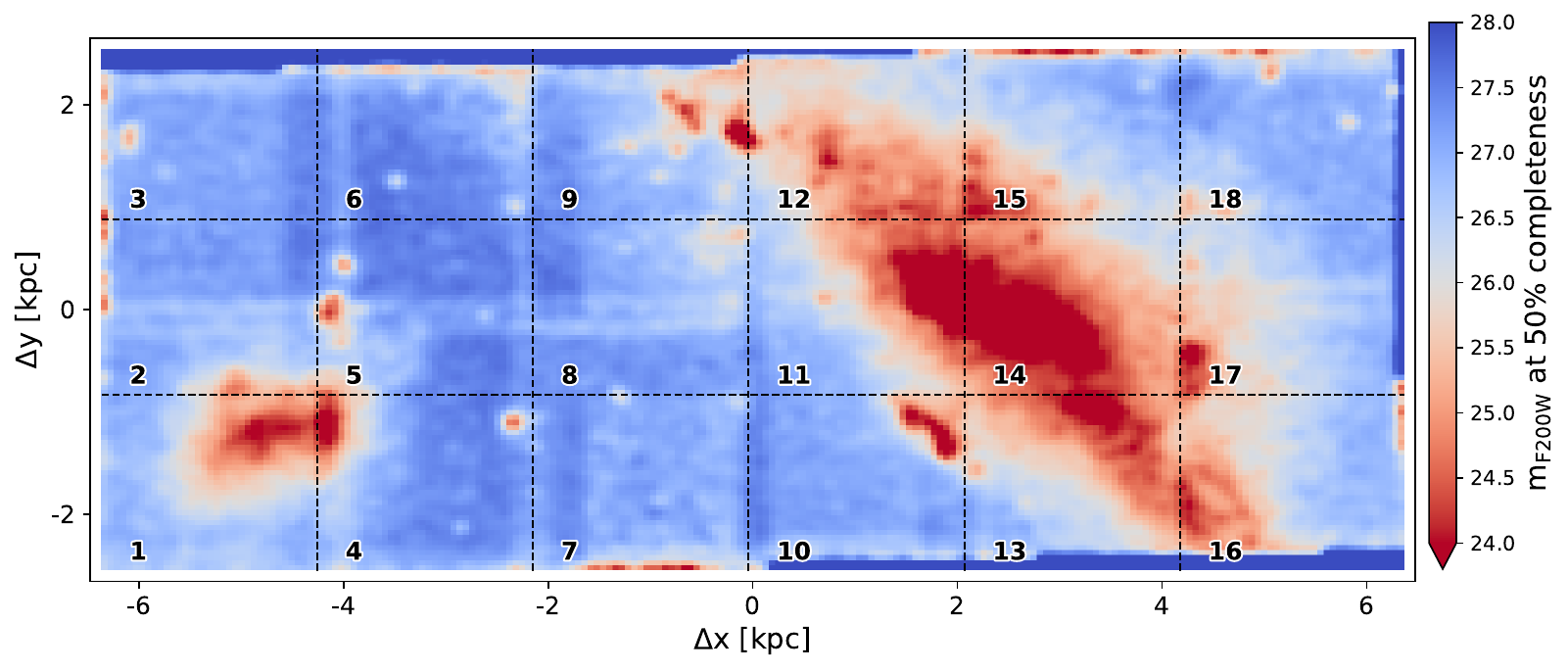}
    \caption{Map of our PSF-fitting photometry completeness across the entire JWST/NIRCam mosaic field of view, computed in a $\sim 200 \times 100$ pixel grid, $(\sim 60 \times 50 \, \mathrm{pc})$ per pixel, from our catalogs of artificial stars ( $\sim 1$ million per tile, $\sim 18$ million in total). Each pixel is color coded by the magnitude reached at the $50\%$ completeness limit in the F200W filter, going from red (i.e., shallower photometry) to blue (i.e., deeper photometry).}
    \label{fig:completeness_map_total}
\end{figure*}

\begin{figure*}
    \centering
    \includegraphics[width=0.90\linewidth]{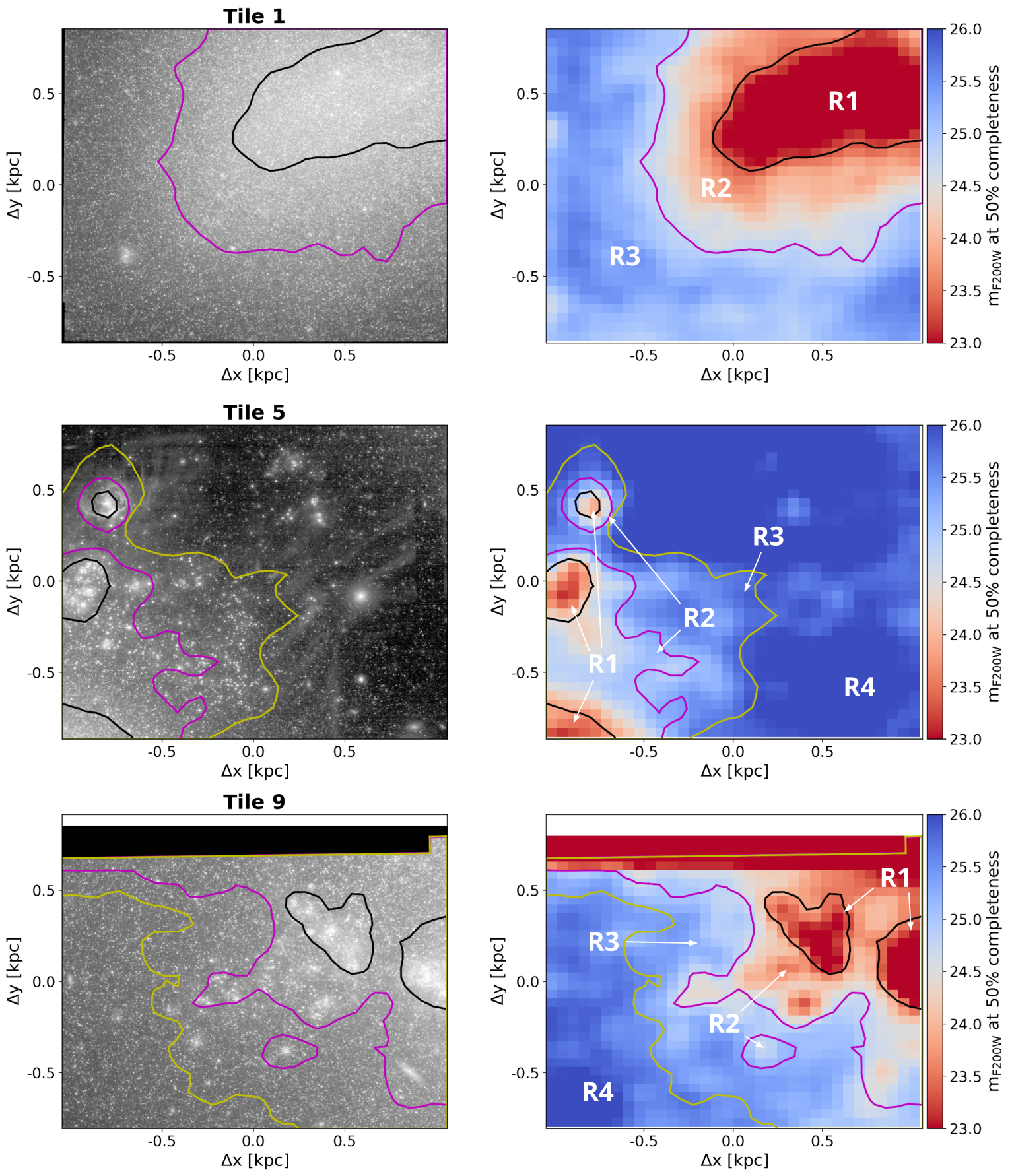}
    \caption{Top panels: zoom-in of tile 1, corresponding to the center of NGC~4485, showing the F200W image (left) and the completeness map derived from ASTs from Figure~\ref{fig:completeness_map_total} (right). The tile is divided into three sub-regions (labeled R1, R2, and R3) with increasing completeness, as delimited by the color contours. Middle panels: same as the top panels, but for tile 5, located on the NGC~4485 side of the tidal bridge. In this case, the tile is divided into four sub-regions (labeled R1, R2, R3, and R4) according to their average completeness. Bottom panels: same as the top and middle panels, but for tile 9, located on the NGC~4490 side of the tidal bridge. In this case, the tile is divided into four sub-regions (labeled R1, R2, R3, and R4) according to their average completeness.}
    \label{fig:tiles_completeness_map_zoomin}
\end{figure*}

The artificial stars allow us to probe how the photometric errors and completeness of our data vary not only in the color-magnitude space, but also (and crucially) spatially across the image. Using our $\sim 18$ million artificial stars we can construct a spatially resolved completeness map of our observation field of view, presented in Figure \ref{fig:completeness_map_total}. Each pixel of the map covers an area of $1.7'' \times 1.4''$ $(\sim 60 \times 50 \, \mathrm{pc})$, and is color coded according to the magnitude in the F200W filter at which the $50\%$ completeness limit is reached (see colorbar on the right hand side). Figure \ref{fig:tiles_completeness_map_zoomin} shows some zoom-ins of the regions covered by tile 1 (i.e., encompassing the center of NGC~4485), tile 5 (i.e., NGC~4485's side of the tidal bridge), and tile 9 (i.e., NGC~4490's side of the tidal bridge) at the same pixel scale of Figure \ref{fig:completeness_map_total}. As can be seen, the completeness maps have a complex morphology, going from very incomplete regions ($m_{50\% comp} \sim 22\mbox{--}24$), that nicely match with regions of high star density and star-forming regions in the F200W image, to more complete ones ($m_{50\% comp} \sim 26\mbox{--}28$), as we move towards less crowded regions of the galaxy. Such a complex spatial morphology makes the CMD fitting process very challenging. To ensure an optimal fit of the CMD, and therefore a reliable recovery of the SFH, we divided each of the 18 tiles into several sub-regions, delimited by the solid lines in Figure \ref{fig:tiles_completeness_map_zoomin}. The sub-regions are defined such that the magnitude corresponding to the $50\%$ completeness level varies by no more than $\sim 1.5$ mag within each region, thereby ensuring a relatively uniform completeness within each region. For example, tile 5 and tile 9 are divided into four sub-regions because they contain both highly crowded (and therefore more incomplete) star-forming regions and more sparsely populated areas. Region 1 (R1) is enclosed by the solid black contours; Region 2 (R2) lies between the black and magenta contours; Region 3 (R3) is located between the magenta and yellow contours; and Region 4 (R4) corresponds to the remaining area outside the yellow contour. In contrast, tile 1 is divided into only three sub-regions (R1, R2, and R3), reflecting the more gradual variation in completeness across the tile. The subdivision scheme adopted for the remaining tiles is presented in Section \ref{sec:Tiles completeness maps} of the Appendix. To ensure a consistent and robust analysis, we derive the SFH of each sub-region independently with \texttt{SFERA 2.0}, and then combine the results to recover the total SFH of the corresponding tile.

\section{Resolved Stellar Populations} \label{sec: resolved stellar pops}
In this Section, we briefly describe the main features of the stellar populations in NGC~4485 and NGC~4490 revealed by the near-infrared CMDs. For a detailed description of the NIRCam CMDs and the different stellar populations present in each tile of our NGC~4485/NGC~4490 mosaic we refer to \citet{Bortolini2025}. 

\begin{figure*}
    \centering
    \includegraphics[width=1.0\linewidth]{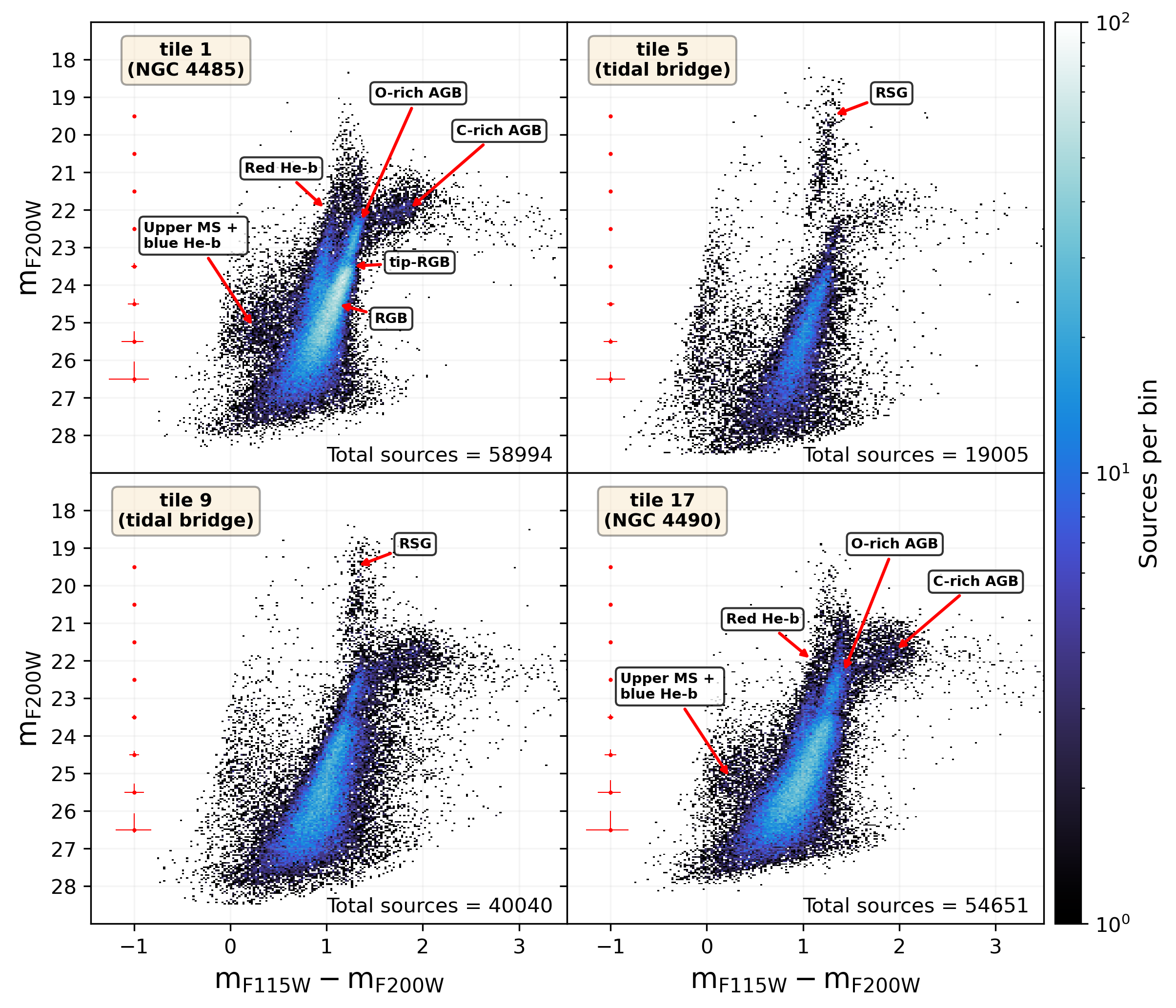}
    \caption{$\mathrm{m_{F200W}}$ vs. $\mathrm{m_{F115W}}-\mathrm{m_{F200W}}$ CMDs of tile 1 (i.e., center of NGC~4485), tile 5 (NGC~4485's side of the tidal bridge), tile 9 (NGC~4490's side of the tidal bridge), and tile 17 (outer region of NGC~4490) for the sources that passed the quality cuts discussed in Section \ref{sec:PSF-fitting photometry}. The diagrams are all binned and mapped to the same logarithmic scale according to the number of stars in each bin (see colorbar on the right-hand side). Some of the stellar evolutionary phases discussed in the text are marked by red arrows and labeled accordingly. The total number of sources in each CMD is reported in the bottom-right side. The average photometric errors, estimated from our ASTs, at $\mathrm{m_{F115W}}-\mathrm{m_{F200W}} = 1$ and for different magnitudes are shown as red error-bars on the left hand-side of each panel.}
    \label{fig:tiles_CMDs}
\end{figure*}

Figure \ref{fig:tiles_CMDs} shows the CMDs for the sources that passed our photometry quality cuts (see Section \ref{sec:PSF-fitting photometry}) for tile 1, 5, 9 and 17, probing parts of the main bodies of the two galaxies (i.e., tile 1 and 17) and their tidal bridge (i.e., tile 5 and 9, see Figure \ref{fig:F200W_image_tiles}). The main stellar evolutionary phases present and described below are marked by red arrows and labeled in the Figure. The total number of sources recovered in each tile is also reported in the bottom-right corner of each panel. The average photometric errors, estimated from our ASTs, at $\mathrm{m_{F115W}}-\mathrm{m_{F200W}} = 1$ and for different magnitude values are shown as red error-bars.  
 
A blue vertical sequence at $\mathrm{m_{F115W}}-\mathrm{m_{F200W}} \sim 0$, composed of young upper main-sequence (MS) stars and more evolved core He-burning (He-b) stars in the blue phase of their evolutionary loops, is clearly visible in each tile. Unfortunately, the intrinsically similar colors in the near-infrared of these stars, convolved with the photometric errors and distance of the target, do not allow a confident separation between these two evolutionary phases. In tiles 1 and 17, these stars are concentrated around $\mathrm{m_{F200W}} \sim 25$, whereas in tiles 5 and 9 the sequence is more uniformly populated, extending up to magnitudes as bright as $\mathrm{m_{F200W}} \sim 19$. 

A red vertical plume, populated by young (age $\lesssim30$ Myr), massive ($\mathrm{8 \, \mathrm{M_{\odot}} \lesssim M \lesssim 30 \, \mathrm{M_{\odot}}}$) He-b red supergiant (RSG) stars, is clearly visible in the tidal bridge (tiles 5 and 9), confirming the presence of a strong and recent starburst. In tiles 1 and 17, a vertical sequence of red He-b stars is still present, although its bright end is significantly less populated. 

A red sequence parallel to, but fainter than the red He-b sequence is also distinguishable in all four tiles, around $\mathrm{m_{F115W}}-\mathrm{m_{F200W}} \sim 1.25$ and $22.3 \lesssim \mathrm{m_{F200W}} \lesssim 23.5$. These are intermediate-mass ($4 \, \mathrm{M_{\odot}} \lesssim M \lesssim 8 \, \mathrm{M_{\odot}}$) oxygen-rich asymptotic giant branch (AGB) stars. At $\mathrm{m_{F115W}}-\mathrm{m_{F200W}} \sim 2$ and $\mathrm{m_{F200W}} \sim 22$ a red cloud of lower mass ($1.2 \, \mathrm{M_{\odot}} \lesssim M \lesssim 4 \, \mathrm{M_{\odot}}$) AGB stars in their thermally pulsing phase (TPAGB) are also visible in all four tiles. During this phase, inner-shell convective episodes known as the third dredge-up bring a significant amount of carbon to the stellar surface. This carbon is responsible for a significant decrease of the effective temperature of the stars and is subsequently expelled into the interstellar medium through strong stellar winds, where it contributes predominantly to the formation of carbonaceous dust. As a consequence, these stars appear redder than their oxygen-rich (O-rich) counterparts. For this reason, they are also referred to as carbon-rich (C-rich) AGB stars. Thanks to the sensitivity and resolution of JWST, these stars can now be resolved with high accuracy in star-forming galaxies outside the Local Group \citep{Hirschauer2024,Bortolini2024b,correnti2025,Bortolini2025,Boyer2025}, and have been also used to measure distances \citep{Lee2023,Li2024,Lee2025,Li2025,Gavetti2026,Gavetti2026b}. For a more detailed description of this complex evolutionary phase we refer the reader to \citet[][]{Vassiliadis1993,Herwig2005,Marigo2008,Nanni2013,Ventura2014,Marigo2017,Ventura2018,Gavetti2025} among others.

The AGB stars described above have a wide range of ages, that go from $\sim 100$ Myr to $\sim 5$ Gyr. We also note the presence of a gap in the distribution of AGB stars between the vertical sequence of O-rich stars and the C-rich cloud, and a vertical finger brighter than $\mathrm{m_{F200W}} \sim 22$, that lies above the O-rich sequence and runs parallel but slightly redder of the sequence of more massive RSG stars. For a detailed description and interpretation of these features from a stellar evolution point of view, we refer the reader to \citet{correnti2025} and Dell'Agli et al. (in prep), in which the authors use the state-of-the-art AGB stellar evolution and dust formation code ATON \citep{Ventura1998,Ventura2014,Ventura2018,Ventura2020} these features in NGC~4449 (a starburst dwarf galaxy in the FEAST sample).

Last but not least, a population of old (age $\geq 1$ Gyr) and low mass ($M \lesssim 1.0 \, \mathrm{M_{\odot}}$) RGB stars dominate the faint and red region of the CMDs of all four tiles. The indicative location of the tip of the RGB is marked at $\mathrm{m_{F200W}} \sim 23.5$, and seems to not change from tile to tile within the uncertainties of our photometry (see also discussion in Section \ref{sec:SFHs}).

\begin{figure*}
    \centering
    \includegraphics[width=1.0\linewidth]{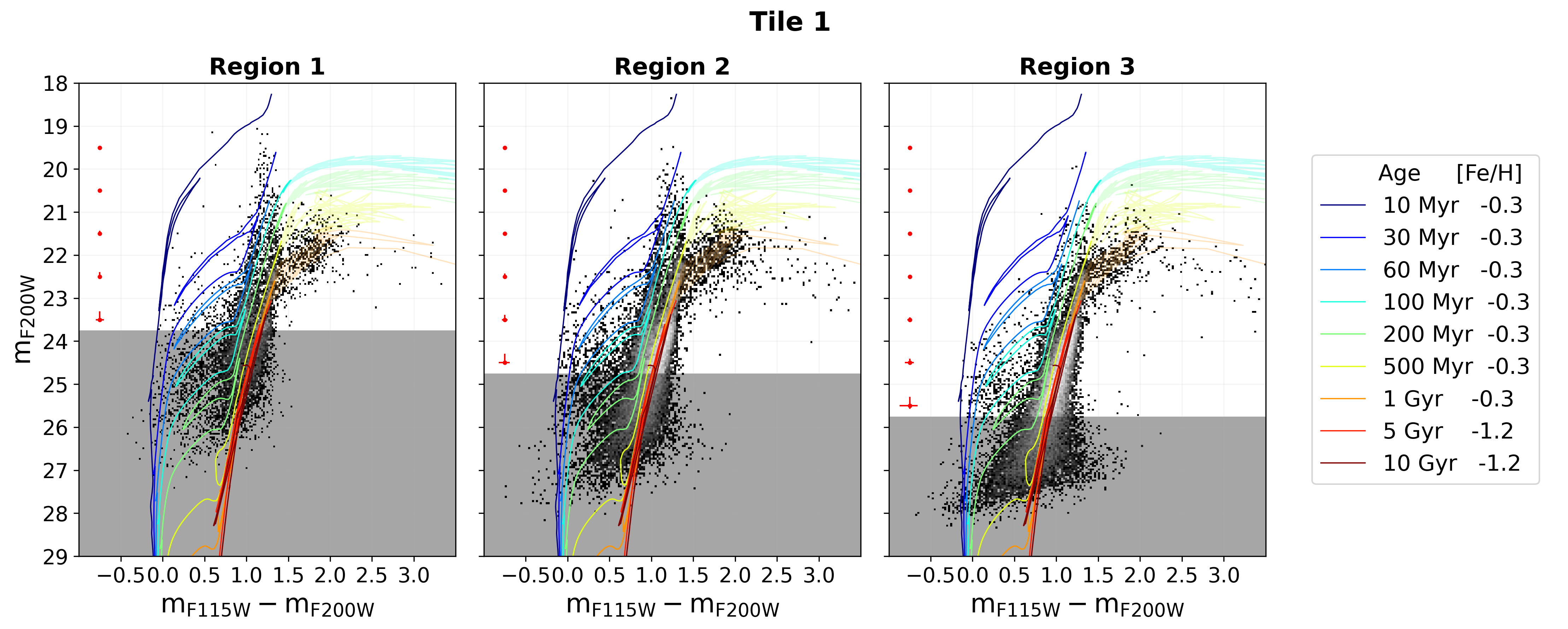}
    \includegraphics[width=1.0\linewidth]{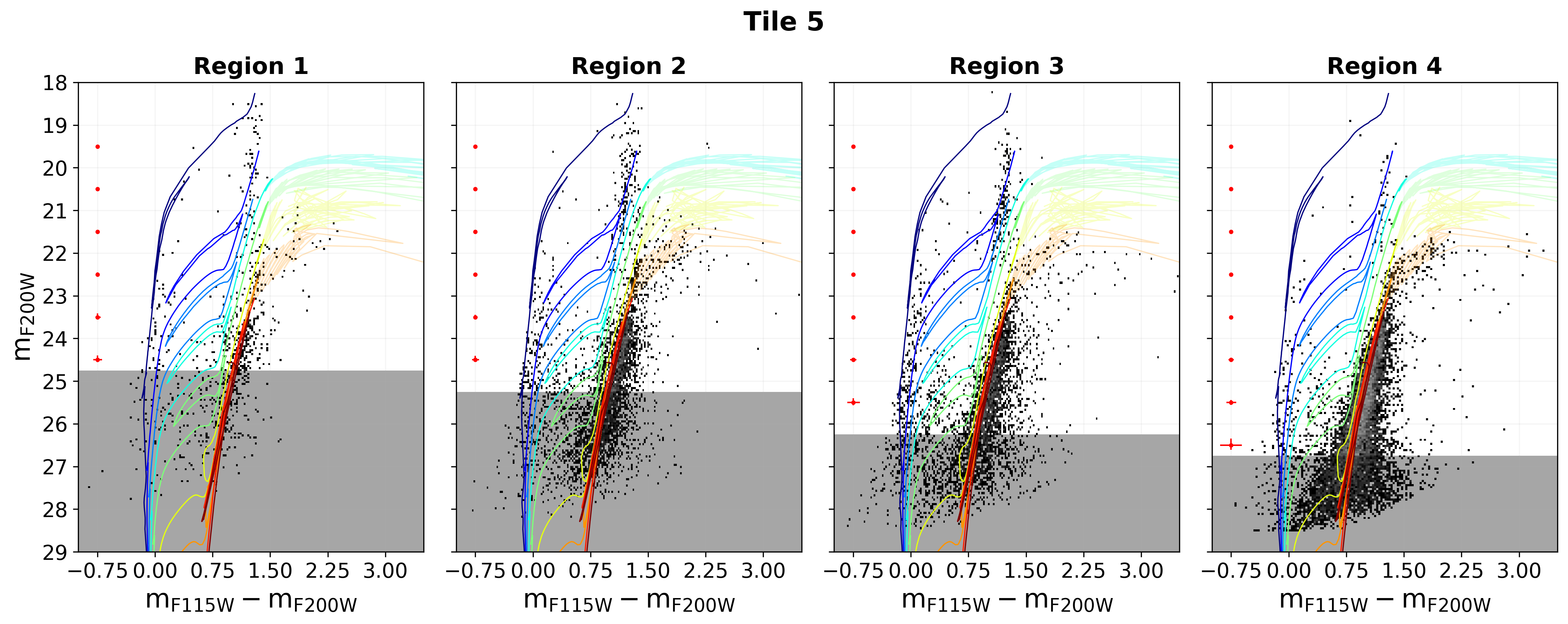}
    \caption{Top panels: CMDs of the three sub-regions of tile 1 (see Figure~\ref{fig:tiles_completeness_map_zoomin}), shown in order of increasing photometric depth from left to right. The CMDs are binned and color-coded, with the colors mapped to a logarithmic scale according to the stellar density in each bin (black: low density; white: high density). PARSEC-COLIBRI \citep{Bressan2012,Marigo2017} isochrones spanning ages from 10 Myr to 10 Gyr are overplotted, adopting metallicities of $[{\rm Fe/H}] = -0.30$ for the young isochrones and $[{\rm Fe/H}] = -1.20$ for isochrones older than 1 Gyr (see legend). Sources located above the shaded region are recovered with a completeness $\geq 50\%$, as estimated from artificial star tests. Bottom panels: same as the top panels, but for the four sub-regions of tile 5 (see Figure~\ref{fig:tiles_completeness_map_zoomin}). The average photometric errors of each sub-region, estimated from our ASTs, at $\mathrm{m_{F115W}}-\mathrm{m_{F200W}} = 1$ and for different magnitude values are shown as red error-bars on the left hand-side of each panel.}
    \label{fig:subregions_CMDs}
\end{figure*}

To illustrate how photometric completeness affects the different sub-regions within each tile, Figure \ref{fig:subregions_CMDs} shows as an example the CMDs for the three different sub-regions of tile 1 and the four sub-regions of tile 5, going from most crowded (i.e., less complete) to least crowded (i.e., more complete) from left to right. Over-plotted on each CMD are PARSEC–COLIBRI stellar isochrones \citep{Bressan2012,Marigo2017,Pastorelli2020}, color-coded by age going from 10 Myr to 10 Gyr. The isochrones with age $\leq 1$ Gyr have a metallicity of $[{\rm Fe/H}] = -0.30$, whereas the older ones are more metal-poor to match the color of the RGB ($[{\rm Fe/H}] = -1.20$). The TPAGB phase tracks for stars older than $100$ Myr are shown with a lower opacity to improve readability. Absolute magnitudes have been calculated from theoretical luminosities and temperatures using the bolometric corrections from \citet{Chen2019}, and then converted to apparent magnitudes assuming a distance modulus of $\mathrm{(m-M)_{0}} = 29.50$ and a foreground reddening of $\mathrm{E(B-V)} = 0.10$ \citep{Calzetti2015,Cignoni2019}. The shaded region of the CMDs in each panel indicates $< 50\%$ completeness, derived from artificial stars within that specific sub-region. The R1 sub-region is the one with the shallowest CMD, being the most crowded area (see Figure \ref{fig:tiles_completeness_map_zoomin}), while R2, R3, and R4 CMDs become progressively deeper, reaching down to $\mathrm{m_{F200W}} \sim 26$ and $\mathrm{m_{F200W}} \sim 27$ in region 3 of tile 1 and region 4 of tile 5, respectively.

Most importantly for the analysis of the SFH of the system, this variation in the photometry completeness affects the age of the faintest He-b stars detectable in each region. Indeed, during the core He burning phase intermediate mass stars ($\mathrm{M \gtrsim 2 \, M_{\odot}}$) exhibit large loops in colors (the so-called blue loops) and their luminosities are proportional to their mass. Because of this characteristic, the loops become fainter with increasing age between 100 and 500 Myr, which makes these loops an excellent indicator of stellar age in distant star-forming galaxies. In the most crowded regions of tile 1 (i.e., R1), the oldest resolved He-b loop stars have ages around $\sim 100$ Myr, while in the least crowded regions (i.e., R3) have ages of $\sim 300$ Myr. For tile 5 we see that in the most crowded region (i.e., R1), the oldest resolved He-b loop stars have ages around $\sim 200$ Myr, while in the least crowded region (i.e., R4) have ages of $\sim 500$ Myr. 

Where He-b stars in the $100\mbox{--}500$ Myr age range are not resolved (i.e. under our completeness limit), we must rely on more massive stars in later evolutionary stages, like AGBs, that trace the same stellar ages. Unfortunately, despite being bright and thus easy to detect in distant galaxies, AGB stars are less sensitive to age than He-b. As shown by the isochrone in Figure \ref{fig:subregions_CMDs}, AGB stars of different ages are harder to disentangle due to being confined in a tight color and magnitude region of the CMD. Moreover, the TPAGB phase is still the most uncertain among all the single stars main evolutionary phases \citep{Marigo2008,Marigo2013,Marigo2017}. In this phase, stars go through a series of processes that are highly complex to model from first principles and that are mutually dependent, like stochastic mass losses, long period pulsations, hot-bottom burning, and the many third dredge-up episodes \citep[see e.g.,][]{Frost1996,Mowlavi1999,Ventura2005,Herwig2005,Marigo2013,Ventura2014}. These uncertainties and difficulties translate into substantial differences in the predictions of different stellar evolution models \citep{Marigo2017}, and in turn makes the recovered SFH based only on this phase more uncertain \citep{Gallart2005,Cignoni2010}. Due to the distance of the galaxies from us, we do not resolve He-b stars older than $500$ Myr in any of the sub-regions. Therefore, to probe the epoch between $500$ Myr and $1$ Gyr we have no other choice than to rely on AGB stars. 

For epochs older than $\sim 1$ Gyr, our best tracers are instead RGB stars, that can be reliably resolved below the tip in each sub-region due to their brightness. Naturally, the variations in crowding conditions affect what populations can be resolved with confidence in similar ways in all the other tiles. Particular attention should be given to the central region of NGC~4490 (i.e., tiles 11 and 14). Owing to the severe stellar crowding in this area, the average completeness limit is only $\sim 23$ mag, causing only BL stars younger than $\sim 60$ Myr to lie above our detection threshold, hindering our ability to safely recover the SFH at intermediate ages. In addition, RGB stars associated with older stellar populations are also largely below the completeness limit, making it impossible to reliably recover the SFH at old epochs. For these reasons, we exclude these tiles from the SFH analysis (see discussion below).


\section{CMD-fitting routine}\label{sec:SFHs}

With both the photometric and artificial star catalogs in hand, we can recover the SFH of the interacting system NGC~4485/NGC~4490 using the synthetic CMD method. This technique, pioneered by \citet{Maeder1974}, \citet{Chiosi1988}, \citet{tosi1991}, \citet{Bertelli1992} among others, has since become a widely adopted and extensively tested approach over the past three decades to study the formation history of dwarf galaxies both inside and outside the Local Group \citep[e.g.,][among many others]{Gallart1999,Annibali2003,Dolphin2005,Cole2007,Monelli2010,McQuinn2010,Weisz2014,Sacchi2016,Skillman2017,Sacchi2018,Cignoni2019,Sacchi2021,Savino2023,McQuinn2024,Fernandez-Alvar2025,Burhenne2026}. 

Several codes that implement this technique have been developed over the years, including \texttt{StarFISH} \citep{Harris2002}, \texttt{MATCH} \citep{Dolphin2002}, \texttt{IAC-pop} \citep{Aparicio2009}, \texttt{SFERA} \citep{Cignoni2015,Sacchi2018}, \texttt{Talos} \citep{deBoer2016}, \texttt{TheStorm} \citep{Bernard2018}, \texttt{ChronoGal} \citep{Gallart2024}, \texttt{StarFormationHistories.jl} \citep{Garling2025}, and \texttt{PANCAKE} \citep{Zheng2025}. Although these codes differ in their specific implementations and optimization strategies, they share the same fundamental principles and overall methodology. In this work, we employ an updated version of the stellar population synthesis routine \texttt{SFERA} (Star Formation Evolution Recovery Algorithm), named \texttt{SFERA 2.0}, which was developed and optimized to fit the bright stellar populations (e.g., upper MS, He-b and AGB stars) that can be observed in the CMDs of star-forming dwarf galaxies outside the Local Group \citep{Bortolini2024b,Bortolini2024a}. 

The first step of the routine is to generate a library of synthetic CMDs for simple stellar populations, covering a small range in age and metallicity, and interpolating an adopted library of stellar isochrones. We refer the interested reader to \citet{Bortolini2024a} for an in depth description on how the interpolation in mass and age of the isochrones and how unresolved binary systems are implemented in the code to uniformly sample all stellar evolutionary phases. In this paper we used the PARSEC-COLIBRI stellar isochrones library \citep{Bressan2012,Marigo2013,Marigo2017,Pastorelli2020}, which calculates in great detail all evolutionary phases for stars in a wide range of masses and metallicities \footnote{The PARSEC-COLIBRI isochrones can be downloaded at: \url{https://stev.oapd.inaf.it/cgi-bin/cmd}}.

The ``simple'' stellar population synthetic CMDs are generated to cover all the possible combinations of 9 non uniform age bins, between $\log(\mathrm{age/yr}) = 7.00$ and $10.13$, and 13 uniform metallicity bins between $\mathrm{[Fe/H]} = -1.50$ and $-0.30$ in steps of $0.1$ dex, for a total of $117$ synthetic CMDs. We adopted an age resolution $\Delta \rm{\log(age/yr)} = 0.25$ for $\rm{\log(age/yr)} < 8.00$, 0.35 for $8.00 \lesssim \rm{\log(age/yr)} < 9.05$, 0.45 for $9.05 \lesssim \rm{\log(age/yr)} < 9.50$, and one last bin for older epochs. The different lengths of the age bins are carefully selected to account for the natural loss of time resolution that occurs for older epochs (see discussion in Section \ref{sec: resolved stellar pops}), and at the same time minimize the stochastic correlation between adjacent bins (see also Appendix \ref{sec:appendixA}). We chose not to include ages younger than $10$ Myr because the $\mathrm{m_{F115W}}-\mathrm{m_{F200W}}$ color is not particularly efficient at separating stars with those ages, plus we expect to miss most of such young stars due to them being still associated in dense, unresolved young star clusters. The synthetic CMDs are then populated assuming a Kroupa IMF \citep{Kroupa2001} from $0.2$ and $300 \, \mathrm{M_{\odot}}$, while assuming a $30\%$  binary fraction \citep{Cignoni2009}, and a set of bolometric corrections to go from theoretical luminosities and temperatures to absolute magnitudes in the F115W and F200W filters \citep{Chen2019}.

The second, and crucial step is to simulate the observational conditions. First the absolute magnitudes are converted to apparent ones by placing the synthetic stellar populations at the distance of the target galaxy, and correcting for reddening due to different star to star line of sight. The reddening distribution is modeled as a Gaussian (with negative values excluded) with a mean value representing the foreground reddening ($\mathrm{E(B-V)}$) and dispersion representing the internal/differential reddening ($\mathrm{\sigma_{E(B-V)}}$) as free parameters in the code. The synthetic CMDs are then convolved with the photometric uncertainties and completeness characterizing the sub-regions of each tile under analysis, obtained through our extensive ASTs.

The last step is to compare a linear combination of our synthetic CMDs, that with appropriate weights can reproduce the complex stellar population, to the galaxy's observed CMD. To do so, we binned both synthetic and observed CMDs in color and magnitude and compare the star counts in each CMD cell. For all tiles and relative sub-regions we adopted a bin size of $0.125$ in color and $0.05$ in magnitude.   

To find the best set of weights, we define the `distance' between our models and the data using the following Poissonian likelihood function:
\begin{equation}
    \chi_{P} = \sum_{k=1}^{N_{cells}} obs_{k} \ln{\left( \frac{obs_{k}}{mod_{k}} \right) - obs_{k} + mod_{k}}
\end{equation}
where $obs_{k}$ and $mod_{k}$ are the number of stars in the data and model $k$ CMD cell. This Poissonian likelihood has been demonstrated by several authors to be the optimal choice in CMD fitting problems \citep{Mighell1999,Dolphin2002}. To minimize this likelihood, \texttt{SFERA 2.0} uses a hybrid-genetic minimization algorithm, i.e., a mix between a classical FORTRAN-based genetic algorithm called \texttt{Pikaia} \citep{Charbonneau1995}\footnote{\texttt{Pikaia} can be freely downloaded from this link: \url{https://www.hao.ucar.edu/modeling/pikaia/pikaia.php}} and a local-search algorithm based on the Simulated Annealing method \citep{Kirkpatrick1983}. The best model SFH is represented by the set of weights that produced the model CMD that minimizes the Poissonian likelihood when compared to the observed one. 

We performed SFH calculations over a coarse grid of distance, $\mathrm{E(B-V)}$, and $\mathrm{\sigma_{E(B-V)}}$ for the sub-regions of each tile of our mosaic. The combinations of values that returned the highest likelihood value are assumed as our best-fit values. In our particular case, we explored values of $\mathrm{E(B-V)} = [0.0,0.10,0.20]$ and $\mathrm{\sigma_{E(B-V)}} = [0,0.10,0.20,0.30]$ \citep{Calzetti2015,Cignoni2019}. Regarding the distance modulus, we explore different values within the range $[29.06\mbox{--}29.71]$ reported by \citet{Sabbi2018}. Somewhat unsurprisingly given the interacting nature of the system, we obtained the best-fit results in each tile with the same value of $\mathrm{(m-M)_{0}} = 29.50$ and foreground reddening $\mathrm{E(B-V)} = 0.10$. The best-fit values of differential reddening varies instead from one sub-region to the next. In the fit, we did not impose any particular age-metallicity relation.

The statistical uncertainties are calculated through a bootstrap technique. We randomly re-sample the stars in the observed CMDs multiple times, and derive again the SFH for each data re-sampling. We then calculate the $16$th-$84$th percentiles of the solution distributions as confidence intervals.

\subsection{The Adequacy of AGB Stellar Evolution Models} \label{sec:The Adequacy of AGB Stellar Evolution Models}
\begin{figure}
    \centering
    \includegraphics[width=1.0\linewidth]{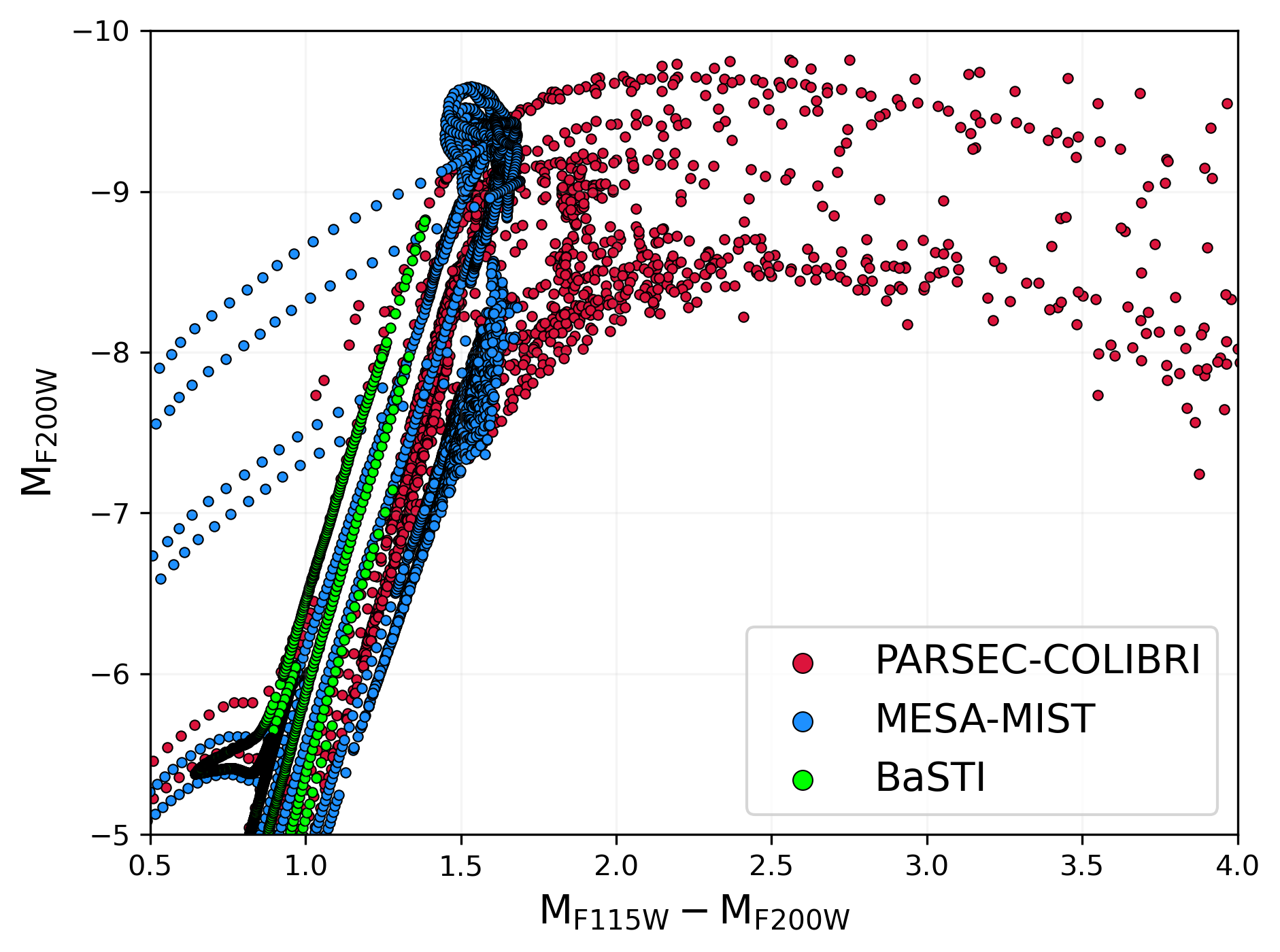}
    \caption{PARSEC-COLIBRI (\citealt{Bressan2012,Marigo2017}, red dots), MESA-MIST (\citealt{Paxton2011,Choi2016,Dotter2016}, blue dots), and BaSTI (\citealt{Cordier2007,Hidalgo2018}, green dots) stellar isochrones with ages of $[100, 200, 400, 600, 800]$ Myr and $Z = 0.008$ (Large Magellanic Cloud range), in the NIRCam $\mathrm{M_{F200W}}$ vs. $\mathrm{M_{F115W}}-\mathrm{M_{F200W}}$ CMD.}
    \label{fig:AGB_iso_comparison}
\end{figure}

It is common practice in CMD-fitting SFH studies to estimate systematic uncertainties by generating synthetic CMDs with multiple sets of stellar evolution models and comparing the recovered SFHs, while keeping all other parameters fixed. PARSEC-COLIBRI \citep{Bressan2012,Marigo2017}, MESA-MIST \citep{Paxton2011,Choi2016,Dotter2016}, and BaSTI \citep{Cordier2007,Hidalgo2018} are among the most widely used stellar isochrone libraries in CMD analyses of resolved galaxies \citep[see, e.g.,][]{Gallart2005,Weisz2014,Sacchi2018,Cignoni2019,McQuinn2024,Bortolini2024b}. Although these models generally provide consistent predictions across most evolutionary phases, they differ substantially in their treatment of the TPAGB phase. As discussed above, the TPAGB is one of the most challenging stages of stellar evolution to model self-consistently because of the complex interplay between mass loss, third dredge-up, thermal pulses, hot-bottom burning, dust production, and radiative transfer through circumstellar envelopes. Different modeling groups adopt different physical prescriptions, approximations and calibration strategies, resulting in significant variations in the predicted properties of these stars.

Among the available models cited above, COLIBRI \citep{Marigo2008,Marigo2013,Marigo2017} has the most comprehensive set of isochrones, including the TPAGB evolution. Its calculations include detailed prescriptions for third dredge-up, hot-bottom burning, changes in low-temperature opacities, stellar pulsation properties, mass-loss, dust formation, and the interaction of stellar radiation with circumstellar material through radiative transfer calculations. In contrast, the BaSTI and MESA-MIST isochrones do not model several of these processes. As a result, their TPAGB predictions are generally hotter and bluer than those of COLIBRI and fail to reproduce many of the complex features observed in the CMDs of star-forming galaxies (see Section 3.8 and Figure 11 of \citealt{Marigo2017} and the discussion in \citealt{Lee2025b}).

For instance, in Figure \ref{fig:AGB_iso_comparison} predictions from PARSEC-COLIBRI (red dots), MESA-MIST (blue dots), and BaSTI (green dots) stellar isochrones with ages of $[100, 200, 400, 600, 800]$ Myr and $Z = 0.008$ (close to that of the LMC's young stellar population) are shown in the specific NIRCam filters used in this paper. In the observational CMD, the portion of the PARSEC-COLIBRI isochrones populated by C-rich AGB stars is substantially redder compared to the region of the CMD populated by the respective MESA-MIST models that do not predict stars redder than $\mathrm{m_{F115W}}-\mathrm{m_{F200W}} = 1.5$, while the BaSTI models do not include at all C-rich AGB stars.

Although there are several codes that calculate self-consistently the entire structure of the stars during the AGB phase, e.g., MONASH \citep{Karakas2014,Karakas2022}, ATON \citep{Ventura1998,Ventura2014,Ventura2018,Ventura2020}, FRUITY \citep{Cristallo2009,Cristallo2011,Cristallo2015}, they do not provide isochrones that cover the full stellar mass range required for our analysis, as they do not include massive stars. Since our galaxies host ongoing star formation, we adopt a homogeneous set of stellar evolution models spanning the entire mass range needed for the SFH recovery. For an interested reader, the application of the ATON models to the population of AGB stars of the irregular dwarf galaxy NGC 4449, is presented in \citet{correnti2025} and in a forthcoming study by Dell'Agli et al. (in prep).

Given the significant population of red AGB stars in our CMDs, PARSEC-COLIBRI is the only stellar isochrones set used for the SFH recovery presented in this work. Nevertheless, for completeness, we provide a comparison between the SFHs recovered using the PARSEC-COLIBRI and MESA-MIST models with and without the inclusion of AGB stars in Section \ref{sec:appendixA} of the Appendix. While we emphasize that the latter lacks the detailed TPAGB treatment required for a realistic description of these stellar populations, this comparison illustrates the systematic differences that arise from adopting alternative stellar evolution libraries and including or not the AGB stars in the fitting.

\begin{figure*}
    \centering
    \includegraphics[width=0.85\linewidth]{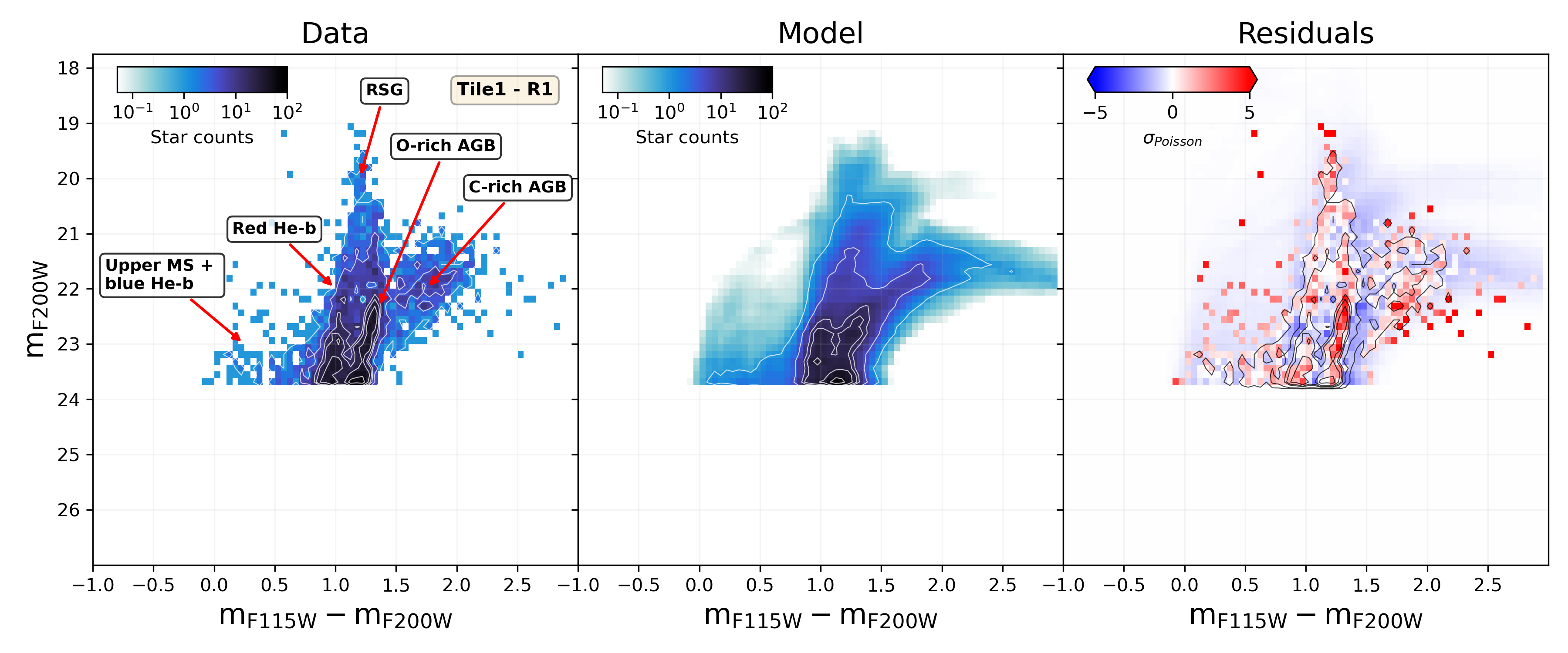}
    \includegraphics[width=0.85\linewidth]{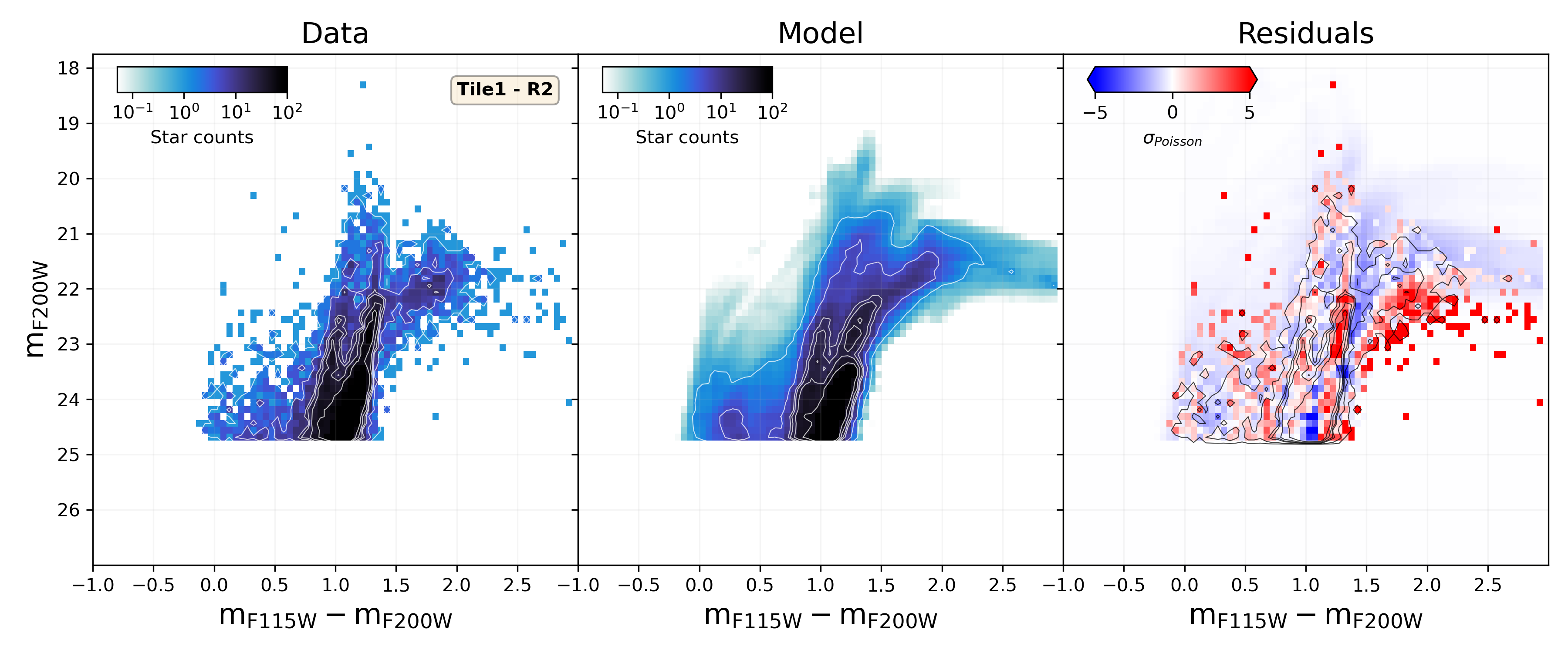}
    \includegraphics[width=0.85\linewidth]{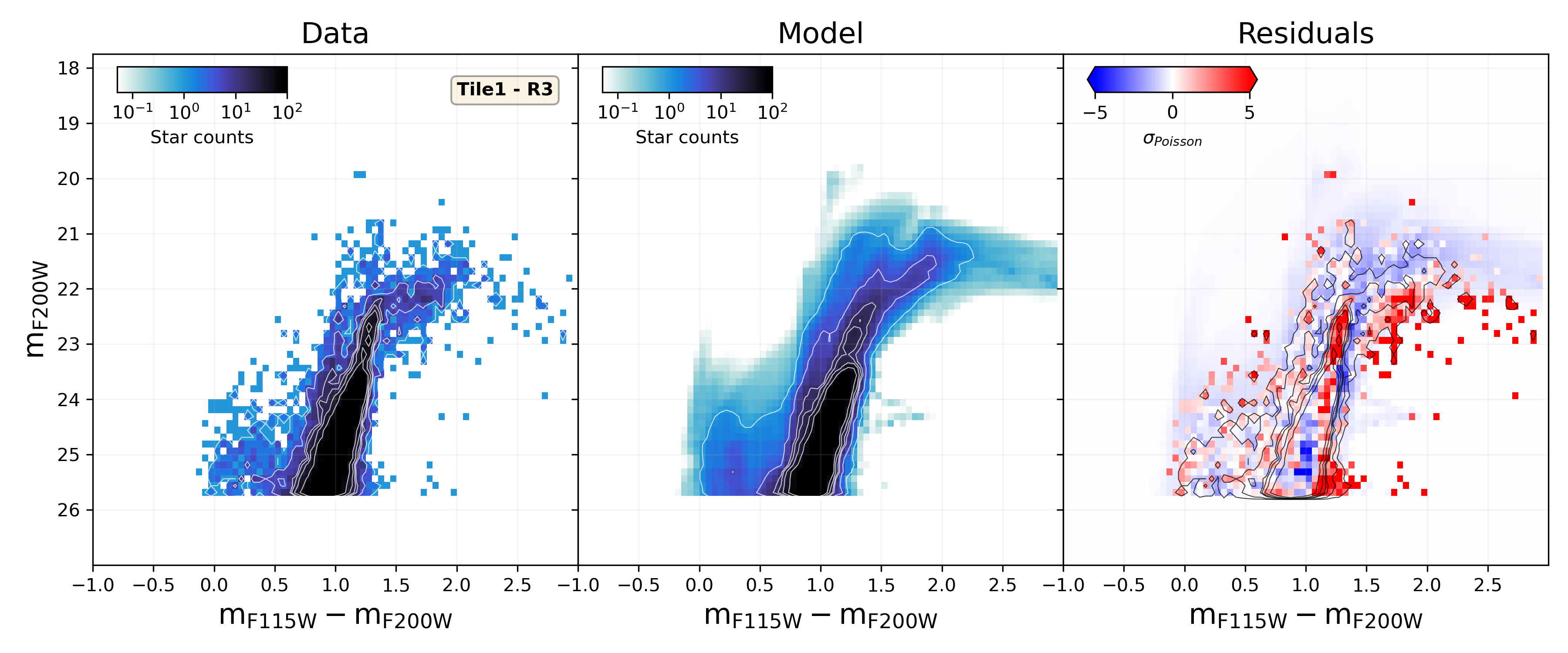}
    \caption{ Example of observed CMDs, best-fitting \texttt{SFERA 2.0} model CMDs, and stellar counts residuals normalized by the Poissonian error for regions R1, R2, and R3 of tile 1 are shown in the top, middle, and bottom panels, respectively. The CMDs are binned using the same binning scheme adopted by \texttt{SFERA 2.0} during the fitting process and are color-coded according to the logarithm of the number of stars in each bin. For easier comparison, white iso-density contours corresponding to stellar counts of [1, 5, 15, 20, 30, 50] are overlaid on top of the observed and best-fit model CMDs. The contours of the observed CMDs are also plotted (in black) on top of the residual map. The positions of some of the main stellar evolutionary phases are marked by red arrows and labeled accordingly.}
    \label{fig:data_model_hess_comparison}
\end{figure*}

\section{Star Formation Histories} \label{sec:Results and Discussion}
In Figure \ref{fig:data_model_hess_comparison}, we compare the observed CMDs with \texttt{SFERA 2.0} best-fit model CMDs for the three sub-regions of tile 1: region 1 (R1, upper panels), region 2 (R2, middle panels), and region 3 (R3, lower panels; see also Figure \ref{fig:tiles_completeness_map_zoomin}). The CMDs are binned using the same color and magnitude bin sizes adopted in the fitting routine. Each bin is color-coded according to the number of stars, ranging from white (low density) to black (high density). White iso-density contours corresponding to stellar counts of $[1, 5, 15, 20, 30, 50]$ are also overlaid to help the reader identify and compare the main stellar features in the CMDs.

For each CMD bin, we show the residuals between the observed stellar counts and the predictions of the \texttt{SFERA 2.0} best-fit model, expressed in units of the Poisson uncertainty (i.e., $(obs_{k} - mod_{k})/ \sqrt{mod_{k}}$). Sources lying below the $50\%$ completeness limit in each sub-region are excluded from the fit and are therefore not shown in the figure, producing the sharp cutoff visible at the faint end of the diagrams.

The \texttt{SFERA 2.0} best-fit models successfully reproduce the main features of the observed CMDs in all three sub-regions. This includes the young upper-MS and He-burning populations, the intermediate-age AGB stars, and the RGB population. The most noticeable discrepancy is found in the locus of the O- and C-rich AGB stars, which appears systematically brighter in the models than in the observations, in particular for the deeper CMDs of region 2 and 3 which provide stronger constraints. This offset is likely related to the complex effects of circumstellar dust produced during the intense stochastic mass-loss episodes experienced by these stars throughout their thermal-pulsing evolution. The resulting dust envelopes can significantly attenuate and redden the stellar flux. As discussed in Section \ref{sec:The Adequacy of AGB Stellar Evolution Models}, this evolutionary phase remains particularly challenging to model self-consistently. Nevertheless, the PARSEC-COLIBRI stellar evolution models reproduce well the color and overall number of stars populating this region of the CMD. Similar fit quality is obtained for all other tiles and their respective sub-regions, indicating that the \texttt{SFERA 2.0} best-fit models provide a consistently good representation of the observed CMDs across the entire dataset.

\begin{figure*}
    \centering
    \includegraphics[width=1.0\linewidth]{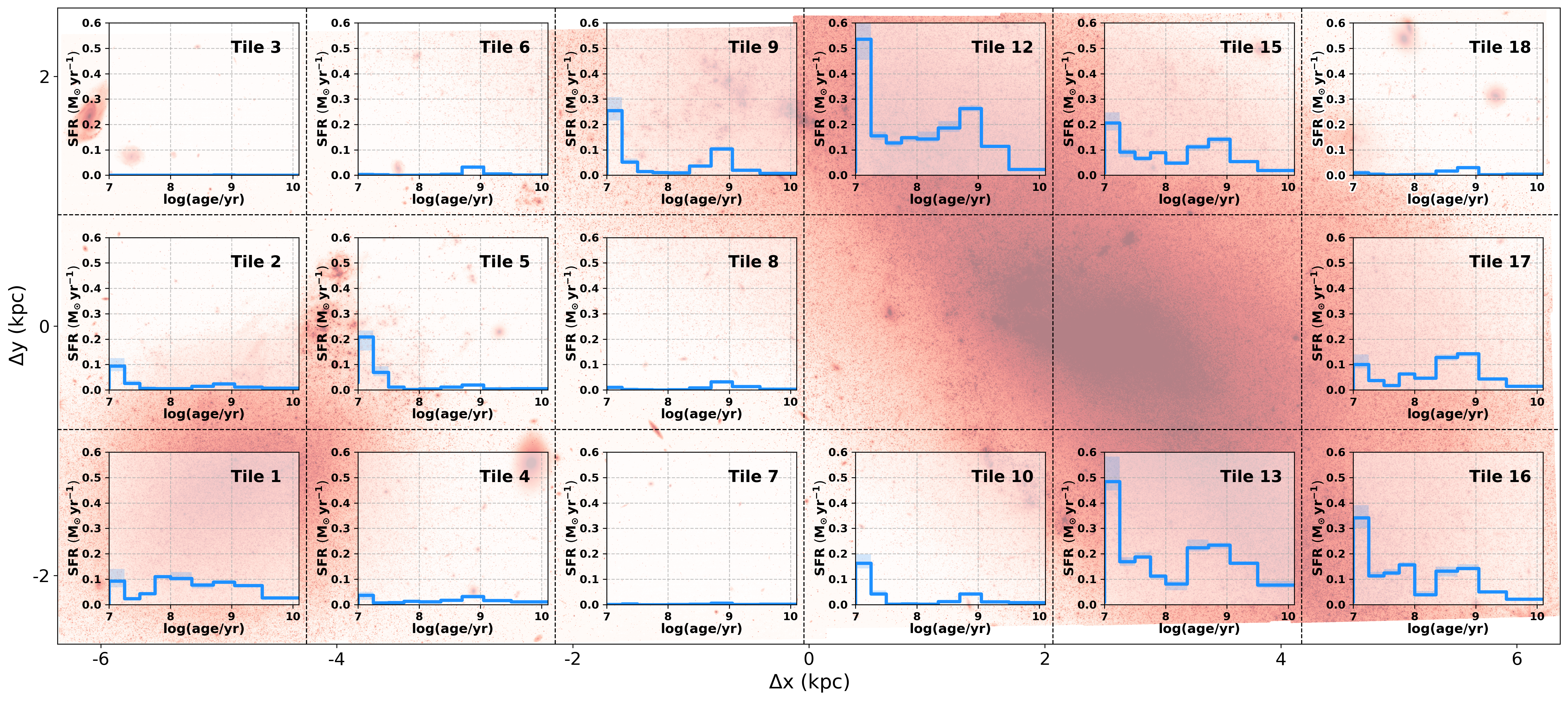}
    \caption{Star-formation rate vs. log(age/yr), i.e., the SFH, as a function of spatial position (i.e., for each tile of our NGC4485/NGC4490 mosaic). In the background, we show the JWST/NIRCam F200W mosaic of NGC~4485/NGC~4490. The SFHs for the central regions of NGC~4490 (tiles 11 and 14) are omitted because the severe crowding prevents a reliable recovery of their SFHs (see discussion in the text). All plots share the same y-axis scale and limits for easier comparison of the relative strength of the star formation activity across different regions of the interacting galaxy pair.}
    \label{fig:SFHs_spatialmap}
\end{figure*}

Figure \ref{fig:SFHs_spatialmap} presents the \texttt{SFERA 2.0} derived SFHs for each tile of our NIRCam F200W mosaic, excluding the central tiles (11 and 14) of NGC~4490. Note that the x-axis shows the logarithm of the age, while the y-axis shows the star-formation rate in units of $\mathrm{M_{\odot} \, yr^{-1}}$, using the same scale and limits for all tiles to facilitate a direct comparison of the SFRs across different regions of the system.

The first feature that catches the eye is a strong enhancement in the recent star formation activity (ages between $\sim 10\mbox{--}30$ Myr) in the tidal bridge region (tiles 5 and 9) and throughout much of the main body of NGC~4490 (e.g., tiles 10, 12, 13, 15, and 16). In contrast, the center of NGC~4485 (e.g., tile 1) shows recent star formation, but at substantially lower rate. Significant star formation activity is also observed at intermediate ages ($200\ \mathrm{Myr} \lesssim \mathrm{age} \lesssim 1\ \mathrm{Gyr}$), particularly in the central region of NGC~4485 (tile 1), the tidal bridge (tiles 5 and 9), and across most of NGC~4490. This burst of star formation is especially evident when compared with the generally lower SFRs inferred at lookback-times older than $\sim 1$ Gyr. Tiles 3 and 7 exhibit significantly lower star formation activity over the entire lookback time than the other tiles analyzed, indicating that they have formed substantially less stellar mass.

\begin{figure*}
    \centering
    \includegraphics[width=1.0\linewidth]{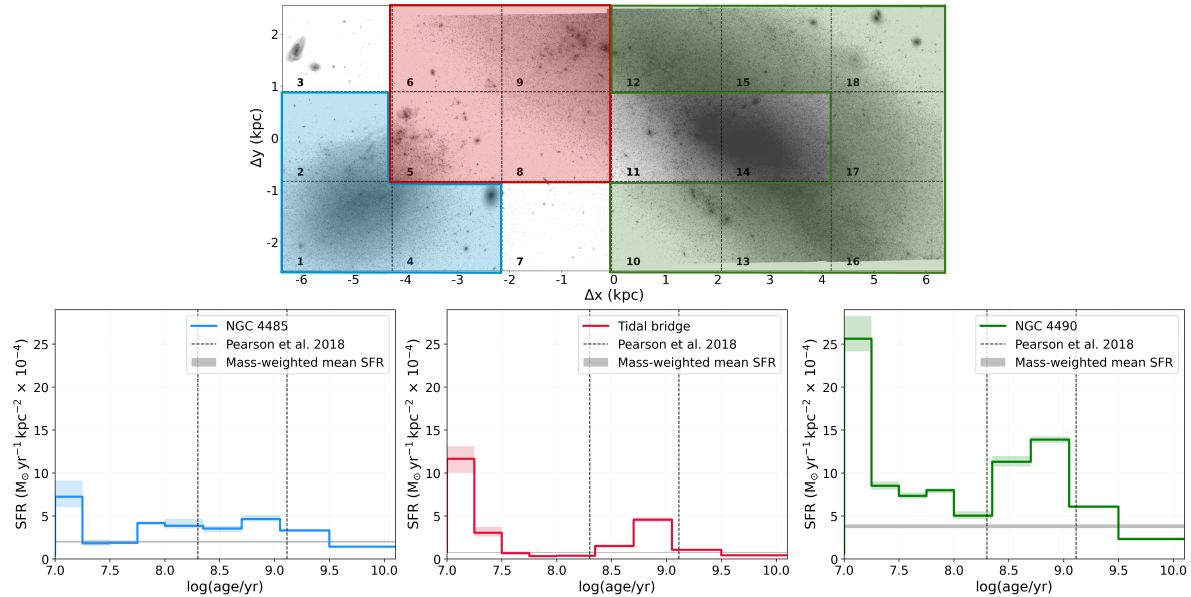}
    \caption{Upper panel: JWST/F200W image of NGC~4485/NGC~4490, divided into three regions: a blue region dominated by NGC~4485, a red region encompassing the tidal bridge connecting the two galaxies, and a green region covering the tiles associated with NGC~4490. Tiles 3 and 7 are not assigned to any specific region because their association is ambiguous; however, their very low average SFRs imply that including them would not affect the results. Tiles 11 and 14 are excluded from the analysis due to severe stellar crowding, which prevents a reliable recovery of their SFHs. Bottom panels: Recovered SFHs of the three regions shown in the upper panel, using the same color scheme. In each panel, the gray horizontal band indicates the mass-weighted average SFR over the full look-back time. The two vertical gray-dashed lines mark the predicted epochs of the first and second perigalactic passages of the system according to \citet{Pearson2018}.}
    \label{fig:NGC4485_NGC4490_bridge_SFHs_area_tilemap}
\end{figure*}

To highlight these trends more clearly, Figure \ref{fig:NGC4485_NGC4490_bridge_SFHs_area_tilemap} shows the combined SFHs of three distinct regions: NGC~4485, represented by tiles 1, 2, and 4 (in blue); the tidal bridge, represented by tiles 5, 6, 8, and 9 (in red); and the main body of NGC~4490, excluding its central region, represented by tiles 10, 12, 13, 15, 16, 17, and 18 (in green). Tiles 3 and 7 are excluded because their association with any of these regions is ambiguous. However, their contribution to the combined SFHs is negligible owing to their relatively low SFRs. To enable a meaningful comparison between the three regions, the SFHs were normalized by the total area in $\mathrm{kpc^{2}}$ covered by each group of tiles, thereby accounting for their different spatial extents.

All three regions exhibit SFHs broadly similar in overall shape. They are characterized by relatively low SFRs at times older than $\sim 1$ Gyr, followed by a pronounced enhancement in star formation at intermediate ages, between $200\ \mathrm{Myr} \lesssim \mathrm{age} \lesssim 1\ \mathrm{Gyr}$, with the peak around $500\mbox{--}800$ Myr ago. After this episode, the SFR declines, reaching a local minimum, before rising again to reach its highest value during the most recent time bin at ages between $\sim 10\mbox{--}30$ Myr ago. Despite these similarities, the relative strengths of the star forming episodes differ significantly when compared to the average SFR of each region, weighted by its total stellar mass, indicated by the gray-shaded area in each panel. Indeed the older SFR peak in NGC~4485 is $\sim 2$ times stronger than the average SFR, while the youngest peak reaches values up to $\sim 4$ times the average. NGC~4490 has an older peak $\sim 4$ times stronger than the average, while the youngest peak is $\sim 7$ times stronger. The tidal bridge is the region in which the differences between the peaks of SFRs are the highest if compared to the averaged SFR of the region, with the oldest and youngest peaks in star-formation activity $\sim 6$ and $\sim 15$ times higher, respectively.

Another notable difference is that the enhancement in SFR at intermediate-ages in NGC~4485 and NGC~4490 appears to have persisted for a longer period, extending to epochs around $60$ Myr ago. In contrast, the corresponding activity in the tidal bridge seems to have begun declining already $\sim 200$ Myr ago, before increasing again in the two youngest age bins. For reference, the predicted epochs of the first ($\sim 1.3$ Gyr ago) and second ($\sim 200$ Myr ago) pericenter passages from the independent N-body simulation of \citet{Pearson2018} are indicated by the vertical gray-dashed lines. In all three regions, the two recovered peaks of star formation in our SFHs occur in time bins corresponding to a few hundred million years after the predicted times of the close encounters between the galaxies. This suggests a temporal offset between the encounters and the subsequent star formation activity, although the coarse time resolution of our SFHs, together with the uncertainties in the predicted encounter times, prevents us from precisely quantifying this delay (see the discussion in Section \ref{sec: resolved stellar pops} and Appendix \ref{sec:appendixA}).

\begin{figure}
    \centering
    \includegraphics[width=1.0\linewidth]{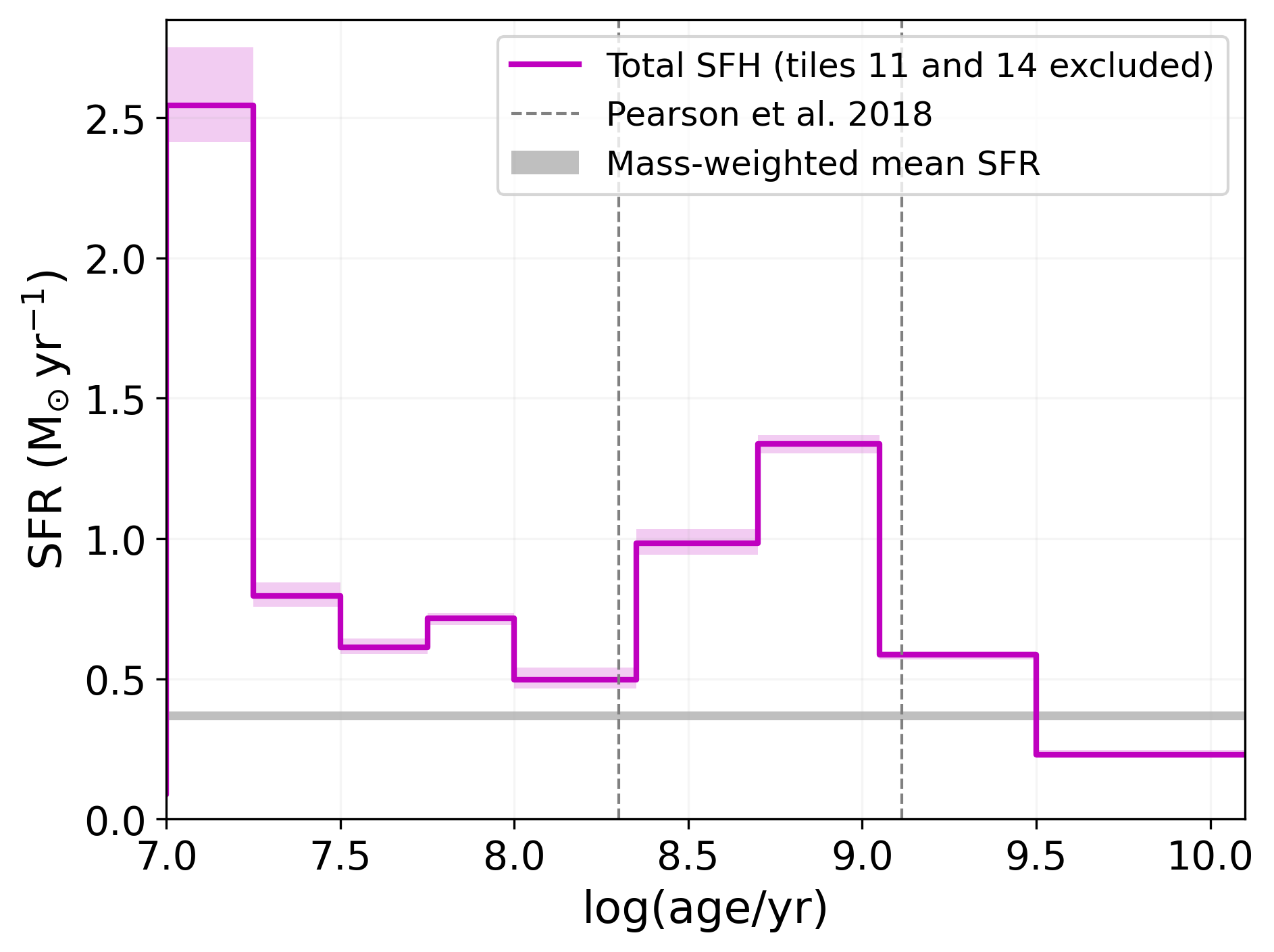}
    \caption{Total SFH of NGC~4485/NGC~4490, excluding the central part of NGC~4490 (i.e., tile 11 and 14). The horizontal shaded area marks the mass-weighted average SFR experienced by the interacting system across the entire lookback time. The vertical dashed lines mark the positions of the first and second perigalactic passages, which happened $\sim1.3$ Gyr and $\sim 200$ Myr ago, according to \citet{Pearson2018}.}
    \label{fig:all_tiles_sum_SFHs}
\end{figure}

Finally, in Figure \ref{fig:all_tiles_sum_SFHs} we show the total SFH of the NGC~4485/NGC~4490 system, after summing all the tiles in our field of view, again with the important caveat that the center of NGC~4490 (tile 11 and 14) is excluded. As for the SFHs in Figure \ref{fig:NGC4485_NGC4490_bridge_SFHs_area_tilemap}, the mass-weighted average SFR is shown by the gray-shaded area, while the position of the two perigalactic passages predicted by \citet{Pearson2018} are marked by the gray-dashed vertical lines. The two peaks in star formation activity are still clearly visible in the global SFH of the system, with the oldest peak $\sim 4$ times higher than the averaged SFR, and the youngest one $\sim 7$ times stronger than the average. 

\section{Discussion and Conclusions} \label{sec:summary and conclusions}

The spatially resolved star-formation histories of NGC~4485 and NGC~4490 (see Figure \ref{fig:SFHs_spatialmap}) nicely demonstrate how star formation in low-mass and gas-rich interacting galaxies is widely spread across the two galaxies. In particular, NGC~4485 shows prominent sites of enhanced young star formation activity displaced from its central region (i.e., tiles 2 and 5), in agreement with that found by recent UV observations \citep{Cignoni2019}, and co-spatial with the presence of dense gas also displaced from its center (Cannon et al. in prep, see Figure \ref{fig:F200W_image_tiles}). This behavior contrasts markedly with the centrally concentrated nuclear starbursts commonly associated with mergers of massive spiral galaxies \citep{Mihos1994,Barnes1996,Ellison2013,Thorp2019}. Tile 9, 12, 13, and 16 in NGC~4490 exhibit some of the highest gas densities (see Figure \ref{fig:F200W_image_tiles}), that match with the strongest signatures of recent star-formation activity recovered by \textsc{SFERA 2.0} (see Figure \ref{fig:SFHs_spatialmap}). However, because the two central tiles of NGC~4490 are not included in our spatially resolved SFH analysis, we cannot assess the recent star-formation activity in its central regions directly.

Interestingly, the combined SFHs of NGC~4485, the tidal bridge, and NGC~4490 shown in Figure \ref{fig:NGC4485_NGC4490_bridge_SFHs_area_tilemap}, as well as the total SFH presented in Figure \ref{fig:all_tiles_sum_SFHs}, exhibit two distinct peaks in star formation, several times stronger than the average SFR. These peaks occur a few hundred million years after the time of first and second perigalactic passages predicted by the N-body simulations of \citet{Pearson2018} (gray-dashed vertical lines). This behavior is consistent with state-of-the-art hydrodynamical simulations of interacting, gas-rich, low-mass galaxies resolved down to massive stars \citep{Lahen2019,Lahen2020}. These models predict enhancements in the global star-formation rate following pericenter passages, particularly within the tidal bridge, where interaction-induced gas compression can promote elevated star-formation activity. They also predict the formation of star-forming regions displaced from the galactic centers (see Figures 2 and 5 of \citealt{Lahen2020}), in good agreement with the spatial distribution of the recent star-formation clumps observed in the NGC~4485/4490 system (i.e., tiles 5, 9, 10, 13, 15, 16).

Taken together, our spatially resolved SFHs and the observed morphology of recent star-forming clumps are consistent with current theoretical predictions for dwarf–dwarf interactions, suggesting that repeated tidal interactions may have played a key role in triggering the major episodes of enhanced star formation observed in the system.

The results described above allow us to suggest the following qualitative scenario for the evolution of this system. At the earliest epochs the two galaxies were not interacting, and had a rather low SFR. The bridge did not exist, and its gas was still part of the two galaxies. As shown in Figure \ref{fig:NGC4485_NGC4490_bridge_SFHs_area_tilemap}, around 3 Gyr ago, the two galaxies experienced a factor of $\sim$2 increase in the star formation rate per unit area (SFR/area). At this stage, the two galaxies had not yet interacted and were forming stars primarily within their main bodies. The gradual increase in their SFR can therefore be understood as a natural consequence of their growing stellar mass, as expected for star-forming galaxies evolving along the star-forming main sequence. As gas accretion continued, the galaxies increased in mass, leading to a corresponding increase in their star-formation activity. The values of the SFR/area ranged from $10^{-3} \, \rm{M_{\odot}\, yr^{-1} \, kpc^{-2}}$ for NGC~4490 to $4 \times10^{-4} \, \rm{M_{\odot}\, yr^{-1} \, kpc^{-2}}$ for NGC~4485, or $1 \, \rm{M_{\odot} \, Gyr^{-1} \, pc^{-2}}$ and 0.4 $\rm{M_{\odot} \, Gyr^{-1} \, pc^{-2}}$, respectively. Assuming a typical gas consumption timescale of 1 Gyr, and a gas surface density of $\sim 1 \, \rm{M_{\odot} \, pc^{-2}}$, typical for dwarfs, the SFR/area in NGC~4490 was then rather standard \citep[][]{Elmegreen2026a,Elmegreen2026b}. Taking into account that the consumption timescale tends to increase for the smallest systems because their low metallicities and pressures prevent an efficient formation of giant molecular clouds \citep[see, e.g.,][]{Bigiel2010}, NGC~4485 appears to have had a rather normal SFR/area at that early epoch as well.

About 1 Gyr ago the first interaction occurred \citep{Pearson2018}, and likely triggered the formation of the bridge and the first burst. According to Figure \ref{fig:NGC4485_NGC4490_bridge_SFHs_area_tilemap}, the duration of this burst was different in the two galaxies and their bridge. We interpret this difference again in terms of the different gas consumption timescales, which allowed the burst to last $\sim0.8$ Gyr in NGC~4490, and about twice as long in NGC~4485. The drop of the SFR/area in the bridge was much faster than in the two galaxies, presumably because its surface gas density was still too low to support much star formation on its own.  

About $200$ Myr ago, when the second interaction probably occurred \citep{Pearson2018}, the three regions reacted in different ways: the bigger galaxy, NGC~4490, showed a slow increase in the SFR/area, while the smaller galaxy, NGC~4485, was still experiencing the high SFR of the first burst and did not change its SFR significantly, nor did the bridge show significant SFR enhancements. We can speculate that there was no new gas accreted by the small galaxy and the bridge between the interactions, so there was no fuel for a new burst after the previous episode exhausted their gas. Instead, the bigger galaxy had still sufficient gas, and/or some new gas could have been accreted by it, that could let the SFR increase. The latter possibility is plausible as the center of mass of the system is closer to (or inside) the big galaxy given their factor of 10 difference in stellar mass.  

At the current epoch, all the three regions show a clear peak in SFR started around 30 Myr ago and still ongoing. Almost all star-forming dwarfs are characterized by ongoing, or recent, bursts, that make them visible in the first place even at quite large distances, mostly thanks to the brightness of their HII regions and massive stars and young clusters \citep[see e.g.,][]{McQuinn2010,Cignoni2019}. While NGC~4490 and the bridge do present concentrations of neutral and molecular gas able to sustain efficient SF (see Figure \ref{fig:F200W_image_tiles}), what is it that allows a SF burst at the border between NGC~4485 and the bridge? Are the current components of tile 5 still attributable to NGC~4485 or have they eventually become part of the bridge and exploit its gas richness? More detailed hydrodynamical simulations of the systems would be needed to answer this question.
 
The key results of our analysis are summarized as follows.
\begin{itemize}
    
    \item We resolved stellar populations across a wide range of evolutionary phases, from the youngest ages probed by upper-MS, RSG, and blue-loop stars, to intermediate-age AGB stars and an old population traced by RGB stars.

     \item We showed that, among the set of isochrones most commonly used in studies of resolved stellar populations and star formation histories, only the PARSEC-COLIBRI \citep{Bressan2012,Marigo2008,Marigo2013,Marigo2017} models provide a sufficiently detailed treatment of the complex physical processes occurring during the AGB evolutionary phase. Owing to their high luminosities, AGB stars dominate the observed near-infrared CMDs of star-forming galaxies beyond the Local Group, where fainter evolutionary features that trace similar ages, such as the red clump and the oldest main-sequence turn-off, are generally inaccessible. Nevertheless, significant discrepancies between the observed and predicted distributions of AGB stars in our CMDs remain, highlighting the need for further improvements in the modeling of AGB stars.

    \item The spatially resolved SFHs reveal spatially extended recent star formation across NGC~4485 and NGC~4490, with prominent young star-forming complexes located in the tidal bridge away from the central region of NGC~4485 (i.e. tiles 2 and 5), and reaching the outskirts of NGC~4490 (tiles 9, 10, 12, 13, 15, 16). However, the lack of SFHs for the two central tiles of NGC~4490 prevents us from directly characterizing its star-formation activity in the central regions.

    \item The SFHs of NGC~4485, NGC~4490, the tidal bridge, and the system as a whole exhibit two significant bursts of star formation, one occurring $\sim10\mbox{--}30$ Myr ago and a second peaking at approximately $500\mbox{--}800$ Myr ago. These episodes of enhanced star formation occur a few hundred million years after the epochs of the first and second perigalactic passages predicted by the independent N-body simulations of \citet{Pearson2018}, which took place approximately $1.3$ Gyr and $\sim200$ Myr ago, respectively. The tidal bridge shows the highest increase with respect to the averaged SFR during the bursts, followed by NGC~4490 and NGC~4485.

    \item Taken together, our results support a scenario in which the NGC~4485/4490 system has experienced at least two major interaction events. These encounters appear to have played a fundamental role in shaping the recent SFH of both galaxies and in producing the actively star-forming tidal bridge observed today. These findings are consistent with recent numerical simulations of interacting gas-rich low-mass galaxies \citep[e.g.,][]{Lahen2019,Lahen2020,Ejdetjarn2025}. In particular, these simulations predict enhanced star formation following perigalactic passages, both within tidal bridges and in off-center star-forming clumps, as a consequence of interaction-induced gas redistribution and compression.
\end{itemize}

In conclusion, the superior spatial resolution and sensitivity of the JWST/NIRCam observations \citep{gregg2024,pedrini2024,Knutas2025,Bortolini2025}, combined with our CMD-fitting routine \texttt{SFERA 2.0}, enable one of the most detailed spatially resolved studies to date of the stellar populations and star formation history of NGC~4485/NGC~4490, an interacting, gas-rich, low-mass galaxy pair at a distance of $\sim7.4$ Mpc. The SFHs presented in this work provide robust constraints on the evolutionary history of the system and represent valuable inputs for future dynamical and chemical-evolution models. More broadly, our results highlight the potential role of galaxy interactions in driving bursty star formation in gas-rich, low-mass galaxies, a process that is expected to be particularly important in the early Universe, when such systems and mergers were more prevalent. In this context, our resolved view of a nearby interacting galaxy pair provides a valuable benchmark for interpreting recent JWST observations of the earliest galaxies \citep{Adamo2025,Curti2025,Stark2026}.

\begin{acknowledgments}

This work is based in part on observations made with the NASA/ESA/CSA James Webb Space Telescope, which is operated by the Association of Universities for Research in Astronomy, Inc., under NASA contract NAS 5-03127.
These observations are associated with program \# 1783. Support for program \# 1783 was provided by NASA through a grant from the Space Telescope Science Institute,
which is operated by the Association of Universities for Research in Astronomy, Inc., under NASA contract NAS 5-03127.
AA acknowledges support from Vetenskapsr\aa det 2021-05559. A.A and A.P. acknowledge support from the Swedish National Space Agency (SNSA) through the grant 2021- 00108. A.A. and H.F.V. acknowledges support from (SNSA) 2023-00260.
KG is supported by the Australian Research Council through the Discovery Early Career Researcher Award (DECRA) Fellowship (project number DE220100766) funded by the Australian Government. 

\end{acknowledgments}

\begin{contribution}

GB performed the artificial star tests, wrote the \texttt{SFERA 2.0} routine and performed the analysis presented in the paper. He wrote the bulk of the text and produced all the figures. M. Correnti performed the PSF-fitting photometry with DOLPHOT. AA, M. Correnti, M. Cignoni, MT, and GO helped with the analysis, interpretations of the results and in drafting the paper. The remaining authors provided comments and helped in the writing and proofreading of the draft.


\end{contribution}

\section*{Data Availability}
The data supporting the findings of this study are openly available at the following MAST DOI: \url{http://dx.doi.org/10.17909/ybh0-r149}. All our data products are available at MAST as a High Level Science Product via \url{https://doi.org/10.17909/6dc1-9h53} and DOI: 10.17909/6dc1-9h53 and on the FEAST webpage \url{https://feast-survey.github.io/}.

%
\facilities{JWST/NIRCam.}

\software{DOLPHOT \citep[\url{http://americano.dolphinsim.com/dolphot/};][]{Dolphin2000, Dolphin2016}, {\tt Astropy} \citep{astropy:2013,astropy:2018,astropy:2022}, {\tt Numpy} \citep{Numpy2020}, {\tt Pandas} \citep{Pandas2010}, {\tt SciPy} \citep{Scipy2020}, {\tt Matplotlib} \citep{Matplotlib2007}, {\tt SAOImageDS9} (developed by the Smithsonian Astrophysical Observatory 2000, \citealt{Joye2003}), {\tt TOPCAT} \citep{Topcat2005}, \texttt{SFERA 2.0} \citep{Bortolini2024a}, \texttt{Pikaia} \citep{Charbonneau1995}.
          }


\appendix

\section{Exploring sources of systematic errors in the star formation histories} \label{sec:appendixA}

\begin{figure*}
    \centering
    \includegraphics[width=1.0\linewidth]{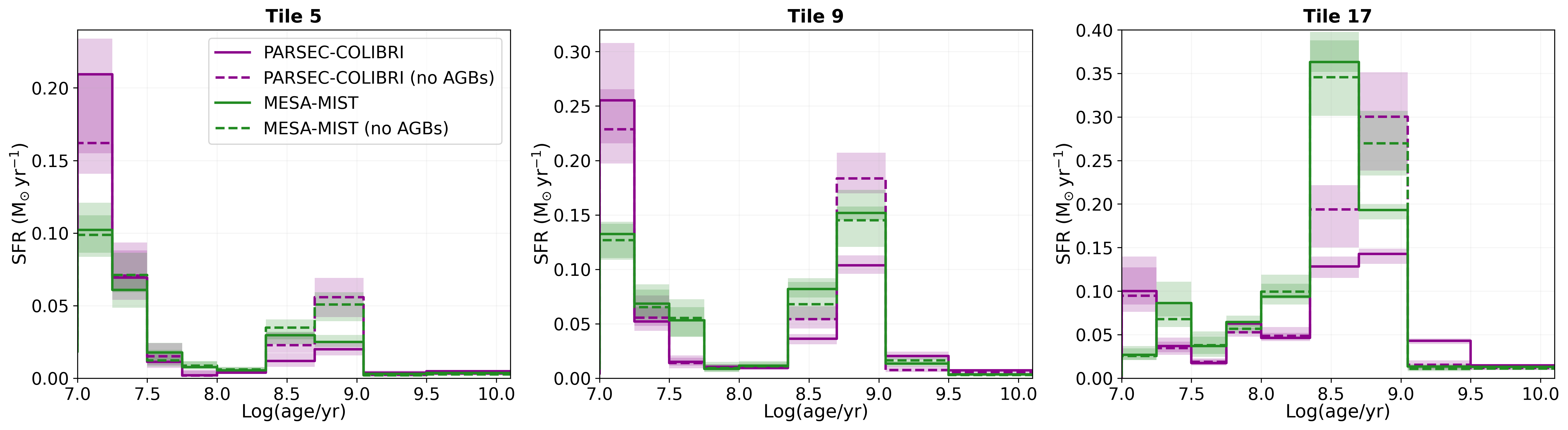}
    \caption{Comparison between the \texttt{SFERA 2.0} recovered SFHs for tile 5, 9 and 17 using PARSEC-COLIBRI (dark violet solid line) and MESA-MIST (green solid line) stellar evolutionary models to generate the synthetic CMDs, but with all the other parameters fixed. As discussed in Section \ref{sec:The Adequacy of AGB Stellar Evolution Models} PARSEC-COLIBRI models self-consistently include TPAGB AGB stars, while MESA-MIST models lack a comprehensive treatment of these complex evolutionary phases.  The differences observed in the SFHs reflect the systematic uncertainties introduced by the intrinsic differences in the stellar models. The dashed lines are the SFHs recovered with the two sets of stellar models, but excluding the region of the CMD occupied by AGB stars from the fitting routine.}
    \label{fig:PARSEC_vs_MIST_SFHs.png}
\end{figure*}

As discussed in Section \ref{sec:The Adequacy of AGB Stellar Evolution Models}, the PARSEC-COLIBRI stellar models provide a more realistic treatment of the complex AGB evolutionary phase than the MESA-MIST models. Nevertheless, to assess the systematic uncertainties introduced by adopting a different set of stellar models, we recovered the SFHs of Tiles 5, 9, and 17 using \texttt{SFERA 2.0} with the same set of best-fitting parameters presented in the main text, but employing MESA-MIST isochrones to construct the synthetic CMDs. The results are shown in Figure \ref{fig:PARSEC_vs_MIST_SFHs.png}, where the SFHs derived with PARSEC-COLIBRI are represented by dark-violet solid lines, while those obtained with MESA-MIST are shown as green solid lines.

As expected, the SFHs for the three tiles differ significantly in the age range traced by AGB stars ($100$ Myr $\lesssim$ age $\lesssim 5$ Gyr), with MESA-MIST predicting higher star formation rates than those inferred from PARSEC-COLIBRI models. Another notable difference is apparent in Tiles 5 and 9, where the inferred strength of the youngest burst of star formation (age $\lesssim 30$ Myr) differs between the two sets of models, with PARSEC-COLIBRI systematically recovering higher rates. These epochs are traced by massive and luminous red supergiants and blue-loop helium-burning stars. This difference is not surprising, given that our understanding of massive helium-burning stellar evolution is still subject to significant uncertainties. While not as uncertain as the TP-AGB phase, this evolutionary stage remains a major source of disagreement between stellar evolution codes, due to differing prescriptions for convective mixing and overshooting, rotation, nuclear reaction rates, and mass loss \citep{Maeder2000,Eldridge2004,Maeder2001,Gallart2005,Bressan2012}. Despite these differences, the predicted ages of the star formation peaks are consistent between the two stellar models, demonstrating the robustness of the synthetic CMD method. 

\begin{figure*}
    \centering
    \includegraphics[width=1.0\linewidth]{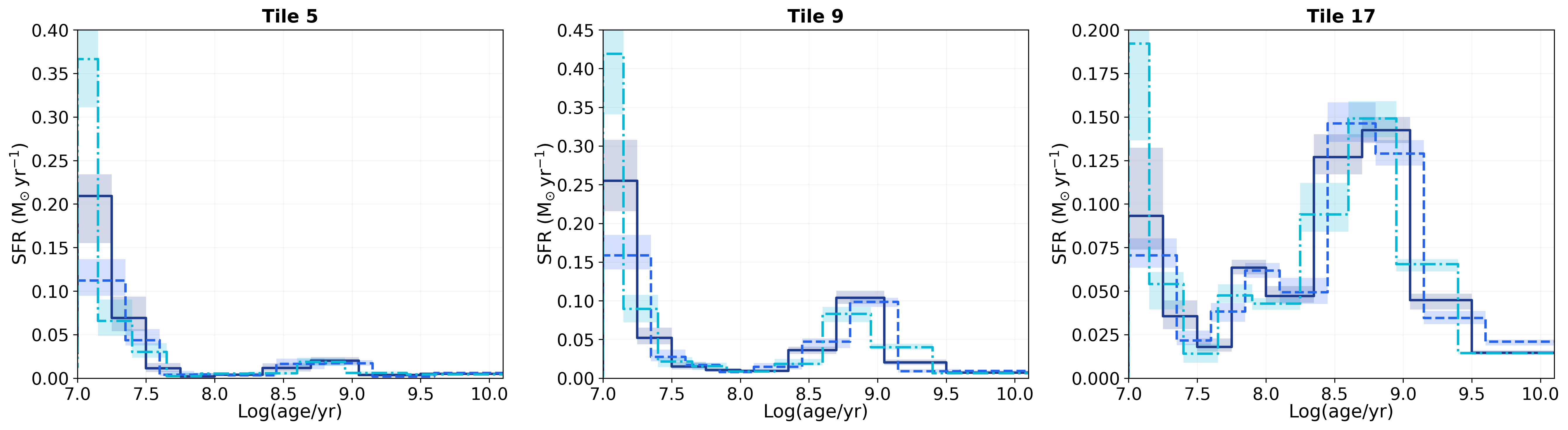}
    \caption{Star formation histories of tile 5, 9, and 17 derived with \texttt{SFERA 2.0} but with three different sets of age bins for the generation of the synthetic CMDs. The best-fit SFHs from the main text are shown as dark-blue solid lines, while the SFHs recovered with the $\pm 0.1,\mathrm{dex}$ in $\log(\mathrm{age/yr})$ age binnings are shown as dashed blue and dashed-dotted cyan lines, respectively.}
    \label{fig:SFHs_agebinning.png}
\end{figure*}

In Figure \ref{fig:PARSEC_vs_MIST_SFHs.png}, we also show the \texttt{SFERA 2.0} SFHs obtained using PARSEC-COLIBRI (dashed dark-violet line) and MESA-MIST (dashed green line) stellar models, excluding from the fitting procedure the region of the CMD populated by AGB stars. As expected, these SFHs are consistent within the uncertainties with those obtained when AGB stars are included, in the age ranges not constrained by these stars (i.e., younger than $\sim 100$ Myr and older than $\sim 5$ Gyr). In contrast, in the age interval dominated by AGB stars in our CMDs, the PARSEC-COLIBRI solutions without AGB constraints yield star formation rates more similar to those inferred using MESA-MIST models, which remain unchanged within the uncertainties when AGB stars are excluded from the fit. These results confirm the crucial importance of detailed AGB evolutionary models to properly fit the CMD of star-forming galaxies beyond the Local Group, and derive an unbiased SFH.

To explore the impact of the adopted age binning for the simple stellar population synthetic CMDs on the recovered SFHs, we ran \texttt{SFERA 2.0} on the CMDs of Tiles 5, 9, and 17 using the same best-fitting parameters as presented in the main text, but assuming two alternative age binnings. In particular, we shifted the age bins by $\pm 0.1,\mathrm{dex}$ in $\log(\mathrm{age/yr})$ relative to those adopted in the main text. The results are shown in Figure \ref{fig:SFHs_agebinning.png}, where the best-fit SFHs from the main text are shown as dark-blue solid lines, while the SFHs recovered with the perturbed age binnings are shown as dashed blue and dashed-dotted cyan lines, respectively. The timing and overall strength of the recovered SFHs remain unchanged, with the exception of the youngest and oldest bins at the edges of the age range, which are more sensitive to the adopted binning.

The adopted IMF and unresolved binary fraction may affect the recovered SFHs. However, given the limited depth of our CMDs, which are dominated by evolved stars and do not reach the low-mass main sequence, we expect these effects to be negligible compared with other sources of uncertainty, such as photometric errors at the distance of our target galaxy ($\sim 7$ Mpc) and differences among stellar evolution models \citep[see, e.g.,][]{Cignoni2010}.

\section{Tiles completeness maps} \label{sec:Tiles completeness maps}
\begin{figure*}
    \centering
    \includegraphics[width=0.8\linewidth]{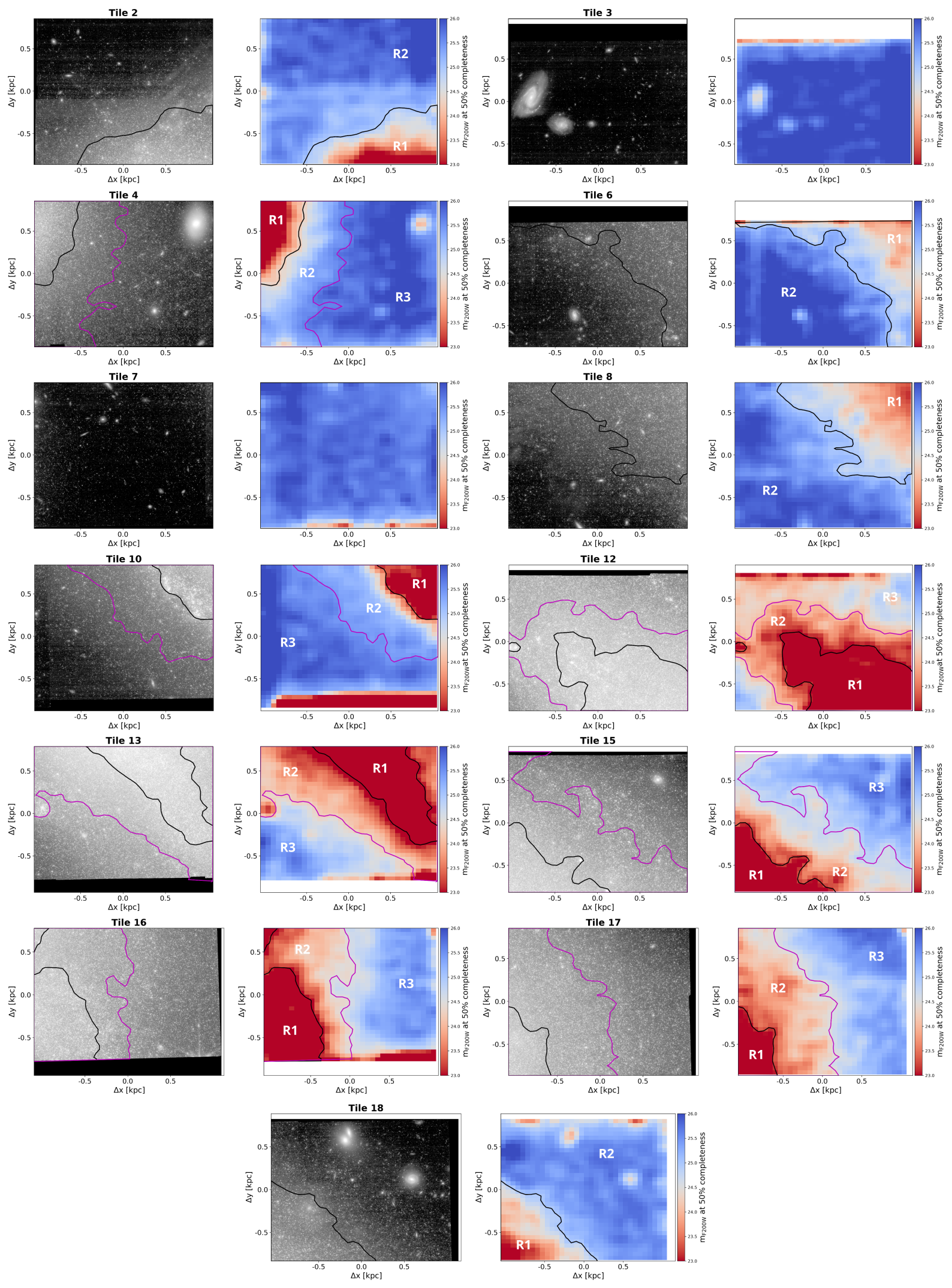}
    \caption{Same as Figure~\ref{fig:tiles_completeness_map_zoomin}, but for the remaining tiles analyzed in this paper, namely tiles 2, 3, 4, 6, 7, 8, 10, 12, 13, 15, 16, 17, and 18. The different sub-regions into which each tile is divided, based on the completeness maps, are labeled R1, R2, and R3. Owing to the relatively uniform completeness maps of tiles 3 and 7, we do not further subdivide these tiles into sub-regions.}
    \label{fig:rest_of_tiles_completenessmaps.png}
\end{figure*}

Figure~\ref{fig:rest_of_tiles_completenessmaps.png} shows zoom-ins of the NIRCam F200W filter mosaic and completeness maps for the remaining tiles for which we recovered the star formation history that are not shown in Figure \ref{fig:tiles_completeness_map_zoomin} in the main text. Each pixel in the completeness maps covers an area of $1.7'' \times 1.4''$ $(\sim 60 \times 50 \, \mathrm{pc})$ and is color-coded according to the F200W magnitude at which the $50\%$ completeness limit is reached, ranging from red (shallower CMDs and lower completeness) to blue (deeper CMDs and higher completeness). The different sub-regions into which each tile is divided are delimited by black and magenta solid lines and are labeled R1, R2, and R3, in order of increasing completeness. Tiles 3 and 7 exhibit relatively uniform completeness across their entire area; therefore, they were not further subdivided into sub-regions. Most of the rest of the tiles required no more than two sub-regions (for instance tiles 2, 6, 8), while more crowded tiles required up to three sub-regions (see for instance tiles 10, 12, or 15).

\bibliography{main}{}

@ARTICLE{correnti2025,
       author = {{Correnti}, Matteo and {Bortolini}, Giacomo and {Dell'Agli}, Flavia and {Adamo}, Angela and {Cignoni}, Michele and {Sacchi}, Elena and {Tosi}, Monica and {Pedrini}, Alex and {Buckner}, Anne S.~M. and {Calzetti}, Daniela and {Duarte-Cabral}, Ana and {Elmegreen}, Bruce G. and {Faustino Vieira}, Helena and {Gallagher}, John S. and {Grasha}, Kathryn and {Gregg}, Benjamin and {Johnson}, Kelsey E. and {Lai}, Thomas S. -Y. and {Lapeer}, Drew and {Linden}, Sean T. and {Messa}, Matteo and {Ostlin}, Goran and {Sabbi}, Elena and {Smith}, Linda J. and {Ventura}, Paolo},
        title = "{FEAST: Probing the Stellar Population of the Starburst Dwarf Galaxy NGC 4449 with JWST/NIRCam}",
      journal = {\apj},
         year = 2025,
        month = sep,
       volume = {990},
       number = {1},
          eid = {72},
        pages = {72},
          doi = {10.3847/1538-4357/adec74},
archivePrefix = {arXiv},
       eprint = {2507.03420},
 primaryClass = {astro-ph.GA},
       adsurl = {https://ui.adsabs.harvard.edu/abs/2025ApJ...990...72C}
}

@ARTICLE{Bortolini2025,
       author = {{Bortolini}, Giacomo and {Correnti}, Matteo and {Adamo}, Angela and {Cignoni}, Michele and {Sacchi}, Elena and {Tosi}, Monica and {{\"O}stlin}, G{\"o}ran and {Kapodistrias}, Anastasios and {Bik}, Arjan and {Calzetti}, Daniela and {Duarte-Cabral}, Ana and {Dell'Agli}, Flavia and {Gallagher}, John S. and {Gregg}, Benjamin and {Grasha}, Kathryn and {Lai}, Thomas S. -Y. and {Lapeer}, Drew and {Linden}, Sean T. and {Messa}, Matteo and {Pedrini}, Alex and {Sabbi}, Elena and {Smith}, Linda J. and {Faustino Vieira}, Helena and {Cannon}, John M. and {Duarte Puertas}, Salvador and {Robert}, Carmelle},
        title = "{FEAST: JWST/NIRCam View of the Resolved Stellar Populations of the Interacting Dwarf Galaxies NGC 4485 and NGC 4490}",
      journal = {\apj},
         year = 2025,
        month = oct,
       volume = {991},
       number = {2},
          eid = {212},
        pages = {212},
          doi = {10.3847/1538-4357/adfccc},
archivePrefix = {arXiv},
       eprint = {2509.01740},
 primaryClass = {astro-ph.GA},
       adsurl = {https://ui.adsabs.harvard.edu/abs/2025ApJ...991..212B}
}

@ARTICLE{pedrini2024,
       author = {{Pedrini}, Alex and {Adamo}, Angela and {Calzetti}, Daniela and {Bik}, Arjan and {Gregg}, Benjamin and {Linden}, Sean T. and {Bajaj}, Varun and {Ryon}, Jenna E. and {Ali}, Ahmad A. and {Bortolini}, Giacomo and {Correnti}, Matteo and {Elmegreen}, Bruce G. and {Elmegreen}, Debra Meloy and {Gallagher}, John S. and {Grasha}, Kathryn and {Gutermuth}, Robert A. and {Johnson}, Kelsey E. and {Melinder}, Jens and {Messa}, Matteo and {{\"O}stlin}, G{\"o}ran and {Sabbi}, Elena and {Smith}, Linda J. and {Tosi}, Monica and {Faustino Vieira}, Helena},
        title = "{FEAST: Feedback in Emerging extragAlactic Star ClusTers: JWST Spots Polycyclic Aromatic Hydrocarbon Destruction in NGC 628 during the Emerging Phase of Star Formation}",
      journal = {\apj},
         year = 2024,
        month = aug,
       volume = {971},
       number = {1},
          eid = {32},
        pages = {32},
          doi = {10.3847/1538-4357/ad534d},
archivePrefix = {arXiv},
       eprint = {2406.01666},
 primaryClass = {astro-ph.GA},
       adsurl = {https://ui.adsabs.harvard.edu/abs/2024ApJ...971...32P}
}

@ARTICLE{gregg2024,
       author = {{Gregg}, Benjamin and {Calzetti}, Daniela and {Adamo}, Angela and {Bajaj}, Varun and {Ryon}, Jenna E. and {Linden}, Sean T. and {Correnti}, Matteo and {Cignoni}, Michele and {Messa}, Matteo and {Sabbi}, Elena and {Gallagher}, John S. and {Grasha}, Kathryn and {Pedrini}, Alex and {Gutermuth}, Robert A. and {Melinder}, Jens and {Kotulla}, Ralf and {P{\'e}rez}, Gustavo and {Krumholz}, Mark R. and {Bik}, Arjan and {{\"O}stlin}, G{\"o}ran and {Johnson}, Kelsey E. and {Bortolini}, Giacomo and {Smith}, Linda J. and {Tosi}, Monica and {Maji}, Subhransu and {Faustino Vieira}, Helena},
        title = "{Feedback in Emerging Extragalactic Star Clusters, FEAST: The Relation between 3.3 {\ensuremath{\mu}}m Polycyclic Aromatic Hydrocarbon Emission and Star Formation Rate Traced by Ionized Gas in NGC 628}",
      journal = {\apj},
         year = 2024,
        month = aug,
       volume = {971},
       number = {1},
          eid = {115},
        pages = {115},
          doi = {10.3847/1538-4357/ad54b4},
archivePrefix = {arXiv},
       eprint = {2405.09667},
 primaryClass = {astro-ph.GA},
       adsurl = {https://ui.adsabs.harvard.edu/abs/2024ApJ...971..115G}
}

@ARTICLE{Rieke2023,
       author = {{Rieke}, Marcia J. and {Kelly}, Douglas M. and {Misselt}, Karl and {Stansberry}, John and {Boyer}, Martha and {Beatty}, Thomas and {Egami}, Eiichi and {Florian}, Michael and {Greene}, Thomas P. and {Hainline}, Kevin and {Leisenring}, Jarron and {Roellig}, Thomas and {Schlawin}, Everett and {Sun}, Fengwu and {Tinnin}, Lee and {Williams}, Christina C. and {Willmer}, Christopher N.~A. and {Wilson}, Debra and {Clark}, Charles R. and {Rohrbach}, Scott and {Brooks}, Brian and {Canipe}, Alicia and {Correnti}, Matteo and {DiFelice}, Audrey and {Gennaro}, Mario and {Girard}, Julien H. and {Hartig}, George and {Hilbert}, Bryan and {Koekemoer}, Anton M. and {Nikolov}, Nikolay K. and {Pirzkal}, Norbert and {Rest}, Armin and {Robberto}, Massimo and {Sunnquist}, Ben and {Telfer}, Randal and {Wu}, Chi Rai and {Ferry}, Malcolm and {Lewis}, Dan and {Baum}, Stefi and {Beichman}, Charles and {Doyon}, Ren{\'e} and {Dressler}, Alan and {Eisenstein}, Daniel J. and {Ferrarese}, Laura and {Hodapp}, Klaus and {Horner}, Scott and {Jaffe}, Daniel T. and {Johnstone}, Doug and {Krist}, John and {Martin}, Peter and {McCarthy}, Donald W. and {Meyer}, Michael and {Rieke}, George H. and {Trauger}, John and {Young}, Erick T.},
        title = "{Performance of NIRCam on JWST in Flight}",
      journal = {\pasp},
         year = 2023,
        month = feb,
       volume = {135},
       number = {1044},
          eid = {028001},
        pages = {028001},
          doi = {10.1088/1538-3873/acac53},
archivePrefix = {arXiv},
       eprint = {2212.12069},
 primaryClass = {astro-ph.IM},
       adsurl = {https://ui.adsabs.harvard.edu/abs/2023PASP..135b8001R}
}

@ARTICLE{Dolphin2000,
       author = {{Dolphin}, Andrew E.},
        title = "{WFPC2 Stellar Photometry with HSTPHOT}",
      journal = {\pasp},
         year = 2000,
        month = oct,
       volume = {112},
       number = {776},
        pages = {1383-1396},
          doi = {10.1086/316630},
archivePrefix = {arXiv},
       eprint = {astro-ph/0006217},
 primaryClass = {astro-ph},
       adsurl = {https://ui.adsabs.harvard.edu/abs/2000PASP..112.1383D}
}

@software{Dolphin2016,
       author = {{Dolphin}, Andrew},
        title = "{DOLPHOT: Stellar photometry}",
 howpublished = {Astrophysics Source Code Library, record ascl:1608.013},
         year = 2016,
        month = aug,
          eid = {ascl:1608.013},
archivePrefix = {ascl},
       eprint = {1608.013},
       adsurl = {https://ui.adsabs.harvard.edu/abs/2016ascl.soft08013D}
}

@ARTICLE{Weisz2024,
       author = {{Weisz}, Daniel R. and {Dolphin}, Andrew E. and {Savino}, Alessandro and {McQuinn}, Kristen B.~W. and {Newman}, Max J.~B. and {Williams}, Benjamin F. and {Kallivayalil}, Nitya and {Anderson}, Jay and {Boyer}, Martha L. and {Correnti}, Matteo and {Geha}, Marla C. and {Sandstrom}, Karin M. and {Cole}, Andrew A. and {Warfield}, Jack T. and {Skillman}, Evan D. and {Cohen}, Roger E. and {Beaton}, Rachael and {Bressan}, Alessandro and {Bolatto}, Alberto and {Boylan-Kolchin}, Michael and {Brooks}, Alyson M. and {Bullock}, James S. and {Conroy}, Charlie and {Cooper}, Michael C. and {Dalcanton}, Julianne J. and {Dotter}, Aaron L. and {Fritz}, Tobias K. and {Garling}, Christopher T. and {Gennaro}, Mario and {Gilbert}, Karoline M. and {Girardi}, Leo and {Johnson}, Benjamin D. and {Johnson}, L. Clifton and {Kalirai}, Jason and {Kirby}, Evan N. and {Lang}, Dustin and {Marigo}, Paola and {Richstein}, Hannah and {Schlafly}, Edward F. and {Tollerud}, Erik J. and {Wetzel}, Andrew},
        title = "{The JWST Resolved Stellar Populations Early Release Science Program. V. DOLPHOT Stellar Photometry for NIRCam and NIRISS}",
      journal = {\apjs},
         year = 2024,
        month = apr,
       volume = {271},
       number = {2},
          eid = {47},
        pages = {47},
          doi = {10.3847/1538-4365/ad2600},
archivePrefix = {arXiv},
       eprint = {2402.03504},
 primaryClass = {astro-ph.GA},
       adsurl = {https://ui.adsabs.harvard.edu/abs/2024ApJS..271...47W}
}

@ARTICLE{Cignoni2010,
       author = {{Cignoni}, Michele and {Tosi}, Monica},
        title = "{Star Formation Histories of Dwarf Galaxies from the Colour-Magnitude Diagrams of Their Resolved Stellar Populations}",
      journal = {Advances in Astronomy},
         year = 2010,
        month = jan,
       volume = {2010},
          eid = {158568},
        pages = {158568},
          doi = {10.1155/2010/158568},
archivePrefix = {arXiv},
       eprint = {0909.4234},
 primaryClass = {astro-ph.GA},
       adsurl = {https://ui.adsabs.harvard.edu/abs/2010AdAst2010E...3C}
}

@ARTICLE{tosi1991,
       author = {{Tosi}, M. and {Greggio}, L. and {Marconi}, G. and {Focardi}, P.},
        title = "{Star Formation in Dwarf Irregular Galaxies: Sextans B}",
      journal = {\aj},
         year = 1991,
        month = sep,
       volume = {102},
        pages = {951},
          doi = {10.1086/115925},
       adsurl = {https://ui.adsabs.harvard.edu/abs/1991AJ....102..951T}
}

@ARTICLE{Annibali2003,
       author = {{Annibali}, F. and {Greggio}, L. and {Tosi}, M. and {Aloisi}, A. and {Leitherer}, Claus},
        title = "{The Star Formation History of NGC 1705: A Poststarburst Galaxy on the Verge of Activity}",
      journal = {\aj},
         year = 2003,
        month = dec,
       volume = {126},
       number = {6},
        pages = {2752-2773},
          doi = {10.1086/379556},
archivePrefix = {arXiv},
       eprint = {astro-ph/0309265},
 primaryClass = {astro-ph},
       adsurl = {https://ui.adsabs.harvard.edu/abs/2003AJ....126.2752A}
}

@ARTICLE{Gallart1999,
       author = {{Gallart}, Carme and {Freedman}, Wendy L. and {Aparicio}, Antonio and {Bertelli}, Giampaolo and {Chiosi}, Cesare},
        title = "{The Star Formation History of the Local Group Dwarf Galaxy Leo I}",
      journal = {\aj},
         year = 1999,
        month = nov,
       volume = {118},
       number = {5},
        pages = {2245-2261},
          doi = {10.1086/301078},
archivePrefix = {arXiv},
       eprint = {astro-ph/9906121},
 primaryClass = {astro-ph},
       adsurl = {https://ui.adsabs.harvard.edu/abs/1999AJ....118.2245G}
}

@ARTICLE{Dolphin2005,
       author = {{Dolphin}, Andrew E. and {Weisz}, Daniel R. and {Skillman}, Evan D. and {Holtzman}, Jon A.},
        title = "{Star Formation Histories of Local Group Dwarf Galaxies}",
      journal = {arXiv e-prints},
         year = 2005,
        month = jun,
          eid = {astro-ph/0506430},
        pages = {astro-ph/0506430},
          doi = {10.48550/arXiv.astro-ph/0506430},
archivePrefix = {arXiv},
       eprint = {astro-ph/0506430},
 primaryClass = {astro-ph},
       adsurl = {https://ui.adsabs.harvard.edu/abs/2005astro.ph..6430D}
}

@ARTICLE{McQuinn2010,
       author = {{McQuinn}, Kristen B.~W. and {Skillman}, Evan D. and {Cannon}, John M. and {Dalcanton}, Julianne and {Dolphin}, Andrew and {Hidalgo-Rodr{\'\i}guez}, Sebastian and {Holtzman}, Jon and {Stark}, David and {Weisz}, Daniel and {Williams}, Benjamin},
        title = "{The Nature of Starbursts. I. The Star Formation Histories of Eighteen Nearby Starburst Dwarf Galaxies}",
      journal = {\apj},
         year = 2010,
        month = sep,
       volume = {721},
       number = {1},
        pages = {297-317},
          doi = {10.1088/0004-637X/721/1/297},
archivePrefix = {arXiv},
       eprint = {1008.1589},
 primaryClass = {astro-ph.CO},
       adsurl = {https://ui.adsabs.harvard.edu/abs/2010ApJ...721..297M}
}

@ARTICLE{Weisz2014,
       author = {{Weisz}, Daniel R. and {Dolphin}, Andrew E. and {Skillman}, Evan D. and {Holtzman}, Jon and {Gilbert}, Karoline M. and {Dalcanton}, Julianne J. and {Williams}, Benjamin F.},
        title = "{The Star Formation Histories of Local Group Dwarf Galaxies. I. Hubble Space Telescope/Wide Field Planetary Camera 2 Observations}",
      journal = {\apj},
         year = 2014,
        month = jul,
       volume = {789},
       number = {2},
          eid = {147},
        pages = {147},
          doi = {10.1088/0004-637X/789/2/147},
archivePrefix = {arXiv},
       eprint = {1404.7144},
 primaryClass = {astro-ph.GA},
       adsurl = {https://ui.adsabs.harvard.edu/abs/2014ApJ...789..147W}
}

@ARTICLE{Sacchi2016,
       author = {{Sacchi}, E. and {Annibali}, F. and {Cignoni}, M. and {Aloisi}, A. and {Sohn}, T. and {Tosi}, M. and {van der Marel}, R.~P. and {Grocholski}, A.~J. and {James}, B.},
        title = "{Stellar Populations and Star Formation History of the Metal-poor Dwarf Galaxy DDO 68}",
      journal = {\apj},
         year = 2016,
        month = oct,
       volume = {830},
       number = {1},
          eid = {3},
        pages = {3},
          doi = {10.3847/0004-637X/830/1/3},
archivePrefix = {arXiv},
       eprint = {1604.06239},
 primaryClass = {astro-ph.GA},
       adsurl = {https://ui.adsabs.harvard.edu/abs/2016ApJ...830....3S}
}

@ARTICLE{Cignoni2019,
       author = {{Cignoni}, M. and {Sacchi}, E. and {Tosi}, M. and {Aloisi}, A. and {Cook}, D.~O. and {Calzetti}, D. and {Lee}, J.~C. and {Sabbi}, E. and {Thilker}, D.~A. and {Adamo}, A. and {Dale}, D.~A. and {Elmegreen}, B.~G. and {Gallagher}, III, J.~S. and {Grebel}, E.~K. and {Johnson}, K.~E. and {Messa}, M. and {Smith}, L.~J. and {Ubeda}, L.},
        title = "{Star Formation Histories of the LEGUS Dwarf Galaxies. III. The Nonbursty Nature of 23 Star-forming Dwarf Galaxies}",
      journal = {\apj},
         year = 2019,
        month = dec,
       volume = {887},
       number = {2},
          eid = {112},
        pages = {112},
          doi = {10.3847/1538-4357/ab53d5},
archivePrefix = {arXiv},
       eprint = {1911.04496},
 primaryClass = {astro-ph.GA},
       adsurl = {https://ui.adsabs.harvard.edu/abs/2019ApJ...887..112C}
}

@ARTICLE{Sacchi2018,
       author = {{Sacchi}, E. and {Cignoni}, M. and {Aloisi}, A. and {Tosi}, M. and {Calzetti}, D. and {Lee}, J.~C. and {Adamo}, A. and {Annibali}, F. and {Dale}, D.~A. and {Elmegreen}, B.~G. and {Gouliermis}, D.~A. and {Grasha}, K. and {Grebel}, E.~K. and {Hunter}, D.~A. and {Sabbi}, E. and {Smith}, L.~J. and {Thilker}, D.~A. and {Ubeda}, L. and {Whitmore}, B.~C.},
        title = "{Star Formation Histories of the LEGUS Dwarf Galaxies. II. Spatially Resolved Star Formation History of the Magellanic Irregular NGC 4449}",
      journal = {\apj},
         year = 2018,
        month = apr,
       volume = {857},
       number = {1},
          eid = {63},
        pages = {63},
          doi = {10.3847/1538-4357/aab844},
archivePrefix = {arXiv},
       eprint = {1803.08041},
 primaryClass = {astro-ph.GA},
       adsurl = {https://ui.adsabs.harvard.edu/abs/2018ApJ...857...63S}
}

@ARTICLE{McQuinn2024,
       author = {{McQuinn}, Kristen. B.~W. and {B. Newman}, Max J. and {Savino}, Alessandro and {Dolphin}, Andrew E. and {Weisz}, Daniel R. and {Williams}, Benjamin F. and {Boyer}, Martha L. and {Cohen}, Roger E. and {Correnti}, Matteo and {Cole}, Andrew A. and {Geha}, Marla C. and {Gennaro}, Mario and {Kallivayalil}, Nitya and {Sandstrom}, Karin M. and {Skillman}, Evan D. and {Anderson}, Jay and {Bolatto}, Alberto and {Boylan-Kolchin}, Michael and {Garling}, Christopher T. and {Gilbert}, Karoline M. and {Girardi}, L{\'e}o and {Kalirai}, Jason S. and {Mazzi}, Alessandro and {Pastorelli}, Giada and {Richstein}, Hannah and {Warfield}, Jack T.},
        title = "{The JWST Resolved Stellar Populations Early Release Science Program. IV. The Star Formation History of the Local Group Galaxy WLM}",
      journal = {\apj},
         year = 2024,
        month = jan,
       volume = {961},
       number = {1},
          eid = {16},
        pages = {16},
          doi = {10.3847/1538-4357/ad1105},
archivePrefix = {arXiv},
       eprint = {2312.03060},
 primaryClass = {astro-ph.GA},
       adsurl = {https://ui.adsabs.harvard.edu/abs/2024ApJ...961...16M}
}

@ARTICLE{Dolphin2002,
       author = {{Dolphin}, A.~E.},
        title = "{Numerical methods of star formation history measurement and applications to seven dwarf spheroidals}",
      journal = {\mnras},
         year = 2002,
        month = may,
       volume = {332},
       number = {1},
        pages = {91-108},
          doi = {10.1046/j.1365-8711.2002.05271.x},
archivePrefix = {arXiv},
       eprint = {astro-ph/0112331},
 primaryClass = {astro-ph},
       adsurl = {https://ui.adsabs.harvard.edu/abs/2002MNRAS.332...91D}
}

@ARTICLE{Aparicio2009,
       author = {{Aparicio}, Antonio and {Hidalgo}, Sebastian L.},
        title = "{IAC-pop: Finding the Star Formation History of Resolved Galaxies}",
      journal = {\aj},
         year = 2009,
        month = aug,
       volume = {138},
       number = {2},
        pages = {558-567},
          doi = {10.1088/0004-6256/138/2/558},
archivePrefix = {arXiv},
       eprint = {0906.0712},
 primaryClass = {astro-ph.CO},
       adsurl = {https://ui.adsabs.harvard.edu/abs/2009AJ....138..558A}
}

@ARTICLE{Cole2007,
       author = {{Cole}, Andrew A. and {Skillman}, Evan D. and {Tolstoy}, Eline and {Gallagher}, III, John S. and {Aparicio}, Antonio and {Dolphin}, Andrew E. and {Gallart}, Carme and {Hidalgo}, Sebastian L. and {Saha}, Abhijit and {Stetson}, Peter B. and {Weisz}, Daniel R.},
        title = "{Leo A: A Late-blooming Survivor of the Epoch of Reionization in the Local Group}",
      journal = {\apjl},
         year = 2007,
        month = apr,
       volume = {659},
       number = {1},
        pages = {L17-L20},
          doi = {10.1086/516711},
archivePrefix = {arXiv},
       eprint = {astro-ph/0702646},
 primaryClass = {astro-ph},
       adsurl = {https://ui.adsabs.harvard.edu/abs/2007ApJ...659L..17C}
}

@ARTICLE{Monelli2010,
       author = {{Monelli}, M. and {Hidalgo}, S.~L. and {Stetson}, P.~B. and {Aparicio}, A. and {Gallart}, C. and {Dolphin}, A.~E. and {Cole}, A.~A. and {Weisz}, D.~R. and {Skillman}, E.~D. and {Bernard}, E.~J. and {Mayer}, L. and {Navarro}, J.~F. and {Cassisi}, S. and {Drozdovsky}, I. and {Tolstoy}, E.},
        title = "{The ACS LCID Project. III. The Star Formation History of the Cetus dSph Galaxy: A Post-reionization Fossil}",
      journal = {\apj},
         year = 2010,
        month = sep,
       volume = {720},
       number = {2},
        pages = {1225-1245},
          doi = {10.1088/0004-637X/720/2/1225},
archivePrefix = {arXiv},
       eprint = {1002.4300},
 primaryClass = {astro-ph.CO},
       adsurl = {https://ui.adsabs.harvard.edu/abs/2010ApJ...720.1225M}
}

@ARTICLE{Skillman2017,
       author = {{Skillman}, Evan D. and {Monelli}, Matteo and {Weisz}, Daniel R. and {Hidalgo}, Sebastian L. and {Aparicio}, Antonio and {Bernard}, Edouard J. and {Boylan-Kolchin}, Michael and {Cassisi}, Santi and {Cole}, Andrew A. and {Dolphin}, Andrew E. and {Ferguson}, Henry C. and {Gallart}, Carme and {Irwin}, Mike J. and {Martin}, Nicolas F. and {Mart{\'\i}nez-V{\'a}zquez}, Clara E. and {Mayer}, Lucio and {McConnachie}, Alan W. and {McQuinn}, Kristen B.~W. and {Navarro}, Julio F. and {Stetson}, Peter B.},
        title = "{The ISLAndS Project. II. The Lifetime Star Formation Histories of Six Andomeda dSphS}",
      journal = {\apj},
         year = 2017,
        month = mar,
       volume = {837},
       number = {2},
          eid = {102},
        pages = {102},
          doi = {10.3847/1538-4357/aa60c5},
archivePrefix = {arXiv},
       eprint = {1606.01207},
 primaryClass = {astro-ph.GA},
       adsurl = {https://ui.adsabs.harvard.edu/abs/2017ApJ...837..102S}
}

@ARTICLE{Sacchi2021,
       author = {{Sacchi}, Elena and {Aloisi}, Alessandra and {Correnti}, Matteo and {Annibali}, Francesca and {Tosi}, Monica and {Garofalo}, Alessia and {Clementini}, Gisella and {Cignoni}, Michele and {James}, Bethan and {Marconi}, Marcella and {Muraveva}, Tatiana and {van der Marel}, Roeland},
        title = "{Reaching the Oldest Stars beyond the Local Group: Ancient Star Formation in UGC 4483}",
      journal = {\apj},
         year = 2021,
        month = apr,
       volume = {911},
       number = {1},
          eid = {62},
        pages = {62},
          doi = {10.3847/1538-4357/abea16},
archivePrefix = {arXiv},
       eprint = {2102.13119},
 primaryClass = {astro-ph.GA},
       adsurl = {https://ui.adsabs.harvard.edu/abs/2021ApJ...911...62S}
}

@ARTICLE{Zheng2025,
       author = {{Zheng}, Yun and {Yang}, Yujiao and {Zhang}, Yong-Kun and {Zheng}, Zheng and {Wang}, Jing and {Staveley-Smith}, Lister and {Tsai}, Chao-Wei and {Li}, Di and {Liu}, Chao and {Hu}, Jingjing and {Chen}, Huaxi and {Quan}, Donghui and {Zheng}, Yinghui and {Li}, Hangyuan},
        title = "{PANCAKE: A Python-based Numerical Color{\textendash}Magnitude Diagram Analysis Package}",
      journal = {\apjs},
         year = 2025,
        month = jul,
       volume = {279},
       number = {1},
          eid = {12},
        pages = {12},
          doi = {10.3847/1538-4365/add88c},
archivePrefix = {arXiv},
       eprint = {2505.04534},
 primaryClass = {astro-ph.GA},
       adsurl = {https://ui.adsabs.harvard.edu/abs/2025ApJS..279...12Z}
}

@ARTICLE{Garling2025,
       author = {{Garling}, Christopher T. and {Kallivayalil}, Nitya and {McQuinn}, Kristen B.~W. and {Warfield}, Jack T. and {Gennaro}, Mario and {Cohen}, Roger E.},
        title = "{Measuring Resolved Star Formation Histories from High-precision Color{\textendash}Magnitude Diagrams with StarFormationHistories.jl}",
      journal = {\apjs},
         year = 2025,
        month = apr,
       volume = {277},
       number = {2},
          eid = {61},
        pages = {61},
          doi = {10.3847/1538-4365/adbb64},
archivePrefix = {arXiv},
       eprint = {2407.19534},
 primaryClass = {astro-ph.GA},
       adsurl = {https://ui.adsabs.harvard.edu/abs/2025ApJS..277...61G}
}

@ARTICLE{Gallart2024,
       author = {{Gallart}, Carme and {Surot}, Francisco and {Cassisi}, Santi and {Fern{\'a}ndez-Alvar}, Emma and {Mirabal}, David and {Rivero}, Alicia and {Ruiz-Lara}, Tom{\'a}s and {Santos-Torres}, Judith and {Aznar-Menargues}, Guillem and {Battaglia}, Giuseppina and {Queiroz}, Anna B. and {Monelli}, Matteo and {Vasiliev}, Eugene and {Chiappini}, Cristina and {Helmi}, Amina and {Hill}, Vanessa and {Massari}, Davide and {Thomas}, Guillaume F.},
        title = "{Chronology of our Galaxy from Gaia colour-magnitude diagram fitting (ChronoGal). I. The formation and evolution of the thin disc from the Gaia Catalogue of Nearby Stars}",
      journal = {\aap},
         year = 2024,
        month = jul,
       volume = {687},
          eid = {A168},
        pages = {A168},
          doi = {10.1051/0004-6361/202349078},
archivePrefix = {arXiv},
       eprint = {2402.09399},
 primaryClass = {astro-ph.GA},
       adsurl = {https://ui.adsabs.harvard.edu/abs/2024A&A...687A.168G}
}

@ARTICLE{Bortolini2024a,
       author = {{Bortolini}, G. and {Cignoni}, M. and {Sacchi}, E. and {Tosi}, M. and {Annibali}, F. and {Pascale}, R. and {Bellazzini}, M. and {Calzetti}, D. and {Adamo}, A. and {Dale}, Daniel A. and {Fumagalli}, M. and {Gallagher}, John S. and {Grasha}, K. and {Johnson}, Kelsey E. and {Linden}, Sean T. and {Messa}, M. and {{\"O}stlin}, G. and {Sabbi}, E. and {Wofford}, A.},
        title = "{The spatially resolved star formation history of the dwarf spiral galaxy NGC 5474}",
      journal = {\mnras},
         year = 2024,
        month = jan,
       volume = {527},
       number = {3},
        pages = {5339-5355},
          doi = {10.1093/mnras/stad3524},
archivePrefix = {arXiv},
       eprint = {2311.08160},
 primaryClass = {astro-ph.GA},
       adsurl = {https://ui.adsabs.harvard.edu/abs/2024MNRAS.527.5339B}
}

@ARTICLE{Bressan2012,
       author = {{Bressan}, Alessandro and {Marigo}, Paola and {Girardi}, L{\'e}o. and {Salasnich}, Bernardo and {Dal Cero}, Claudia and {Rubele}, Stefano and {Nanni}, Ambra},
        title = "{PARSEC: stellar tracks and isochrones with the PAdova and TRieste Stellar Evolution Code}",
      journal = {\mnras},
         year = 2012,
        month = nov,
       volume = {427},
       number = {1},
        pages = {127-145},
          doi = {10.1111/j.1365-2966.2012.21948.x},
archivePrefix = {arXiv},
       eprint = {1208.4498},
 primaryClass = {astro-ph.SR},
       adsurl = {https://ui.adsabs.harvard.edu/abs/2012MNRAS.427..127B}
}

@ARTICLE{Marigo2017,
       author = {{Marigo}, Paola and {Girardi}, L{\'e}o and {Bressan}, Alessandro and {Rosenfield}, Philip and {Aringer}, Bernhard and {Chen}, Yang and {Dussin}, Marco and {Nanni}, Ambra and {Pastorelli}, Giada and {Rodrigues}, Tha{\'\i}se S. and {Trabucchi}, Michele and {Bladh}, Sara and {Dalcanton}, Julianne and {Groenewegen}, Martin A.~T. and {Montalb{\'a}n}, Josefina and {Wood}, Peter R.},
        title = "{A New Generation of PARSEC-COLIBRI Stellar Isochrones Including the TP-AGB Phase}",
      journal = {\apj},
         year = 2017,
        month = jan,
       volume = {835},
       number = {1},
          eid = {77},
        pages = {77},
          doi = {10.3847/1538-4357/835/1/77},
archivePrefix = {arXiv},
       eprint = {1701.08510},
 primaryClass = {astro-ph.SR},
       adsurl = {https://ui.adsabs.harvard.edu/abs/2017ApJ...835...77M}
}

@ARTICLE{Paxton2011,
       author = {{Paxton}, Bill and {Bildsten}, Lars and {Dotter}, Aaron and {Herwig}, Falk and {Lesaffre}, Pierre and {Timmes}, Frank},
        title = "{Modules for Experiments in Stellar Astrophysics (MESA)}",
      journal = {\apjs},
         year = 2011,
        month = jan,
       volume = {192},
       number = {1},
          eid = {3},
        pages = {3},
          doi = {10.1088/0067-0049/192/1/3},
archivePrefix = {arXiv},
       eprint = {1009.1622},
 primaryClass = {astro-ph.SR},
       adsurl = {https://ui.adsabs.harvard.edu/abs/2011ApJS..192....3P}
}

@ARTICLE{Choi2016,
       author = {{Choi}, Jieun and {Dotter}, Aaron and {Conroy}, Charlie and {Cantiello}, Matteo and {Paxton}, Bill and {Johnson}, Benjamin D.},
        title = "{Mesa Isochrones and Stellar Tracks (MIST). I. Solar-scaled Models}",
      journal = {\apj},
         year = 2016,
        month = jun,
       volume = {823},
       number = {2},
          eid = {102},
        pages = {102},
          doi = {10.3847/0004-637X/823/2/102},
archivePrefix = {arXiv},
       eprint = {1604.08592},
 primaryClass = {astro-ph.SR},
       adsurl = {https://ui.adsabs.harvard.edu/abs/2016ApJ...823..102C}
}

@ARTICLE{Dotter2016,
       author = {{Dotter}, Aaron},
        title = "{MESA Isochrones and Stellar Tracks (MIST) 0: Methods for the Construction of Stellar Isochrones}",
      journal = {\apjs},
         year = 2016,
        month = jan,
       volume = {222},
       number = {1},
          eid = {8},
        pages = {8},
          doi = {10.3847/0067-0049/222/1/8},
archivePrefix = {arXiv},
       eprint = {1601.05144},
 primaryClass = {astro-ph.SR},
       adsurl = {https://ui.adsabs.harvard.edu/abs/2016ApJS..222....8D}
}

@ARTICLE{Cignoni2015,
       author = {{Cignoni}, M. and {Sabbi}, E. and {van der Marel}, R.~P. and {Tosi}, M. and {Zaritsky}, D. and {Anderson}, J. and {Lennon}, D.~J. and {Aloisi}, A. and {de Marchi}, G. and {Gouliermis}, D.~A. and {Grebel}, E.~K. and {Smith}, L.~J. and {Zeidler}, P.},
        title = "{Hubble Tarantula Treasury Project. II. The Star-formation History of the Starburst Region NGC 2070 in 30 Doradus}",
      journal = {\apj},
         year = 2015,
        month = oct,
       volume = {811},
       number = {2},
          eid = {76},
        pages = {76},
          doi = {10.1088/0004-637X/811/2/76},
archivePrefix = {arXiv},
       eprint = {1505.04799},
 primaryClass = {astro-ph.SR},
       adsurl = {https://ui.adsabs.harvard.edu/abs/2015ApJ...811...76C}
}

@ARTICLE{Bortolini2024b,
       author = {{Bortolini}, Giacomo and {{\"O}stlin}, G{\"o}ran and {Habel}, Nolan and {Hirschauer}, Alec S. and {Jones}, Olivia C. and {Justtanont}, Kay and {Meixner}, Margaret and {Boyer}, Martha L. and {Blommaert}, Joris A.~D.~L. and {Crouzet}, Nicolas and {Lenki{\'c}}, Laura and {Nally}, Conor and {Sargent}, Beth A. and {van der Werf}, Paul and {G{\"u}del}, Manuel and {Henning}, Thomas and {Lagage}, Pierre O.},
        title = "{Imaging of I Zw 18 by JWST: II. Spatially resolved star formation history}",
      journal = {\aap},
         year = 2024,
        month = sep,
       volume = {689},
          eid = {A146},
        pages = {A146},
          doi = {10.1051/0004-6361/202450632},
archivePrefix = {arXiv},
       eprint = {2406.17429},
 primaryClass = {astro-ph.GA},
       adsurl = {https://ui.adsabs.harvard.edu/abs/2024A&A...689A.146B}
}

@ARTICLE{Kroupa2001,
       author = {{Kroupa}, Pavel},
        title = "{On the variation of the initial mass function}",
      journal = {\mnras},
         year = 2001,
        month = apr,
       volume = {322},
       number = {2},
        pages = {231-246},
          doi = {10.1046/j.1365-8711.2001.04022.x},
archivePrefix = {arXiv},
       eprint = {astro-ph/0009005},
 primaryClass = {astro-ph},
       adsurl = {https://ui.adsabs.harvard.edu/abs/2001MNRAS.322..231K}
}

@ARTICLE{Sabbi2018,
       author = {{Sabbi}, E. and {Calzetti}, D. and {Ubeda}, L. and {Adamo}, A. and {Cignoni}, M. and {Thilker}, D. and {Aloisi}, A. and {Elmegreen}, B.~G. and {Elmegreen}, D.~M. and {Gouliermis}, D.~A. and {Grebel}, E.~K. and {Messa}, M. and {Smith}, L.~J. and {Tosi}, M. and {Dolphin}, A. and {Andrews}, J.~E. and {Ashworth}, G. and {Bright}, S.~N. and {Brown}, T.~M. and {Chandar}, R. and {Christian}, C. and {Clayton}, G.~C. and {Cook}, D.~O. and {Dale}, D.~A. and {de Mink}, S.~E. and {Dobbs}, C. and {Evans}, A.~S. and {Fumagalli}, M. and {Gallagher}, III, J.~S. and {Grasha}, K. and {Herrero}, A. and {Hunter}, D.~A. and {Johnson}, K.~E. and {Kahre}, L. and {Kennicutt}, R.~C. and {Kim}, H. and {Krumholz}, M.~R. and {Lee}, J.~C. and {Lennon}, D. and {Martin}, C. and {Nair}, P. and {Nota}, A. and {{\"O}stlin}, G. and {Pellerin}, A. and {Prieto}, J. and {Regan}, M.~W. and {Ryon}, J.~E. and {Sacchi}, E. and {Schaerer}, D. and {Schiminovich}, D. and {Shabani}, F. and {Van Dyk}, S.~D. and {Walterbos}, R. and {Whitmore}, B.~C. and {Wofford}, A.},
        title = "{The Resolved Stellar Populations in the LEGUS Galaxies1}",
      journal = {\apjs},
         year = 2018,
        month = mar,
       volume = {235},
       number = {1},
          eid = {23},
        pages = {23},
          doi = {10.3847/1538-4365/aaa8e5},
archivePrefix = {arXiv},
       eprint = {1801.05467},
 primaryClass = {astro-ph.GA},
       adsurl = {https://ui.adsabs.harvard.edu/abs/2018ApJS..235...23S}
}

@ARTICLE{Charbonneau1995,
       author = {{Charbonneau}, P.},
        title = "{Genetic Algorithms in Astronomy and Astrophysics}",
      journal = {\apjs},
         year = 1995,
        month = dec,
       volume = {101},
        pages = {309},
          doi = {10.1086/192242},
       adsurl = {https://ui.adsabs.harvard.edu/abs/1995ApJS..101..309C}
}

@ARTICLE{Mighell1999,
       author = {{Mighell}, Kenneth J.},
        title = "{Parameter Estimation in Astronomy with Poisson-distributed Data. I.The {\ensuremath{\chi}}$^{2}$$_{{\ensuremath{\gamma}}}$ Statistic}",
      journal = {\apj},
         year = 1999,
        month = jun,
       volume = {518},
       number = {1},
        pages = {380-393},
          doi = {10.1086/307253},
archivePrefix = {arXiv},
       eprint = {astro-ph/9903093},
 primaryClass = {astro-ph},
       adsurl = {https://ui.adsabs.harvard.edu/abs/1999ApJ...518..380M}
}

@ARTICLE{Gallart2005,
       author = {{Gallart}, C. and {Zoccali}, M. and {Aparicio}, A.},
        title = "{The Adequacy of Stellar Evolution Models for the Interpretation of the Color-Magnitude Diagrams of Resolved Stellar Populations}",
      journal = {\araa},
         year = 2005,
        month = sep,
       volume = {43},
       number = {1},
        pages = {387-434},
          doi = {10.1146/annurev.astro.43.072103.150608},
       adsurl = {https://ui.adsabs.harvard.edu/abs/2005ARA&A..43..387G}
}

@ARTICLE{Kirkpatrick1983,
       author = {{Kirkpatrick}, S. and {Gelatt}, C.~D. and {Vecchi}, M.~P.},
        title = "{Optimization by Simulated Annealing}",
      journal = {Science},
         year = 1983,
        month = may,
       volume = {220},
       number = {4598},
        pages = {671-680},
          doi = {10.1126/science.220.4598.671},
       adsurl = {https://ui.adsabs.harvard.edu/abs/1983Sci...220..671K}
}

@ARTICLE{Ventura2018,
       author = {{Ventura}, P. and {Karakas}, A. and {Dell'Agli}, F. and {Garc{\'\i}a-Hern{\'a}ndez}, D.~A. and {Guzman-Ramirez}, L.},
        title = "{Gas and dust from solar metallicity AGB stars}",
      journal = {\mnras},
         year = 2018,
        month = apr,
       volume = {475},
       number = {2},
        pages = {2282-2305},
          doi = {10.1093/mnras/stx3338},
archivePrefix = {arXiv},
       eprint = {1712.08582},
 primaryClass = {astro-ph.SR},
       adsurl = {https://ui.adsabs.harvard.edu/abs/2018MNRAS.475.2282V}
}

@ARTICLE{SDSS2000,
       author = {{York}, Donald G. and {Adelman}, J. and {Anderson}, Jr., John E. and {Anderson}, Scott F. and {Annis}, James and {Bahcall}, Neta A. and {Bakken}, J.~A. and {Barkhouser}, Robert and {Bastian}, Steven and {Berman}, Eileen and {Boroski}, William N. and {Bracker}, Steve and {Briegel}, Charlie and {Briggs}, John W. and {Brinkmann}, J. and {Brunner}, Robert and {Burles}, Scott and {Carey}, Larry and {Carr}, Michael A. and {Castander}, Francisco J. and {Chen}, Bing and {Colestock}, Patrick L. and {Connolly}, A.~J. and {Crocker}, J.~H. and {Csabai}, Istv{\'a}n and {Czarapata}, Paul C. and {Davis}, John Eric and {Doi}, Mamoru and {Dombeck}, Tom and {Eisenstein}, Daniel and {Ellman}, Nancy and {Elms}, Brian R. and {Evans}, Michael L. and {Fan}, Xiaohui and {Federwitz}, Glenn R. and {Fiscelli}, Larry and {Friedman}, Scott and {Frieman}, Joshua A. and {Fukugita}, Masataka and {Gillespie}, Bruce and {Gunn}, James E. and {Gurbani}, Vijay K. and {de Haas}, Ernst and {Haldeman}, Merle and {Harris}, Frederick H. and {Hayes}, J. and {Heckman}, Timothy M. and {Hennessy}, G.~S. and {Hindsley}, Robert B. and {Holm}, Scott and {Holmgren}, Donald J. and {Huang}, Chi-hao and {Hull}, Charles and {Husby}, Don and {Ichikawa}, Shin-Ichi and {Ichikawa}, Takashi and {Ivezi{\'c}}, {\v{Z}}eljko and {Kent}, Stephen and {Kim}, Rita S.~J. and {Kinney}, E. and {Klaene}, Mark and {Kleinman}, A.~N. and {Kleinman}, S. and {Knapp}, G.~R. and {Korienek}, John and {Kron}, Richard G. and {Kunszt}, Peter Z. and {Lamb}, D.~Q. and {Lee}, B. and {Leger}, R. French and {Limmongkol}, Siriluk and {Lindenmeyer}, Carl and {Long}, Daniel C. and {Loomis}, Craig and {Loveday}, Jon and {Lucinio}, Rich and {Lupton}, Robert H. and {MacKinnon}, Bryan and {Mannery}, Edward J. and {Mantsch}, P.~M. and {Margon}, Bruce and {McGehee}, Peregrine and {McKay}, Timothy A. and {Meiksin}, Avery and {Merelli}, Aronne and {Monet}, David G. and {Munn}, Jeffrey A. and {Narayanan}, Vijay K. and {Nash}, Thomas and {Neilsen}, Eric and {Neswold}, Rich and {Newberg}, Heidi Jo and {Nichol}, R.~C. and {Nicinski}, Tom and {Nonino}, Mario and {Okada}, Norio and {Okamura}, Sadanori and {Ostriker}, Jeremiah P. and {Owen}, Russell and {Pauls}, A. George and {Peoples}, John and {Peterson}, R.~L. and {Petravick}, Donald and {Pier}, Jeffrey R. and {Pope}, Adrian and {Pordes}, Ruth and {Prosapio}, Angela and {Rechenmacher}, Ron and {Quinn}, Thomas R. and {Richards}, Gordon T. and {Richmond}, Michael W. and {Rivetta}, Claudio H. and {Rockosi}, Constance M. and {Ruthmansdorfer}, Kurt and {Sandford}, Dale and {Schlegel}, David J. and {Schneider}, Donald P. and {Sekiguchi}, Maki and {Sergey}, Gary and {Shimasaku}, Kazuhiro and {Siegmund}, Walter A. and {Smee}, Stephen and {Smith}, J. Allyn and {Snedden}, S. and {Stone}, R. and {Stoughton}, Chris and {Strauss}, Michael A. and {Stubbs}, Christopher and {SubbaRao}, Mark and {Szalay}, Alexander S. and {Szapudi}, Istvan and {Szokoly}, Gyula P. and {Thakar}, Anirudda R. and {Tremonti}, Christy and {Tucker}, Douglas L. and {Uomoto}, Alan and {Vanden Berk}, Dan and {Vogeley}, Michael S. and {Waddell}, Patrick and {Wang}, Shu-i. and {Watanabe}, Masaru and {Weinberg}, David H. and {Yanny}, Brian and {Yasuda}, Naoki and {SDSS Collaboration}},
        title = "{The Sloan Digital Sky Survey: Technical Summary}",
      journal = {\aj},
         year = 2000,
        month = sep,
       volume = {120},
       number = {3},
        pages = {1579-1587},
          doi = {10.1086/301513},
archivePrefix = {arXiv},
       eprint = {astro-ph/0006396},
 primaryClass = {astro-ph},
       adsurl = {https://ui.adsabs.harvard.edu/abs/2000AJ....120.1579Y}
}

@ARTICLE{Savino2023,
       author = {{Savino}, Alessandro and {Weisz}, Daniel R. and {Skillman}, Evan D. and {Dolphin}, Andrew and {Cole}, Andrew A. and {Kallivayalil}, Nitya and {Wetzel}, Andrew and {Anderson}, Jay and {Besla}, Gurtina and {Boylan-Kolchin}, Michael and {Brown}, Thomas M. and {Bullock}, James S. and {Collins}, Michelle L.~M. and {Cooper}, M.~C. and {Deason}, Alis J. and {Dotter}, Aaron L. and {Fardal}, Mark and {Ferguson}, Annette M.~N. and {Fritz}, Tobias K. and {Geha}, Marla C. and {Gilbert}, Karoline M. and {Guhathakurta}, Puragra and {Ibata}, Rodrigo and {Irwin}, Michael J. and {Jeon}, Myoungwon and {Kirby}, Evan N. and {Lewis}, Geraint F. and {Mackey}, Dougal and {Majewski}, Steven R. and {Martin}, Nicolas and {McConnachie}, Alan and {Patel}, Ekta and {Rich}, R. Michael and {Simon}, Joshua D. and {Sohn}, Sangmo Tony and {Tollerud}, Erik J. and {van der Marel}, Roeland P.},
        title = "{The Hubble Space Telescope Survey of M31 Satellite Galaxies. II. The Star Formation Histories of Ultrafaint Dwarf Galaxies}",
      journal = {\apj},
         year = 2023,
        month = oct,
       volume = {956},
       number = {2},
          eid = {86},
        pages = {86},
          doi = {10.3847/1538-4357/acf46f},
archivePrefix = {arXiv},
       eprint = {2305.13360},
 primaryClass = {astro-ph.GA},
       adsurl = {https://ui.adsabs.harvard.edu/abs/2023ApJ...956...86S}
}

@ARTICLE{Fernandez-Alvar2025,
       author = {{Fern{\'a}ndez-Alvar}, Emma and {Ruiz-Lara}, Tom{\'a}s and {Gallart}, Carme and {Cassisi}, Santi and {Surot}, Francisco and {Gonz{\'a}lez-Koda}, Yllari K. and {Callingham}, Thomas M. and {Queiroz}, Anna B. and {Battaglia}, Giuseppina and {Thomas}, Guillaume and {Chiappini}, Cristina and {Hill}, Vanessa and {Dodd}, Emma and {Helmi}, Amina and {Aznar-Menargues}, Guillem and {de la Cueva}, Alejandro and {Mirabal}, David and {Quintana-Ansaldo}, M{\'o}nica and {Rivero}, Alicia},
        title = "{Chronology of our Galaxy from Gaia colour─magnitude diagram fitting (ChronoGal): II. Unveiling the formation and evolution of the kinematically selected thick and thin discs}",
      journal = {\aap},
         year = 2025,
        month = dec,
       volume = {704},
          eid = {A258},
        pages = {A258},
          doi = {10.1051/0004-6361/202553814},
archivePrefix = {arXiv},
       eprint = {2503.19536},
 primaryClass = {astro-ph.GA},
       adsurl = {https://ui.adsabs.harvard.edu/abs/2025A&A...704A.258F}
}

@ARTICLE{Burhenne2026,
       author = {{Burhenne}, Clare and {McQuinn}, Kristen B.~W. and {Cohen}, Roger E. and {Murray}, Claire E. and {Patel}, Ekta and {Williams}, Benjamin F. and {Lindberg}, Christina W. and {Merica-Jones}, Petia Yanchulova and {Gordon}, Karl D. and {Choi}, Yumi and {Dolphin}, Andrew E. and {Roman-Duval}, Julia C.},
        title = "{Scylla. V. Constraints on the Spatial and Temporal Distribution of Bursts and the Interaction History of the Magellanic Clouds from Their Resolved Stellar Populations}",
      journal = {\apj},
         year = 2026,
        month = jan,
       volume = {997},
       number = {1},
          eid = {23},
        pages = {23},
          doi = {10.3847/1538-4357/ae13af},
archivePrefix = {arXiv},
       eprint = {2511.02947},
 primaryClass = {astro-ph.GA},
       adsurl = {https://ui.adsabs.harvard.edu/abs/2026ApJ...997...23B}
}

@ARTICLE{Gavetti2026,
       author = {{Gavetti}, C. and {Ventura}, P. and {Dell'Agli}, F. and {Correnti}, M. and {La Franca}, F.},
        title = "{An analysis of the JAGB method from the perspective of the asymptotic giant branch evolution}",
      journal = {\aap},
         year = 2026,
        month = mar,
       volume = {707},
          eid = {A384},
        pages = {A384},
          doi = {10.1051/0004-6361/202557743},
       adsurl = {https://ui.adsabs.harvard.edu/abs/2026A&A...707A.384G}
}

@ARTICLE{Gavetti2025,
       author = {{Gavetti}, C. and {Ventura}, P. and {Dell'Agli}, F. and {La Franca}, F. and {Marini}, E. and {Correnti}, M. and {Tailo}, M.},
        title = "{A study of Andromeda to improve our knowledge of the evolution and dust production by AGB stars}",
      journal = {\aap},
         year = 2025,
        month = jul,
       volume = {699},
          eid = {A23},
        pages = {A23},
          doi = {10.1051/0004-6361/202451724},
archivePrefix = {arXiv},
       eprint = {2504.12940},
 primaryClass = {astro-ph.GA},
       adsurl = {https://ui.adsabs.harvard.edu/abs/2025A&A...699A..23G}
}

@ARTICLE{Nanni2013,
       author = {{Nanni}, Ambra and {Bressan}, Alessandro and {Marigo}, Paola and {Girardi}, L{\'e}o},
        title = "{Evolution of thermally pulsing asymptotic giant branch stars - II. Dust production at varying metallicity}",
      journal = {\mnras},
         year = 2013,
        month = sep,
       volume = {434},
       number = {3},
        pages = {2390-2417},
          doi = {10.1093/mnras/stt1175},
archivePrefix = {arXiv},
       eprint = {1306.6183},
 primaryClass = {astro-ph.SR},
       adsurl = {https://ui.adsabs.harvard.edu/abs/2013MNRAS.434.2390N}
}

@ARTICLE{Ventura2014,
       author = {{Ventura}, P. and {Dell'Agli}, F. and {Schneider}, R. and {Di Criscienzo}, M. and {Rossi}, C. and {La Franca}, F. and {Gallerani}, S. and {Valiante}, R.},
        title = "{Dust from asymptotic giant branch stars: relevant factors and modelling uncertainties}",
      journal = {\mnras},
         year = 2014,
        month = mar,
       volume = {439},
       number = {1},
        pages = {977-989},
          doi = {10.1093/mnras/stu028},
archivePrefix = {arXiv},
       eprint = {1401.1332},
 primaryClass = {astro-ph.SR},
       adsurl = {https://ui.adsabs.harvard.edu/abs/2014MNRAS.439..977V}
}

@ARTICLE{Vassiliadis1993,
       author = {{Vassiliadis}, E. and {Wood}, P.~R.},
        title = "{Evolution of Low- and Intermediate-Mass Stars to the End of the Asymptotic Giant Branch with Mass Loss}",
      journal = {\apj},
         year = 1993,
        month = aug,
       volume = {413},
        pages = {641},
          doi = {10.1086/173033},
       adsurl = {https://ui.adsabs.harvard.edu/abs/1993ApJ...413..641V}
}

@ARTICLE{Herwig2005,
       author = {{Herwig}, Falk},
        title = "{Evolution of Asymptotic Giant Branch Stars}",
      journal = {\araa},
         year = 2005,
        month = sep,
       volume = {43},
       number = {1},
        pages = {435-479},
          doi = {10.1146/annurev.astro.43.072103.150600},
       adsurl = {https://ui.adsabs.harvard.edu/abs/2005ARA&A..43..435H}
}

@ARTICLE{Lee2025,
       author = {{Lee}, Abigail J. and {Freedman}, Wendy L. and {Madore}, Barry F. and {Jang}, In Sung and {Owens}, Kayla A. and {Hoyt}, Taylor J.},
        title = "{The Chicago─Carnegie Hubble Program: The JWST J-region Asymptotic Giant Branch Extragalactic Distance Scale}",
      journal = {\apj},
         year = 2025,
        month = jun,
       volume = {985},
       number = {2},
          eid = {182},
        pages = {182},
          doi = {10.3847/1538-4357/adc8a1},
archivePrefix = {arXiv},
       eprint = {2408.03474},
 primaryClass = {astro-ph.GA},
       adsurl = {https://ui.adsabs.harvard.edu/abs/2025ApJ...985..182L}
}

@ARTICLE{Lee2023,
       author = {{Lee}, Abigail J.},
        title = "{Carbon Stars as Standard Candles: An Empirical Test for the Reddening, Metallicity, and Age Sensitivity of the J-region Asymptotic Giant Branch (JAGB) Method}",
      journal = {\apj},
         year = 2023,
        month = oct,
       volume = {956},
       number = {1},
          eid = {15},
        pages = {15},
          doi = {10.3847/1538-4357/acee69},
archivePrefix = {arXiv},
       eprint = {2305.02453},
 primaryClass = {astro-ph.GA},
       adsurl = {https://ui.adsabs.harvard.edu/abs/2023ApJ...956...15L}
}

@ARTICLE{Gavetti2026b,
       author = {{Gavetti}, Claudio and {Ventura}, Paolo and {Dell'Agli}, Flavia and {Graziani}, Luca and {Correnti}, Matteo and {Ventura}, Chiara and {La Franca}, Fabio},
        title = "{The applicability of the JAGB method for measuring the distance of galaxies subject to different metal enrichment rates}",
      journal = {arXiv e-prints},
         year = 2026,
        month = may,
          eid = {arXiv:2605.11198},
        pages = {arXiv:2605.11198},
          doi = {10.48550/arXiv.2605.11198},
archivePrefix = {arXiv},
       eprint = {2605.11198},
 primaryClass = {astro-ph.GA},
       adsurl = {https://ui.adsabs.harvard.edu/abs/2026arXiv260511198G}
}

@ARTICLE{Pastorelli2020,
       author = {{Pastorelli}, Giada and {Marigo}, Paola and {Girardi}, L{\'e}o and {Aringer}, Bernhard and {Chen}, Yang and {Rubele}, Stefano and {Trabucchi}, Michele and {Bladh}, Sara and {Boyer}, Martha L. and {Bressan}, Alessandro and {Dalcanton}, Julianne J. and {Groenewegen}, Martin A.~T. and {Lebzelter}, Thomas and {Mowlavi}, Nami and {Chubb}, Katy L. and {Cioni}, Maria-Rosa L. and {de Grijs}, Richard and {Ivanov}, Valentin D. and {Nanni}, Ambra and {van Loon}, Jacco Th and {Zaggia}, Simone},
        title = "{Constraining the thermally pulsing asymptotic giant branch phase with resolved stellar populations in the Large Magellanic Cloud}",
      journal = {\mnras},
         year = 2020,
        month = nov,
       volume = {498},
       number = {3},
        pages = {3283-3301},
          doi = {10.1093/mnras/staa2565},
archivePrefix = {arXiv},
       eprint = {2008.08595},
 primaryClass = {astro-ph.SR},
       adsurl = {https://ui.adsabs.harvard.edu/abs/2020MNRAS.498.3283P}
}

@ARTICLE{Chen2019,
       author = {{Chen}, Yang and {Girardi}, L{\'e}o and {Fu}, Xiaoting and {Bressan}, Alessandro and {Aringer}, Bernhard and {Dal Tio}, Piero and {Pastorelli}, Giada and {Marigo}, Paola and {Costa}, Guglielmo and {Zhang}, Xing},
        title = "{YBC: a stellar bolometric corrections database with variable extinction coefficients. Application to PARSEC isochrones}",
      journal = {\aap},
         year = 2019,
        month = dec,
       volume = {632},
          eid = {A105},
        pages = {A105},
          doi = {10.1051/0004-6361/201936612},
archivePrefix = {arXiv},
       eprint = {1910.09037},
 primaryClass = {astro-ph.SR},
       adsurl = {https://ui.adsabs.harvard.edu/abs/2019A&A...632A.105C}
}

@ARTICLE{Calzetti2015,
       author = {{Calzetti}, D. and {Lee}, J.~C. and {Sabbi}, E. and {Adamo}, A. and {Smith}, L.~J. and {Andrews}, J.~E. and {Ubeda}, L. and {Bright}, S.~N. and {Thilker}, D. and {Aloisi}, A. and {Brown}, T.~M. and {Chandar}, R. and {Christian}, C. and {Cignoni}, M. and {Clayton}, G.~C. and {da Silva}, R. and {de Mink}, S.~E. and {Dobbs}, C. and {Elmegreen}, B.~G. and {Elmegreen}, D.~M. and {Evans}, A.~S. and {Fumagalli}, M. and {Gallagher}, III, J.~S. and {Gouliermis}, D.~A. and {Grebel}, E.~K. and {Herrero}, A. and {Hunter}, D.~A. and {Johnson}, K.~E. and {Kennicutt}, R.~C. and {Kim}, H. and {Krumholz}, M.~R. and {Lennon}, D. and {Levay}, K. and {Martin}, C. and {Nair}, P. and {Nota}, A. and {{\"O}stlin}, G. and {Pellerin}, A. and {Prieto}, J. and {Regan}, M.~W. and {Ryon}, J.~E. and {Schaerer}, D. and {Schiminovich}, D. and {Tosi}, M. and {Van Dyk}, S.~D. and {Walterbos}, R. and {Whitmore}, B.~C. and {Wofford}, A.},
        title = "{Legacy Extragalactic UV Survey (LEGUS) With the Hubble Space Telescope. I. Survey Description}",
      journal = {\aj},
         year = 2015,
        month = feb,
       volume = {149},
       number = {2},
          eid = {51},
        pages = {51},
          doi = {10.1088/0004-6256/149/2/51},
archivePrefix = {arXiv},
       eprint = {1410.7456},
 primaryClass = {astro-ph.GA},
       adsurl = {https://ui.adsabs.harvard.edu/abs/2015AJ....149...51C}
}

@ARTICLE{Ventura1998,
       author = {{Ventura}, Paolo and {Zeppieri}, Anna and {Mazzitelli}, Italo and {D'Antona}, Francesca},
        title = "{Full spectrum of turbulence convective mixing: I. theoretical main sequences and turn-off for 0.6 /\textbackslash 15 M\_\textbackslashodot}",
      journal = {\aap},
         year = 1998,
        month = jun,
       volume = {334},
        pages = {953-968},
       adsurl = {https://ui.adsabs.harvard.edu/abs/1998A&A...334..953V}
}

@ARTICLE{Ventura2020,
       author = {{Ventura}, P. and {Dell'Agli}, F. and {Lugaro}, M. and {Romano}, D. and {Tailo}, M. and {Yag{\"u}e}, A.},
        title = "{Gas and dust from metal-rich AGB stars}",
      journal = {\aap},
         year = 2020,
        month = sep,
       volume = {641},
          eid = {A103},
        pages = {A103},
          doi = {10.1051/0004-6361/202038289},
archivePrefix = {arXiv},
       eprint = {2007.02120},
 primaryClass = {astro-ph.SR},
       adsurl = {https://ui.adsabs.harvard.edu/abs/2020A&A...641A.103V}
}

@INPROCEEDINGS{Harris2002,
       author = {{Harris}, J. and {Zaritsky}, D.},
        title = "{StarFISH: A Robust Synthetic Color-Magnitude Diagram Package}",
    booktitle = {Observed HR Diagrams and Stellar Evolution},
         year = 2002,
       editor = {{Lejeune}, Thibault and {Fernandes}, Jo{\~a}o},
       series = {Astronomical Society of the Pacific Conference Series},
       volume = {274},
        month = jan,
        pages = {600},
       adsurl = {https://ui.adsabs.harvard.edu/abs/2002ASPC..274..600H}
}

@ARTICLE{Bernard2018,
       author = {{Bernard}, Edouard J. and {Schultheis}, Mathias and {Di Matteo}, Paola and {Hill}, Vanessa and {Haywood}, Misha and {Calamida}, Annalisa},
        title = "{Star formation history of the Galactic bulge from deep HST imaging of low reddening windows}",
      journal = {\mnras},
         year = 2018,
        month = jul,
       volume = {477},
       number = {3},
        pages = {3507-3519},
          doi = {10.1093/mnras/sty902},
archivePrefix = {arXiv},
       eprint = {1801.01426},
 primaryClass = {astro-ph.GA},
       adsurl = {https://ui.adsabs.harvard.edu/abs/2018MNRAS.477.3507B}
}

@ARTICLE{deBoer2016,
       author = {{de Boer}, T.~J.~L. and {Fraser}, M.},
        title = "{Four and one more: The formation history and total mass of globular clusters in the Fornax dSph}",
      journal = {\aap},
         year = 2016,
        month = may,
       volume = {590},
          eid = {A35},
        pages = {A35},
          doi = {10.1051/0004-6361/201527580},
archivePrefix = {arXiv},
       eprint = {1510.05642},
 primaryClass = {astro-ph.GA},
       adsurl = {https://ui.adsabs.harvard.edu/abs/2016A&A...590A..35D}
}

@ARTICLE{Cignoni2009,
       author = {{Cignoni}, M. and {Sabbi}, E. and {Nota}, A. and {Tosi}, M. and {Degl'Innocenti}, S. and {Moroni}, P.~G. Prada and {Angeretti}, L. and {Carlson}, Lynn Redding and {Gallagher}, J. and {Meixner}, M. and {Sirianni}, M. and {Smith}, L.~J.},
        title = "{Star Formation History in the Small Magellanic Cloud: The Case of NGC 602}",
      journal = {\aj},
         year = 2009,
        month = mar,
       volume = {137},
       number = {3},
        pages = {3668-3684},
          doi = {10.1088/0004-6256/137/3/3668},
archivePrefix = {arXiv},
       eprint = {0901.1237},
 primaryClass = {astro-ph.GA},
       adsurl = {https://ui.adsabs.harvard.edu/abs/2009AJ....137.3668C}
}

@ARTICLE{Chiosi1988,
       author = {{Chiosi}, C. and {Bertelli}, G. and {Bressan}, A.},
        title = "{Integrated colours and ages of LMC clusters : the nature of the bimodal distribution of the (B-V) colours.}",
      journal = {\aap},
         year = 1988,
        month = may,
       volume = {196},
        pages = {84-108},
       adsurl = {https://ui.adsabs.harvard.edu/abs/1988A&A...196...84C}
}

@ARTICLE{Maeder1974,
       author = {{Maeder}, A.},
        title = "{Stellar evolution near the main sequence: on some systematic differences between cluster sequences and model calculations.}",
      journal = {\aap},
         year = 1974,
        month = may,
       volume = {32},
        pages = {177-190},
       adsurl = {https://ui.adsabs.harvard.edu/abs/1974A&A....32..177M}
}

@ARTICLE{Bertelli1992,
       author = {{Bertelli}, Gianpaolo and {Mateo}, Mario and {Chiosi}, Cesare and {Bressan}, Alessandro},
        title = "{The Star Formation History of the Large Magellanic Cloud}",
      journal = {\apj},
         year = 1992,
        month = apr,
       volume = {388},
        pages = {400},
          doi = {10.1086/171163},
       adsurl = {https://ui.adsabs.harvard.edu/abs/1992ApJ...388..400B}
}

@ARTICLE{Marigo2008,
       author = {{Marigo}, P. and {Girardi}, L. and {Bressan}, A. and {Groenewegen}, M.~A.~T. and {Silva}, L. and {Granato}, G.~L.},
        title = "{Evolution of asymptotic giant branch stars. II. Optical to far-infrared isochrones with improved TP-AGB models}",
      journal = {\aap},
         year = 2008,
        month = may,
       volume = {482},
       number = {3},
        pages = {883-905},
          doi = {10.1051/0004-6361:20078467},
archivePrefix = {arXiv},
       eprint = {0711.4922},
 primaryClass = {astro-ph},
       adsurl = {https://ui.adsabs.harvard.edu/abs/2008A&A...482..883M}
}

@ARTICLE{Marigo2013,
       author = {{Marigo}, Paola and {Bressan}, Alessandro and {Nanni}, Ambra and {Girardi}, L{\'e}o and {Pumo}, Maria Letizia},
        title = "{Evolution of thermally pulsing asymptotic giant branch stars - I. The COLIBRI code}",
      journal = {\mnras},
         year = 2013,
        month = sep,
       volume = {434},
       number = {1},
        pages = {488-526},
          doi = {10.1093/mnras/stt1034},
archivePrefix = {arXiv},
       eprint = {1305.4485},
 primaryClass = {astro-ph.SR},
       adsurl = {https://ui.adsabs.harvard.edu/abs/2013MNRAS.434..488M}
}

@ARTICLE{Frost1996,
       author = {{Frost}, C.~A. and {Lattanzio}, J.~C.},
        title = "{On the Numerical Treatment and Dependence of the Third Dredge-up Phenomenon}",
      journal = {\apj},
         year = 1996,
        month = dec,
       volume = {473},
        pages = {383},
          doi = {10.1086/178152},
       adsurl = {https://ui.adsabs.harvard.edu/abs/1996ApJ...473..383F}
}

@ARTICLE{Mowlavi1999,
       author = {{Mowlavi}, N.},
        title = "{On the third dredge-up phenomenon in asymptotic giant branch stars}",
      journal = {\aap},
         year = 1999,
        month = apr,
       volume = {344},
        pages = {617-631},
          doi = {10.48550/arXiv.astro-ph/9903473},
archivePrefix = {arXiv},
       eprint = {astro-ph/9903473},
 primaryClass = {astro-ph},
       adsurl = {https://ui.adsabs.harvard.edu/abs/1999A&A...344..617M}
}

@ARTICLE{Ventura2005,
       author = {{Ventura}, P. and {D'Antona}, F.},
        title = "{Full computation of massive AGB evolution. II. The role of mass loss and cross-sections}",
      journal = {\aap},
         year = 2005,
        month = sep,
       volume = {439},
       number = {3},
        pages = {1075-1091},
          doi = {10.1051/0004-6361:20042396},
archivePrefix = {arXiv},
       eprint = {astro-ph/0505221},
 primaryClass = {astro-ph},
       adsurl = {https://ui.adsabs.harvard.edu/abs/2005A&A...439.1075V}
}

@ARTICLE{Hidalgo2018,
       author = {{Hidalgo}, Sebastian L. and {Pietrinferni}, Adriano and {Cassisi}, Santi and {Salaris}, Maurizio and {Mucciarelli}, Alessio and {Savino}, Alessandro and {Aparicio}, Antonio and {Silva Aguirre}, Victor and {Verma}, Kuldeep},
        title = "{The Updated BaSTI Stellar Evolution Models and Isochrones. I. Solar-scaled Calculations}",
      journal = {\apj},
         year = 2018,
        month = apr,
       volume = {856},
       number = {2},
          eid = {125},
        pages = {125},
          doi = {10.3847/1538-4357/aab158},
archivePrefix = {arXiv},
       eprint = {1802.07319},
 primaryClass = {astro-ph.GA},
       adsurl = {https://ui.adsabs.harvard.edu/abs/2018ApJ...856..125H}
}

@ARTICLE{Cordier2007,
       author = {{Cordier}, Daniel and {Pietrinferni}, Adriano and {Cassisi}, Santi and {Salaris}, Maurizio},
        title = "{A Large Stellar Evolution Database for Population Synthesis Studies. III. Inclusion of the Full Asymptotic Giant Branch Phase and Web Tools for Stellar Population Analyses}",
      journal = {\aj},
         year = 2007,
        month = feb,
       volume = {133},
       number = {2},
        pages = {468-478},
          doi = {10.1086/509870},
archivePrefix = {arXiv},
       eprint = {astro-ph/0612669},
 primaryClass = {astro-ph},
       adsurl = {https://ui.adsabs.harvard.edu/abs/2007AJ....133..468C}
}

@ARTICLE{White1978,
       author = {{White}, S.~D.~M. and {Rees}, M.~J.},
        title = "{Core condensation in heavy halos: a two-stage theory for galaxy formation and clustering.}",
      journal = {\mnras},
         year = 1978,
        month = may,
       volume = {183},
        pages = {341-358},
          doi = {10.1093/mnras/183.3.341},
       adsurl = {https://ui.adsabs.harvard.edu/abs/1978MNRAS.183..341W}
}

@ARTICLE{Peebles1982,
       author = {{Peebles}, P.~J.~E.},
        title = "{Large-scale background temperature and mass fluctuations due to scale-invariant primeval perturbations}",
      journal = {\apjl},
         year = 1982,
        month = dec,
       volume = {263},
        pages = {L1-L5},
          doi = {10.1086/183911},
       adsurl = {https://ui.adsabs.harvard.edu/abs/1982ApJ...263L...1P}
}

@ARTICLE{White1991,
       author = {{White}, Simon D.~M. and {Frenk}, Carlos S.},
        title = "{Galaxy Formation through Hierarchical Clustering}",
      journal = {\apj},
         year = 1991,
        month = sep,
       volume = {379},
        pages = {52},
          doi = {10.1086/170483},
       adsurl = {https://ui.adsabs.harvard.edu/abs/1991ApJ...379...52W}
}

@ARTICLE{Belokurov2006,
       author = {{Belokurov}, V. and {Zucker}, D.~B. and {Evans}, N.~W. and {Gilmore}, G. and {Vidrih}, S. and {Bramich}, D.~M. and {Newberg}, H.~J. and {Wyse}, R.~F.~G. and {Irwin}, M.~J. and {Fellhauer}, M. and {Hewett}, P.~C. and {Walton}, N.~A. and {Wilkinson}, M.~I. and {Cole}, N. and {Yanny}, B. and {Rockosi}, C.~M. and {Beers}, T.~C. and {Bell}, E.~F. and {Brinkmann}, J. and {Ivezi{\'c}}, {\v{Z}}. and {Lupton}, R.},
        title = "{The Field of Streams: Sagittarius and Its Siblings}",
      journal = {\apjl},
         year = 2006,
        month = may,
       volume = {642},
       number = {2},
        pages = {L137-L140},
          doi = {10.1086/504797},
archivePrefix = {arXiv},
       eprint = {astro-ph/0605025},
 primaryClass = {astro-ph},
       adsurl = {https://ui.adsabs.harvard.edu/abs/2006ApJ...642L.137B}
}

@ARTICLE{Ibata2001,
       author = {{Ibata}, Rodrigo and {Irwin}, Michael and {Lewis}, Geraint and {Ferguson}, Annette M.~N. and {Tanvir}, Nial},
        title = "{A giant stream of metal-rich stars in the halo of the galaxy M31}",
      journal = {\nat},
         year = 2001,
        month = jul,
       volume = {412},
       number = {6842},
        pages = {49-52},
          doi = {10.1038/35083506},
archivePrefix = {arXiv},
       eprint = {astro-ph/0107090},
 primaryClass = {astro-ph},
       adsurl = {https://ui.adsabs.harvard.edu/abs/2001Natur.412...49I}
}

@ARTICLE{McConnachie2009,
       author = {{McConnachie}, Alan W. and {Irwin}, Michael J. and {Ibata}, Rodrigo A. and {Dubinski}, John and {Widrow}, Lawrence M. and {Martin}, Nicolas F. and {C{\^o}t{\'e}}, Patrick and {Dotter}, Aaron L. and {Navarro}, Julio F. and {Ferguson}, Annette M.~N. and {Puzia}, Thomas H. and {Lewis}, Geraint F. and {Babul}, Arif and {Barmby}, Pauline and {Bienaym{\'e}}, Olivier and {Chapman}, Scott C. and {Cockcroft}, Robert and {Collins}, Michelle L.~M. and {Fardal}, Mark A. and {Harris}, William E. and {Huxor}, Avon and {Mackey}, A. Dougal and {Pe{\~n}arrubia}, Jorge and {Rich}, R. Michael and {Richer}, Harvey B. and {Siebert}, Arnaud and {Tanvir}, Nial and {Valls-Gabaud}, David and {Venn}, Kimberly A.},
        title = "{The remnants of galaxy formation from a panoramic survey of the region around M31}",
      journal = {\nat},
         year = 2009,
        month = sep,
       volume = {461},
       number = {7260},
        pages = {66-69},
          doi = {10.1038/nature08327},
archivePrefix = {arXiv},
       eprint = {0909.0398},
 primaryClass = {astro-ph.CO},
       adsurl = {https://ui.adsabs.harvard.edu/abs/2009Natur.461...66M}
}

@ARTICLE{Martinez-Delgado2010,
       author = {{Mart{\'\i}nez-Delgado}, David and {Gabany}, R. Jay and {Crawford}, Ken and {Zibetti}, Stefano and {Majewski}, Steven R. and {Rix}, Hans-Walter and {Fliri}, J{\"u}rgen and {Carballo-Bello}, Julio A. and {Bardalez-Gagliuffi}, Daniella C. and {Pe{\~n}arrubia}, Jorge and {Chonis}, Taylor S. and {Madore}, Barry and {Trujillo}, Ignacio and {Schirmer}, Mischa and {McDavid}, David A.},
        title = "{Stellar Tidal Streams in Spiral Galaxies of the Local Volume: A Pilot Survey with Modest Aperture Telescopes}",
      journal = {\aj},
         year = 2010,
        month = oct,
       volume = {140},
       number = {4},
        pages = {962-967},
          doi = {10.1088/0004-6256/140/4/962},
archivePrefix = {arXiv},
       eprint = {1003.4860},
 primaryClass = {astro-ph.CO},
       adsurl = {https://ui.adsabs.harvard.edu/abs/2010AJ....140..962M}
}

@ARTICLE{Crnojevic2016,
       author = {{Crnojevi{\'c}}, D. and {Sand}, D.~J. and {Spekkens}, K. and {Caldwell}, N. and {Guhathakurta}, P. and {McLeod}, B. and {Seth}, A. and {Simon}, J.~D. and {Strader}, J. and {Toloba}, E.},
        title = "{The Extended Halo of Centaurus A: Uncovering Satellites, Streams, and Substructures}",
      journal = {\apj},
         year = 2016,
        month = may,
       volume = {823},
       number = {1},
          eid = {19},
        pages = {19},
          doi = {10.3847/0004-637X/823/1/19},
archivePrefix = {arXiv},
       eprint = {1512.05366},
 primaryClass = {astro-ph.GA},
       adsurl = {https://ui.adsabs.harvard.edu/abs/2016ApJ...823...19C}
}

@ARTICLE{Annibali2022,
       author = {{Annibali}, Francesca and {Tosi}, Monica},
        title = "{Chemical and stellar properties of star-forming dwarf galaxies}",
      journal = {Nature Astronomy},
         year = 2022,
        month = jan,
       volume = {6},
        pages = {48-58},
          doi = {10.1038/s41550-021-01575-x},
archivePrefix = {arXiv},
       eprint = {2201.06600},
 primaryClass = {astro-ph.GA},
       adsurl = {https://ui.adsabs.harvard.edu/abs/2022NatAs...6...48A}
}

@ARTICLE{Stierwalt2015,
       author = {{Stierwalt}, S. and {Besla}, G. and {Patton}, D. and {Johnson}, K. and {Kallivayalil}, N. and {Putman}, M. and {Privon}, G. and {Ross}, G.},
        title = "{TiNy Titans: The Role of Dwarf-Dwarf Interactions in Low-mass Galaxy Evolution}",
      journal = {\apj},
         year = 2015,
        month = may,
       volume = {805},
       number = {1},
          eid = {2},
        pages = {2},
          doi = {10.1088/0004-637X/805/1/2},
archivePrefix = {arXiv},
       eprint = {1412.4796},
 primaryClass = {astro-ph.GA},
       adsurl = {https://ui.adsabs.harvard.edu/abs/2015ApJ...805....2S}
}

@ARTICLE{Karachentsev2018,
       author = {{Karachentsev}, I.~D. and {Kaisina}, E.~I. and {Makarov}, D.~I.},
        title = "{Morphological properties of galaxies in different Local Volume environments}",
      journal = {\mnras},
         year = 2018,
        month = sep,
       volume = {479},
       number = {3},
        pages = {4136-4152},
          doi = {10.1093/mnras/sty1774},
archivePrefix = {arXiv},
       eprint = {1806.09822},
 primaryClass = {astro-ph.GA},
       adsurl = {https://ui.adsabs.harvard.edu/abs/2018MNRAS.479.4136K}
}

@ARTICLE{Conselice2014,
       author = {{Conselice}, Christopher J.},
        title = "{The Evolution of Galaxy Structure Over Cosmic Time}",
      journal = {\araa},
         year = 2014,
        month = aug,
       volume = {52},
        pages = {291-337},
          doi = {10.1146/annurev-astro-081913-040037},
archivePrefix = {arXiv},
       eprint = {1403.2783},
 primaryClass = {astro-ph.GA},
       adsurl = {https://ui.adsabs.harvard.edu/abs/2014ARA&A..52..291C}
}

@ARTICLE{Annibali2020,
       author = {{Annibali}, F. and {Beccari}, G. and {Bellazzini}, M. and {Tosi}, M. and {Cusano}, F. and {Paris}, D. and {Cignoni}, M. and {Ciotti}, L. and {Nipoti}, C. and {Sacchi}, E.},
        title = "{The Smallest Scale of Hierarchy survey (SSH) - I. Survey description}",
      journal = {\mnras},
         year = 2020,
        month = feb,
       volume = {491},
       number = {4},
        pages = {5101-5125},
          doi = {10.1093/mnras/stz3185},
archivePrefix = {arXiv},
       eprint = {1911.08543},
 primaryClass = {astro-ph.GA},
       adsurl = {https://ui.adsabs.harvard.edu/abs/2020MNRAS.491.5101A}
}

@ARTICLE{Sacchi2024,
       author = {{Sacchi}, Elena and {Bellazzini}, Michele and {Annibali}, Francesca and {Tosi}, Monica and {Beccari}, Giacomo and {Cannon}, John M. and {Hunter}, Laura C. and {Paris}, Diego and {Roychowdhury}, Sambit and {Schisgal}, Lila and {van Zee}, Liese and {Cignoni}, Michele and {Cusano}, Felice and {de Jong}, Roelof S. and {Hunt}, Leslie and {Pascale}, Raffaele},
        title = "{The Smallest Scale of Hierarchy Survey (SSH): III. Dwarf-dwarf satellite merging phenomena in the low-mass regime}",
      journal = {\aap},
         year = 2024,
        month = nov,
       volume = {691},
          eid = {A65},
        pages = {A65},
          doi = {10.1051/0004-6361/202450106},
archivePrefix = {arXiv},
       eprint = {2406.01683},
 primaryClass = {astro-ph.GA},
       adsurl = {https://ui.adsabs.harvard.edu/abs/2024A&A...691A..65S}
}

@INPROCEEDINGS{Bournaud2011,
       author = {{Bournaud}, F.},
        title = "{Star formation in galaxy interactions and mergers}",
    booktitle = {EAS Publications Series},
         year = 2011,
       editor = {{Charbonnel}, Corinne and {Montmerle}, Thierry},
       series = {EAS Publications Series},
       volume = {51},
        month = nov,
    publisher = {EDP},
        pages = {107-131},
          doi = {10.1051/eas/1151008},
archivePrefix = {arXiv},
       eprint = {1106.1793},
 primaryClass = {astro-ph.CO},
       adsurl = {https://ui.adsabs.harvard.edu/abs/2011EAS....51..107B}
}

@ARTICLE{Barnes1991,
       author = {{Barnes}, Joshua E. and {Hernquist}, Lars E.},
        title = "{Fueling Starburst Galaxies with Gas-rich Mergers}",
      journal = {\apjl},
         year = 1991,
        month = apr,
       volume = {370},
        pages = {L65},
          doi = {10.1086/185978},
       adsurl = {https://ui.adsabs.harvard.edu/abs/1991ApJ...370L..65B}
}

@ARTICLE{Barnes1996,
       author = {{Barnes}, Joshua E. and {Hernquist}, Lars},
        title = "{Transformations of Galaxies. II. Gasdynamics in Merging Disk Galaxies}",
      journal = {\apj},
         year = 1996,
        month = nov,
       volume = {471},
        pages = {115},
          doi = {10.1086/177957},
       adsurl = {https://ui.adsabs.harvard.edu/abs/1996ApJ...471..115B}
}

@ARTICLE{Li2008,
       author = {{Li}, Cheng and {Kauffmann}, Guinevere and {Heckman}, Timothy M. and {Jing}, Y.~P. and {White}, Simon D.~M.},
        title = "{Interaction-induced star formation in a complete sample of {}10$^{5}$ nearby star-forming galaxies}",
      journal = {\mnras},
         year = 2008,
        month = apr,
       volume = {385},
       number = {4},
        pages = {1903-1914},
          doi = {10.1111/j.1365-2966.2008.13000.x},
archivePrefix = {arXiv},
       eprint = {0711.3792},
 primaryClass = {astro-ph},
       adsurl = {https://ui.adsabs.harvard.edu/abs/2008MNRAS.385.1903L}
}

@ARTICLE{Ellison2013,
       author = {{Ellison}, Sara L. and {Mendel}, J. Trevor and {Patton}, David R. and {Scudder}, Jillian M.},
        title = "{Galaxy pairs in the Sloan Digital Sky Survey - VIII. The observational properties of post-merger galaxies}",
      journal = {\mnras},
         year = 2013,
        month = nov,
       volume = {435},
       number = {4},
        pages = {3627-3638},
          doi = {10.1093/mnras/stt1562},
archivePrefix = {arXiv},
       eprint = {1308.3707},
 primaryClass = {astro-ph.CO},
       adsurl = {https://ui.adsabs.harvard.edu/abs/2013MNRAS.435.3627E}
}

@ARTICLE{Pearson2016,
       author = {{Pearson}, Sarah and {Besla}, Gurtina and {Putman}, Mary E. and {Lutz}, Katharina A. and {Fern{\'a}ndez}, Ximena and {Stierwalt}, Sabrina and {Patton}, David R. and {Kim}, Jinhyub and {Kallivayalil}, Nitya and {Johnson}, Kelsey and {Sung}, Eon-Chang},
        title = "{Local Volume TiNy Titans: gaseous dwarf-dwarf interactions in the Local Universe}",
      journal = {\mnras},
         year = 2016,
        month = jun,
       volume = {459},
       number = {2},
        pages = {1827-1846},
          doi = {10.1093/mnras/stw757},
archivePrefix = {arXiv},
       eprint = {1603.09342},
 primaryClass = {astro-ph.GA},
       adsurl = {https://ui.adsabs.harvard.edu/abs/2016MNRAS.459.1827P}
}

@ARTICLE{Pearson2018,
       author = {{Pearson}, Sarah and {Privon}, George C. and {Besla}, Gurtina and {Putman}, Mary E. and {Mart{\'\i}nez-Delgado}, David and {Johnston}, Kathryn V. and {Gabany}, R. Jay and {Patton}, David R. and {Kallivayalil}, Nitya},
        title = "{Modelling the baryon cycle in low-mass galaxy encounters: the case of NGC 4490 and NGC 4485}",
      journal = {\mnras},
         year = 2018,
        month = nov,
       volume = {480},
       number = {3},
        pages = {3069-3090},
          doi = {10.1093/mnras/sty2052},
archivePrefix = {arXiv},
       eprint = {1807.03791},
 primaryClass = {astro-ph.GA},
       adsurl = {https://ui.adsabs.harvard.edu/abs/2018MNRAS.480.3069P}
}

@ARTICLE{Bekki2005,
       author = {{Bekki}, Kenji and {Chiba}, Masashi},
        title = "{Formation and evolution of the Magellanic Clouds - I. Origin of structural, kinematic and chemical properties of the Large Magellanic Cloud}",
      journal = {\mnras},
         year = 2005,
        month = jan,
       volume = {356},
       number = {2},
        pages = {680-702},
          doi = {10.1111/j.1365-2966.2004.08510.x},
archivePrefix = {arXiv},
       eprint = {astro-ph/0412318},
 primaryClass = {astro-ph},
       adsurl = {https://ui.adsabs.harvard.edu/abs/2005MNRAS.356..680B}
}

@ARTICLE{Pilyugin2007,
       author = {{Pilyugin}, Leonid S. and {Thuan}, Trinh X.},
        title = "{The Oxygen Abundance of Nearby Galaxies from Sloan Digital Sky Survey Spectra}",
      journal = {\apj},
         year = 2007,
        month = nov,
       volume = {669},
       number = {1},
        pages = {299-314},
          doi = {10.1086/521597},
archivePrefix = {arXiv},
       eprint = {0707.2856},
 primaryClass = {astro-ph},
       adsurl = {https://ui.adsabs.harvard.edu/abs/2007ApJ...669..299P}
}

@ARTICLE{Paudel2018,
       author = {{Paudel}, Sanjaya and {Smith}, Rory and {Yoon}, Suk Jin and {Calder{\'o}n-Castillo}, Paula and {Duc}, Pierre-Alain},
        title = "{A Catalog of Merging Dwarf Galaxies in the Local Universe}",
      journal = {\apjs},
         year = 2018,
        month = aug,
       volume = {237},
       number = {2},
          eid = {36},
        pages = {36},
          doi = {10.3847/1538-4365/aad555},
archivePrefix = {arXiv},
       eprint = {1807.07195},
 primaryClass = {astro-ph.GA},
       adsurl = {https://ui.adsabs.harvard.edu/abs/2018ApJS..237...36P}
}

@ARTICLE{Privon2017,
       author = {{Privon}, G.~C. and {Stierwalt}, S. and {Patton}, D.~R. and {Besla}, G. and {Pearson}, S. and {Putman}, M. and {Johnson}, K.~E. and {Kallivayalil}, N. and {Liss}, S. and {Titans}, TiNy},
        title = "{A Widespread, Clumpy Starburst in the Isolated Ongoing Dwarf Galaxy Merger dm1647+21}",
      journal = {\apj},
         year = 2017,
        month = sep,
       volume = {846},
       number = {1},
          eid = {74},
        pages = {74},
          doi = {10.3847/1538-4357/aa8560},
archivePrefix = {arXiv},
       eprint = {1708.02587},
 primaryClass = {astro-ph.GA},
       adsurl = {https://ui.adsabs.harvard.edu/abs/2017ApJ...846...74P}
}

@ARTICLE{Besla2012,
       author = {{Besla}, Gurtina and {Kallivayalil}, Nitya and {Hernquist}, Lars and {van der Marel}, Roeland P. and {Cox}, T.~J. and {Kere{\v{s}}}, Du{\v{s}}an},
        title = "{The role of dwarf galaxy interactions in shaping the Magellanic System and implications for Magellanic Irregulars}",
      journal = {\mnras},
         year = 2012,
        month = apr,
       volume = {421},
       number = {3},
        pages = {2109-2138},
          doi = {10.1111/j.1365-2966.2012.20466.x},
archivePrefix = {arXiv},
       eprint = {1201.1299},
 primaryClass = {astro-ph.GA},
       adsurl = {https://ui.adsabs.harvard.edu/abs/2012MNRAS.421.2109B}
}

@ARTICLE{Bullock2017,
       author = {{Bullock}, James S. and {Boylan-Kolchin}, Michael},
        title = "{Small-Scale Challenges to the {\ensuremath{\Lambda}}CDM Paradigm}",
      journal = {\araa},
         year = 2017,
        month = aug,
       volume = {55},
       number = {1},
        pages = {343-387},
          doi = {10.1146/annurev-astro-091916-055313},
archivePrefix = {arXiv},
       eprint = {1707.04256},
 primaryClass = {astro-ph.CO},
       adsurl = {https://ui.adsabs.harvard.edu/abs/2017ARA&A..55..343B}
}

@ARTICLE{Stierwalt2026,
       author = {{Stierwalt}, Sabrina and {Luber}, Nicholas and {Goldberg Little}, Zelie and {Privon}, George C. and {Besla}, Gurtina and {Johnson}, Kelsey E. and {Kallivayalil}, Nitya and {Patton}, David R. and {Putman}, Mary and {Simpson-Heil}, David},
        title = "{TiNy Titans H I: Discovering Satellites via H I Gas in an Isolated, Compact Group of Dwarf Galaxies}",
      journal = {\apj},
         year = 2026,
        month = jan,
       volume = {997},
       number = {1},
          eid = {132},
        pages = {132},
          doi = {10.3847/1538-4357/ae23cc},
archivePrefix = {arXiv},
       eprint = {2512.09174},
 primaryClass = {astro-ph.GA},
       adsurl = {https://ui.adsabs.harvard.edu/abs/2026ApJ...997..132S}
}

@ARTICLE{Armus1989,
       author = {{Armus}, Lee and {Heckman}, Timothy M. and {Miley}, George K.},
        title = "{Long-Slit Optical Spectroscopy of Powerful Far-Infrared Galaxies: The Nature of the Nuclear Energy Source}",
      journal = {\apj},
         year = 1989,
        month = dec,
       volume = {347},
        pages = {727},
          doi = {10.1086/168164},
       adsurl = {https://ui.adsabs.harvard.edu/abs/1989ApJ...347..727A}
}

@ARTICLE{Mihos1994,
       author = {{Mihos}, J. Christopher and {Hernquist}, Lars},
        title = "{Triggering of Starbursts in Galaxies by Minor Mergers}",
      journal = {\apjl},
         year = 1994,
        month = apr,
       volume = {425},
        pages = {L13},
          doi = {10.1086/187299},
       adsurl = {https://ui.adsabs.harvard.edu/abs/1994ApJ...425L..13M}
}

@ARTICLE{Ellison2008,
       author = {{Ellison}, Sara L. and {Patton}, David R. and {Simard}, Luc and {McConnachie}, Alan W.},
        title = "{Galaxy Pairs in the Sloan Digital Sky Survey. I. Star Formation, Active Galactic Nucleus Fraction, and the Mass-Metallicity Relation}",
      journal = {\aj},
         year = 2008,
        month = may,
       volume = {135},
       number = {5},
        pages = {1877-1899},
          doi = {10.1088/0004-6256/135/5/1877},
archivePrefix = {arXiv},
       eprint = {0803.0161},
 primaryClass = {astro-ph},
       adsurl = {https://ui.adsabs.harvard.edu/abs/2008AJ....135.1877E}
}

@ARTICLE{Thorp2019,
       author = {{Thorp}, Mallory D. and {Ellison}, Sara L. and {Simard}, Luc and {S{\'a}nchez}, Sebastian F. and {Antonio}, Braulio},
        title = "{Spatially resolved star formation and metallicity profiles in post-merger galaxies from MaNGA}",
      journal = {\mnras},
         year = 2019,
        month = jan,
       volume = {482},
       number = {1},
        pages = {L55-L59},
          doi = {10.1093/mnrasl/sly185},
archivePrefix = {arXiv},
       eprint = {1810.00897},
 primaryClass = {astro-ph.GA},
       adsurl = {https://ui.adsabs.harvard.edu/abs/2019MNRAS.482L..55T}
}

@ARTICLE{Hani2020,
       author = {{Hani}, Maan H. and {Gosain}, Hayman and {Ellison}, Sara L. and {Patton}, David R. and {Torrey}, Paul},
        title = "{Interacting galaxies in the IllustrisTNG simulations - II: star formation in the post-merger stage}",
      journal = {\mnras},
         year = 2020,
        month = apr,
       volume = {493},
       number = {3},
        pages = {3716-3731},
          doi = {10.1093/mnras/staa459},
archivePrefix = {arXiv},
       eprint = {2001.04472},
 primaryClass = {astro-ph.GA},
       adsurl = {https://ui.adsabs.harvard.edu/abs/2020MNRAS.493.3716H}
}

@ARTICLE{Pan2019,
       author = {{Pan}, Hsi-An and {Lin}, Lihwai and {Hsieh}, Bau-Ching and {Barrera-Ballesteros}, Jorge K. and {S{\'a}nchez}, Sebasti{\'a}n F. and {Hsu}, Chin-Hao and {Keenan}, Ryan and {Tissera}, Patricia B. and {Boquien}, M{\'e}d{\'e}ric and {Dai}, Y. Sophia and {Knapen}, Johan H. and {Riffel}, Rog{\'e}rio and {Argudo-Fern{\'a}ndez}, Maria and {Xiao}, Ting and {Yuan}, Fang-Ting},
        title = "{SDSS-IV MaNGA: Spatial Evolution of Star Formation Triggered by Galaxy Interactions}",
      journal = {\apj},
         year = 2019,
        month = aug,
       volume = {881},
       number = {2},
          eid = {119},
        pages = {119},
          doi = {10.3847/1538-4357/ab311c},
archivePrefix = {arXiv},
       eprint = {1907.04491},
 primaryClass = {astro-ph.GA},
       adsurl = {https://ui.adsabs.harvard.edu/abs/2019ApJ...881..119P}
}

@ARTICLE{Stierwalt2017,
       author = {{Stierwalt}, S. and {Liss}, S.~E. and {Johnson}, K.~E. and {Patton}, D.~R. and {Privon}, G.~C. and {Besla}, G. and {Kallivayalil}, N. and {Putman}, M.},
        title = "{Direct evidence of hierarchical assembly at low masses from isolated dwarf galaxy groups}",
      journal = {Nature Astronomy},
         year = 2017,
        month = jan,
       volume = {1},
          eid = {0025},
        pages = {0025},
          doi = {10.1038/s41550-016-0025},
archivePrefix = {arXiv},
       eprint = {1701.01731},
 primaryClass = {astro-ph.GA},
       adsurl = {https://ui.adsabs.harvard.edu/abs/2017NatAs...1E..25S}
}

@ARTICLE{Mandal2024,
       author = {{Mandal}, Suvadip and {Kerton}, C.~R.},
        title = "{Constraining the interaction history of Arp 269 using stellar population synthesis modelling}",
      journal = {\mnras},
         year = 2024,
        month = apr,
       volume = {529},
       number = {2},
        pages = {953-966},
          doi = {10.1093/mnras/stae500},
       adsurl = {https://ui.adsabs.harvard.edu/abs/2024MNRAS.529..953M}
}

@ARTICLE{Bullock2005,
       author = {{Bullock}, James S. and {Johnston}, Kathryn V.},
        title = "{Tracing Galaxy Formation with Stellar Halos. I. Methods}",
      journal = {\apj},
         year = 2005,
        month = dec,
       volume = {635},
       number = {2},
        pages = {931-949},
          doi = {10.1086/497422},
archivePrefix = {arXiv},
       eprint = {astro-ph/0506467},
 primaryClass = {astro-ph},
       adsurl = {https://ui.adsabs.harvard.edu/abs/2005ApJ...635..931B}
}

@ARTICLE{Merritt2016,
       author = {{Merritt}, Allison and {van Dokkum}, Pieter and {Abraham}, Roberto and {Zhang}, Jielai},
        title = "{The Dragonfly nearby Galaxies Survey. I. Substantial Variation in the Diffuse Stellar Halos around Spiral Galaxies}",
      journal = {\apj},
         year = 2016,
        month = oct,
       volume = {830},
       number = {2},
          eid = {62},
        pages = {62},
          doi = {10.3847/0004-637X/830/2/62},
archivePrefix = {arXiv},
       eprint = {1606.08847},
 primaryClass = {astro-ph.GA},
       adsurl = {https://ui.adsabs.harvard.edu/abs/2016ApJ...830...62M}
}

@ARTICLE{Correnti2025b,
       author = {{Correnti}, Matteo and {Annibali}, Francesca and {Bellazzini}, Michele and {Marinelli}, Mariarosa and {Aloisi}, Alessandra and {Cignoni}, Michele and {Tosi}, Monica and {Pascale}, Raffaele and {Cannon}, John M. and {Schisgal}, Lila and {Hunt}, Leslie K. and {Sacchi}, Elena and {Sohn}, Sangmo Tony},
        title = "{DDO 68-C: HST Confirms Yet Another Companion of the Isolated Dwarf Galaxy DDO 68}",
      journal = {\apj},
         year = 2025,
        month = mar,
       volume = {982},
       number = {1},
          eid = {31},
        pages = {31},
          doi = {10.3847/1538-4357/adb7e6},
archivePrefix = {arXiv},
       eprint = {2502.18171},
 primaryClass = {astro-ph.GA},
       adsurl = {https://ui.adsabs.harvard.edu/abs/2025ApJ...982...31C}
}

@ARTICLE{Lahen2020,
       author = {{Lah{\'e}n}, Natalia and {Naab}, Thorsten and {Johansson}, Peter H. and {Elmegreen}, Bruce and {Hu}, Chia-Yu and {Walch}, Stefanie and {Steinwandel}, Ulrich P. and {Moster}, Benjamin P.},
        title = "{The GRIFFIN Project{\textemdash}Formation of Star Clusters with Individual Massive Stars in a Simulated Dwarf Galaxy Starburst}",
      journal = {\apj},
         year = 2020,
        month = mar,
       volume = {891},
       number = {1},
          eid = {2},
        pages = {2},
          doi = {10.3847/1538-4357/ab7190},
archivePrefix = {arXiv},
       eprint = {1911.05093},
 primaryClass = {astro-ph.GA},
       adsurl = {https://ui.adsabs.harvard.edu/abs/2020ApJ...891....2L}
}

@ARTICLE{Lahen2019,
       author = {{Lah{\'e}n}, Natalia and {Naab}, Thorsten and {Johansson}, Peter H. and {Elmegreen}, Bruce and {Hu}, Chia-Yu and {Walch}, Stefanie},
        title = "{The Formation of Low-metallicity Globular Clusters in Dwarf Galaxy Mergers}",
      journal = {\apjl},
         year = 2019,
        month = jul,
       volume = {879},
       number = {2},
          eid = {L18},
        pages = {L18},
          doi = {10.3847/2041-8213/ab2a13},
archivePrefix = {arXiv},
       eprint = {1905.09840},
 primaryClass = {astro-ph.GA},
       adsurl = {https://ui.adsabs.harvard.edu/abs/2019ApJ...879L..18L}
}

@ARTICLE{Elmegreen1998,
       author = {{Elmegreen}, Debra Meloy and {Chromey}, Frederick R. and {Knowles}, Benjamin D. and {Wittenmyer}, Robert A.},
        title = "{Observations of a Tidal Tail in the Interacting Galaxies NGC 4485/4490}",
      journal = {\aj},
         year = 1998,
        month = apr,
       volume = {115},
       number = {4},
        pages = {1433-1437},
          doi = {10.1086/300301},
       adsurl = {https://ui.adsabs.harvard.edu/abs/1998AJ....115.1433E}
}

@ARTICLE{Clemens1999,
       author = {{Clemens}, M.~S. and {Alexander}, P. and {Green}, D.~A.},
        title = "{The star formation history in the interacting galaxies NGC 4490 and 4485}",
      journal = {\mnras},
         year = 1999,
        month = aug,
       volume = {307},
       number = {3},
        pages = {481-490},
          doi = {10.1046/j.1365-8711.1999.02594.x},
       adsurl = {https://ui.adsabs.harvard.edu/abs/1999MNRAS.307..481C}
}

@ARTICLE{Clemens2000,
       author = {{Clemens}, M.~S. and {Alexander}, P. and {Green}, D.~A.},
        title = "{Ram-pressure stripping of the interstellar medium in NGC 4485}",
      journal = {\mnras},
         year = 2000,
        month = feb,
       volume = {312},
       number = {2},
        pages = {236-246},
          doi = {10.1046/j.1365-8711.2000.03069.x},
       adsurl = {https://ui.adsabs.harvard.edu/abs/2000MNRAS.312..236C}
}

@ARTICLE{Lee2025b,
       author = {{Lee}, Abigail J. and {Weisz}, Daniel R. and {Ren}, Yi and {Savino}, Alessandro and {Dolphin}, Andrew E.},
        title = "{Measuring Star Formation Histories from Asymptotic Giant Branch Stars. I. A Demonstration in M31}",
      journal = {\apj},
         year = 2025,
        month = dec,
       volume = {995},
       number = {2},
          eid = {135},
        pages = {135},
          doi = {10.3847/1538-4357/ae12f0},
archivePrefix = {arXiv},
       eprint = {2410.09256},
 primaryClass = {astro-ph.GA},
       adsurl = {https://ui.adsabs.harvard.edu/abs/2025ApJ...995..135L}
}

@article{astropy:2013,
Adsurl = {http://adsabs.harvard.edu/abs/2013A%26A...558A..33A},
Archiveprefix = {arXiv},
Author = {{Astropy Collaboration} and {Robitaille}, T.~P. and {Tollerud}, E.~J. and {Greenfield}, P. and {Droettboom}, M. and {Bray}, E. and {Aldcroft}, T. and {Davis}, M. and {Ginsburg}, A. and {Price-Whelan}, A.~M. and {Kerzendorf}, W.~E. and {Conley}, A. and {Crighton}, N. and {Barbary}, K. and {Muna}, D. and {Ferguson}, H. and {Grollier}, F. and {Parikh}, M.~M. and {Nair}, P.~H. and {Unther}, H.~M. and {Deil}, C. and {Woillez}, J. and {Conseil}, S. and {Kramer}, R. and {Turner}, J.~E.~H. and {Singer}, L. and {Fox}, R. and {Weaver}, B.~A. and {Zabalza}, V. and {Edwards}, Z.~I. and {Azalee Bostroem}, K. and {Burke}, D.~J. and {Casey}, A.~R. and {Crawford}, S.~M. and {Dencheva}, N. and {Ely}, J. and {Jenness}, T. and {Labrie}, K. and {Lim}, P.~L. and {Pierfederici}, F. and {Pontzen}, A. and {Ptak}, A. and {Refsdal}, B. and {Servillat}, M. and {Streicher}, O.},
Doi = {10.1051/0004-6361/201322068},
Eid = {A33},
Eprint = {1307.6212},
Journal = {\aap},
Month = oct,
Pages = {A33},
Primaryclass = {astro-ph.IM},
Title = {{Astropy: A community Python package for astronomy}},
Volume = 558,
Year = 2013}

@ARTICLE{NumPy2020,
  author  = {Harris, Charles R. and Millman, K. Jarrod and
            van der Walt, Stéfan J and Gommers, Ralf and
            Virtanen, Pauli and Cournapeau, David and
            Wieser, Eric and Taylor, Julian and Berg, Sebastian and
            Smith, Nathaniel J. and Kern, Robert and Picus, Matti and
            Hoyer, Stephan and van Kerkwijk, Marten H. and
            Brett, Matthew and Haldane, Allan and
            Fernández del Río, Jaime and Wiebe, Mark and
            Peterson, Pearu and Gérard-Marchant, Pierre and
            Sheppard, Kevin and Reddy, Tyler and Weckesser, Warren and
            Abbasi, Hameer and Gohlke, Christoph and
            Oliphant, Travis E.},
  title   = {Array programming with {NumPy}},
  journal = {Nature},
  year    = {2020},
  volume  = {585},
  pages   = {357–362},
  doi     = {10.1038/s41586-020-2649-2}
}

@InProceedings{Pandas2010,
  author    = { {W}es {M}c{K}inney },
  title     = { {D}ata {S}tructures for {S}tatistical {C}omputing in {P}ython },
  booktitle = { {P}roceedings of the 9th {P}ython in {S}cience {C}onference },
  pages     = { 56 - 61 },
  year      = { 2010 },
  editor    = { {S}t\'efan van der {W}alt and {J}arrod {M}illman },
  doi       = { 10.25080/Majora-92bf1922-00a }
}

@article{Matplotlib2007,
  title={Matplotlib: A 2D graphics environment},
  author={Hunter, John D},
  journal={Computing in science \& engineering},
  volume={9},
  number={03},
  pages={90--95},
  year={2007},
  publisher={IEEE Computer Society}
}

@inproceedings{Joye2003,
  title={New features of SAOImage DS9},
  author={Joye, W Aꎬ and Mandel, E},
  booktitle={Astronomical data analysis software and systems XII},
  volume={295},
  pages={489},
  year={2003}
}
\bibliographystyle{aasjournalv7}



\end{document}